\documentclass[aps,prb, floatfix,notitlepage,nofootinbib,twocolumn,superscriptaddress,reprint]{revtex4-2}
\usepackage{amsmath}
\usepackage{amssymb}
\usepackage{mathtools}
\usepackage{graphicx}
\usepackage[tight]{subfigure}
\usepackage{enumitem}
\usepackage{soul}
\usepackage{pifont}
\usepackage{xspace}
\usepackage{makecell}
\newcommand{\xmark}{\ding{55}}
\newcommand{\cmark}{\ding{51}}
\usepackage{multirow}

\usepackage[english]{babel}
\makeatletter
\def\bbl@set@language#1{%
  \edef\languagename{%
    \ifnum\escapechar=\expandafter`\string#1\@empty
    \else\string#1\@empty\fi}%
  \@ifundefined{babel@language@alias@\languagename}{}{%
    \edef\languagename{\@nameuse{babel@language@alias@\languagename}}%
  }%
  \select@language{\languagename}%
  \expandafter\ifx\csname date\languagename\endcsname\relax\else
    \if@filesw
      \protected@write\@auxout{}{\string\select@language{\languagename}}%
      \bbl@for\bbl@tempa\BabelContentsFiles{%
        \addtocontents{\bbl@tempa}{\xstring\select@language{\languagename}}}%
      \bbl@usehooks{write}{}%
    \fi
  \fi}
\newcommand{\DeclareLanguageAlias}[2]{%
  \global\@namedef{babel@language@alias@#1}{#2}%
}
\makeatother
\DeclareLanguageAlias{en}{english}

\newcommand{\I}{\mathbb{I}}
\newcommand{\tr}{\text{tr}}

\usepackage{bm}
\renewcommand{\vec}[1]{\boldsymbol{\mathbf{#1}}}

\newcommand{\mone}[0]{Model 1\xspace}
\newcommand{\modelN}[0]{Model $N$\xspace}
\newcommand{\monep}[0]{Model 1${}^{+}$\xspace}

\usepackage[pdfusetitle,colorlinks=true,citecolor=blue,linkcolor=magenta]{hyperref}
\usepackage[capitalise,compress]{cleveref}
\usepackage{braket}
\graphicspath{{./}{./images/}}

\def\bea{\begin{equation}\begin{aligned}}
\def\eea{\end{aligned}\end{equation}}
\usepackage[normalem]{ulem}

\begin{document}

\title{Hydrodynamic theory of scrambling in chaotic long-range interacting systems}

\author{Tianci Zhou}
\email{tzhou13@mit.edu}
\affiliation{Center for Theoretical Physics, Massachusetts Institute of Technology, Cambridge, Massachusetts 02139, USA}
\author{Andrew Guo}
\email{guoa@umd.edu}
\affiliation{Joint Center for Quantum Information and Computer Science, NIST/University of Maryland, College Park, MD 20742, USA}
\affiliation{Joint Quantum Institute, NIST/University of Maryland, College Park, MD 20742, USA}
\author{Shenglong Xu}
\affiliation{Department of Physics \& Astronomy, Texas A\&M University, College Station, Texas 77843, USA}
\author{Xiao Chen}
\email{chenaad@bc.edu}
\affiliation{Department of Physics, Boston College, Chestnut Hill, MA 02467, USA}
\author{Brian Swingle}
\affiliation{Brandeis University, Waltham, MA 02453, USA}

\date{\today}

\begin{abstract}
The Fisher-Kolmogorov-Petrovsky-Piskunov (FKPP) equation provides a mean-field theory of out-of-time-ordered commutators in locally interacting quantum chaotic systems at high energy density; in the systems with power-law interactions, the corresponding fractional-derivative FKPP equation provides an analogous mean-field theory. However, the fractional FKPP description is potentially subject to strong quantum fluctuation effects, so it is not clear a priori if it provides a suitable effective description for generic chaotic systems with power-law interactions. Here we study this problem using a model of coupled quantum dots with interactions decaying as $\frac{1}{r^{\alpha}}$, where each dot hosts $N$ degrees of freedom. The large $N$ limit corresponds to the mean-field description, while quantum fluctuations contributing to the OTOC can be modeled by $\frac{1}{N}$ corrections consisting of a cutoff function and noise. Within this framework, we show that the parameters of the effective theory can be chosen to reproduce the butterfly light cone scalings that we previously found for $N=1$ and generic finite $N$. In order to reproduce these scalings, the fractional index $\mu$ in the FKPP equation needs to be shifted from the na\"ive value of $\mu = 2\alpha - 1$ to a renormalized value $\mu = 2\alpha - 2$. We provide supporting analytic evidence for the cutoff model and numerical confirmation for the full fractional FKPP equation with cutoff and noise.
\end{abstract}

\maketitle
\tableofcontents

\section{Introduction}
\label{sec:intro}

Unitary dynamics in chaotic many-body systems scrambles quantum information, which can then no longer be accessed by local measurements. Recent interest in this physics was stimulated by work on scrambling in black holes~\cite{shenker_black_2014}, which turn out to be the fastest scramblers~\cite{maldacena_bound_2015} with all degrees of freedom strongly interacting with each other. Scrambling there takes the form of exponential growth, resembling the Lyapunov behavior characteristic of classical chaos. The same phenomenology can also be observed in other solvable all-to-all interacting systems such as the Sachdev-Ye-Kitaev model~\cite{sachdev_gapless_1993,kitaev2015}. When spatial structure is present, the scrambling time required to spread quantum information to a remote location may be extensive in the system size. Intuitively, scrambling can be understood as a classical epidemic spreading process in space~\cite{nahum_operator_2018,von_keyserlingk_operator_2018,zhou_operator_2020,xu_locality_2018,chen_quantum_2018,zhou_operator_2018,roberts_operator_2018,qi_quantum_2018}.


A convenient tool to quantify operator spreading is the out-of-time-ordered commutator. On a lattice, it is defined by the formula
\begin{equation}
C(x, t) = \tr( [ W(t) , V]^{\dagger} [ W(t), V] ) / \tr( \I ).
\end{equation}
Here $W(t)$ is a time-evolved operator initially located at site $0$, and $V$ is an operator located at $x$ which probes the component of $W(t)$ at that site. The interacting Hamiltonian evolves the operator $W$ such that its support gradually increases to reach that of $V$, creating a non-zero value for the commutator. Although the OTOC is dramatically different from conventional time-ordered correlators, the research on quantum chaos in the past few years has brought us new tools and viewpoints to understand the operator spreading and quantitatively compute the OTOC. In the following, we first review these tools and viewpoints in systems with local interactions.

One powerful perspective---the stochastic approach---is to treat the operator spreading as a classical stochastic spreading process. Earlier works justified the mapping to the stochastic process at high temperature or high energy density by solving for the randomly averaged value of $C(x,t)$ for evolution made of random unitary gates~\cite{nahum_operator_2018,von_keyserlingk_operator_2018}, or Hamiltonians with noisy interactions~\cite{zhou_operator_2020,xu_locality_2018,chen_quantum_2018,zhou_operator_2018}. When the system has local interactions, the spreading process for the average profile of $C(x,t)$ has a linear propagation speed $v_B$, which is called the butterfly velocity owning to its chaos interpretation.

\begin{figure}[h]
\centering
\includegraphics[width=\columnwidth]{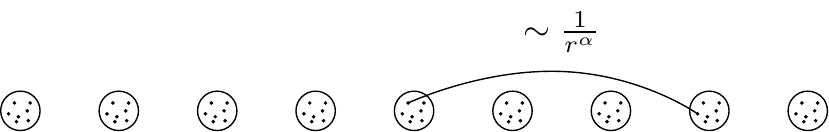}
\caption{Schematic picture for a large $N$ quantum dot model. Each dot hosts $N$ spins. Spins at different dot interact through a power-law decaying interactions.}
\label{fig:q_dot}
\end{figure}

An alternative large-$N$ approach is to view the profile of $C(x,t)$ as a propagating wave. The associated wave equation (for systems with local interactions) is the (noisy) Fisher-Kolmogorov-Petrovsky-Piskunov (FKPP) equation~\cite{fisher_wave_1937,kolmogorov_investigation_1937} (for a recent review, see Ref.~\onlinecite{brunet_aspects_2016})
\begin{equation}
\label{eq:fkpp}
\begin{aligned}
\partial_t h =& D \Delta h + \gamma h ( 1 - h)   + \eta\sqrt{\frac{\gamma h ( 1 - h)}{N}}.
\end{aligned}
\end{equation}
In this equation the height variable $h(x,t)$ is proportional to $C(x,t)$ (to be defined in Sec.~\ref{sec:otoc_finite_N}), $\Delta$ is the Laplacian operator modeling diffusion, $\gamma$ is the reaction strength and $\eta(x,t)$ is the standard Gaussian noise with correlation $\langle \eta(x,t) \eta(x', t') \rangle = \delta( x - x') \delta(t - t')$. The parameter $N$ is assumed to be large (but finite) and can be identified as the number of degrees of freedom on each lattice site (see each quantum dot in Fig.~\ref{fig:q_dot}).
The FKPP equation with a diffusion term for the OTOC has been (heuristically) derived for a variety of models, including electrons with various interactions via augmented Keldysh formalism and random averaging over quantum circuits or noisy evolution~\cite{chen_operator_2018,chen_quantum_2018,nahum_operator_2018,von_keyserlingk_operator_2018,rakovszky_diffusive_2017,khemani_operator_2017}.
With the $\frac{1}{N}$ noise, the FKPP equation in \cref{eq:fkpp} is believed to hold for OTOCs of generic quantum systems with many (but finite) local degrees of freedom \cite{chen_quantum_2018,zhou_operator_2020,aleiner_microscopic_2016}.

In the mean-field limit ($N =\infty$), the deterministic terms represent diffusion (with constant $D$) and local growth (with strength $\gamma$), which combine to create a bilaterally propagating wave with constant front velocity $v_B$~\cite{fisher_wave_1937,kolmogorov_investigation_1937,ablowitz_explicit_1979,brunet_aspects_2016}. At finite $N$, the term with spacetime Gaussian noise $\eta(x,t)$ models the $\frac{1}{N}$ fluctuations. It generates diffusively broadened wave fronts.
Thus, the noisy FKPP equation reproduces the phenomenology observed in the $N = 1$ case described by the stochastic approach~\cite{aleiner_microscopic_2016,chen_operator_2018,chen_quantum_2018,nahum_operator_2018}.

There have been many proposals and experiments to measure the dynamical behaviors of OTOCs \cite{mi_information_2021-1,li_measuring_2017,Bollinger17,landsman_verified_2019,yao_interferometric_2016-1,meier_exploring_2019,schnell_high-resolution_2001,sanchez_clustering_2014,sanchez_perturbation_2020, joshi_quantum_2020} on various quantum simulators. Most analog quantum simulators,
such as the Rydberg atom arrays \cite{Lukin17}, nuclear magnetic resonance (NMR~\cite{schnell_high-resolution_2001,sanchez_clustering_2014,sanchez_perturbation_2020}), trapped ions \cite{Blatt12,Britton12,joshi_quantum_2020}, have long-range power-law decaying interaction $~1/r^\alpha$. In these systems with a wider range of interactions between the constituent degrees of freedom,  scrambling is no longer bounded by a characteristic velocity and can achieve super-ballistic spreading. Also, the number of degrees $N$ per site can be much greater than $1$, depending on the material and its fine structure used for engineering the spins.
For instance, recent experiments have measured OTOCs in NMR systems on solid adamantane with $N = 16$ \cite{sanchez_clustering_2014,sanchez_perturbation_2020}.
The correlated-spin-cluster size has reached $10^3$ or even $10^4$, requiring a many-body analysis.

A minimal model to study scrambling in these systems is a quantum dot model where each site has $N$ spins and the interaction decays as $\frac{1}{r^{\alpha}}$ shown in Fig.~\ref{fig:q_dot}. The key phenomenology in the spreading with long-range interaction is the scaling of the butterfly light cone, which is defined to be the spacetime contour of $C(x,t) $ at a small fixed threshold $\epsilon$.

In our previous work~\cite{zhou_operator_2020}, we analytically solve the average values of the OTOC in a noisy-interacting spin model (using the stochastic approach) with $N = 1$, whose light cone scalings are displayed in Tab.~\ref{tab:lc_scaling}.
\begin{table}
\centering
\begin{tabular}{ |c|c| }
 \hline
 $\alpha$ & butterfly light cone scaling  \\ \hline
  $(\frac{d}{2}, d)$  & $\exp(t^{\eta})$ \\\hline
  $(d, d+ \frac{1}{2})$  & $t^{\frac{1}{\zeta}}$ \\\hline
  $( d+ \frac{1}{2}, \infty )$ & $t$ \\
 \hline
\end{tabular}
\caption{The sizes of the (butterfly) light cones in terms of the $d$-dimensional quantum dot model at $N = 1$, see Fig.~\ref{fig:phase_diagram} for more details.}
\label{tab:lc_scaling}
\end{table}
In the direction of increasing $\alpha$ starting from $\frac{d}{2}$, we have light cones with shapes that are logarithmic (spatial size proportional to a stretched exponential), power-law and linear functions of $x$.

Generalizing to experimentally relevant, finite-$N$ cases follows two routes. Continuing the arguments from the stochastic model~\cite{zhou_operator_2020}, we can imagine grouping $N$ neighboring spins to form a special finite-$N$ model. This model has  light-cone scalings identical to the $N = 1$ case (although the non-universal constants in the precise scaling of the contours may be $N$-dependent). This hints that results in Tab.~\ref{tab:lc_scaling} also apply to $N > 1$.

On the other hand, one can also approach the problem from the infinite-$N$ limit, and then add back the finite-$N$ correction. At infinite $N$, the OTOC obeys a mean-field equation
\begin{equation}
\label{eq:ffkpp}
\partial_t h = D_{\mu} \Delta^{\frac{\mu}{2}}  h + \gamma h ( 1 - h).
\end{equation}
Here $\Delta^{\frac{\mu}{2}}$ is a fractional derivative, generalizing the diffusion term in Eq.~\ref{eq:fkpp}. It has a $-|k^{\mu}|$ kernel in the momentum space (see the precise real-space definition in Eq.~\eqref{eq:f_deri}). This is the fractional FKPP equation. At the infinite-$N$ limit, numerics in Ref.~\cite{chen_quantum_2018} show that Eq.~\ref{eq:ffkpp} has an asymptotic logarithmic light cone~(causal region exponentially large in $t$) even for arbitrarily large exponent $\alpha$, and that there is no linear-light-cone regime.

The discrepancy between the $N=\infty$ results and predictions from $N = 1$ theory for finite $N$ suggests that a $\frac{1}{N}$ correction should play a crucial role in determining the operator spreading phenomenology. And, since it will need to change the light-cone phase diagrams away from the $N = \infty$ case, the $1/N$ correction will not merely be perturbative.

As such, we may ask the question, how can we incorporate the $\frac{1}{N}$ fluctuation in Eq.~\eqref{eq:ffkpp}? Conventional wisdom suggests two ways. First, we note that the variable $h$ thus constructed for the problem of OTOC (and other problems leading to the FKPP equation, more details in Sec.~\ref{sec:otoc_finite_N}) is discrete as an integer multiple of $\frac{1}{N}$. Instances with $h < \frac{1}{N}$ on a given site actually have no activity at that site. This justifies modifying the dynamics to cut off $h$ below $\frac{1}{N}$. This is a $\mathcal{O}(\frac{1}{N})$ correction. Another way is to add a $\frac{1}{N}$ noise term similar to the one in Eq.~\eqref{eq:fkpp}. The noise naturally models the statistical fluctuations in each instance of the stochastic process. In the quantum OTOC problem, these fluctuations represent quantum fluctuations of the size of the operator $W(t)$.

In this work, we provide a noisy hydrodynamic equation that incorporates both $1/N$ corrections, which under a certain identification of $\alpha$ and $\mu$ qualitatively matches the phenomenology of the small-$N$ model for the power-law and linear light cone regimes. We tabulate various theories considered in Tab.~\ref{tab:flowchart}.

First, we find that the butterfly light cone of the finite-$N$ coupled quantum dots is the same as the $N = 1$ case (Tab.~\ref{tab:lc_scaling} and Fig.~\ref{fig:phase_diagram}). We prove this result by squeezing it with the known $N = 1$ light cone scalings (Sec.~\ref{sec:otoc_finite_N}).

Then we investigate how to properly incorporate the finite-$N$ corrections in the mean-field FKPP equation. In Sec.~\ref{sec:mean_field}, we introduce a cutoff term and noise term derived from the leading $\mathcal{O}(\frac{1}{N})$ corrections. The relation between the superdiffusive index $\mu$ and the long-range interaction exponent $\alpha$ is $\mu = 2\alpha - 1$ at this order.
In Sec.~\ref{sec:cut-off_th}, we solve for the light-cone structures of Eq.~\eqref{eq:noisy_f_fkpp} without noise (analytically for $\gamma \rightarrow \infty$ and numerically when $\gamma$ is finite), and show that only the renormalized value $\mu = 2\alpha - 2$ gives consistent results with the $N = 1$ theory. In Sec.~\ref{sec:eff_of_noise}, we study the effects of various forms of the noise term, including spatially local and long-ranged noise. The long-range noise reproduces the wavefront broadening with a slightly smaller broadening exponent. In some choices of long-range noise, the critical point separating the linear and power-law light cone is slightly shifted for the noise obtained from the mean-field calculation. The final form of the equation that {\it partly} reproduces the operator spreading phenomenology is Eq.~\ref{eq:noisy_f_fkpp} (Sec.~\ref{sec:disc}), where we use two different exponents to model the power-law decay of the superdiffusion kernel and strength of the noise.
We discuss the origin of the renormalized relation (Sec.~\ref{sec:disc_renorm}), different noise forms and further directions (Sec.~\ref{sec:disc}).

Our results on the noisy fractional FKPP equation may also be of independent interest~\cite{brunet_phenomenological_2006,brunet_shift_1997,del-castillo-negrete_front_2003,dumortier_critical_2007,coulon_transition_2012,brockmann_front_2007,del-castillo-negrete_front_2003}, as the equation describes a large class of superdiffusive reaction processes. Our analytical results on the cutoff theory and the numerical simulation with the noise term establish a power-law light cone regime that was not explored previously~\cite{brockmann_front_2007,del-castillo-negrete_front_2003}.

\begin{table}[]
\begin{tabular}{|c|c|c|}
\hline
Theories & \makecell{Lightcone \\ scalings} & \makecell{Wavefront \\ broadening} \\ \hline
Exact mapping (Sec.~\ref{sec:otoc_finite_N}) &
\makecell{identical \\to $N = 1$}  & unknown \\ \hline
\makecell{Fokker-Planck \\ perturbation} (Sec.~\ref{sec:mean_field}) &
\xmark  & \xmark \\ \hline
\makecell{FKPP + cutoff\\ + renormalized $\mu$} (Sec.~\ref{sec:cut-off_th})&
\cmark  & \xmark \\ \hline
\makecell{FKPP + cutoff\\ + renormalized $\mu$\\ + local noise} (Sec.~\ref{sec:eff_of_noise}) & \cmark  &\xmark                    \\ \hline
\makecell{FKPP + cutoff\\ + renormalized $\mu$\\ + long-range noise} (Sec.~\ref{sec:eff_of_noise})&
\cmark (almost)  & $\sim$ \\ \hline
\end{tabular}
\label{tab:flowchart}
\caption{Structure of this paper. We tabulate various FKPP theories  and their abilities to predict the OTOC phenomenology for $\alpha > 1$. In the last row, $\sim$ means that the theory can approximately reproduce the results of $N = 1$ in numerics.}
\end{table}

\section{OTOC with finite N}
\label{sec:otoc_finite_N}

In this section, we first review our previous result on the $N = 1$ quantum dot model~\cite{zhou_operator_2020} and deduce the asymptotic light cone scalings for $N > 1$.

\begin{figure}[h]
\centering
\includegraphics[width=\columnwidth]{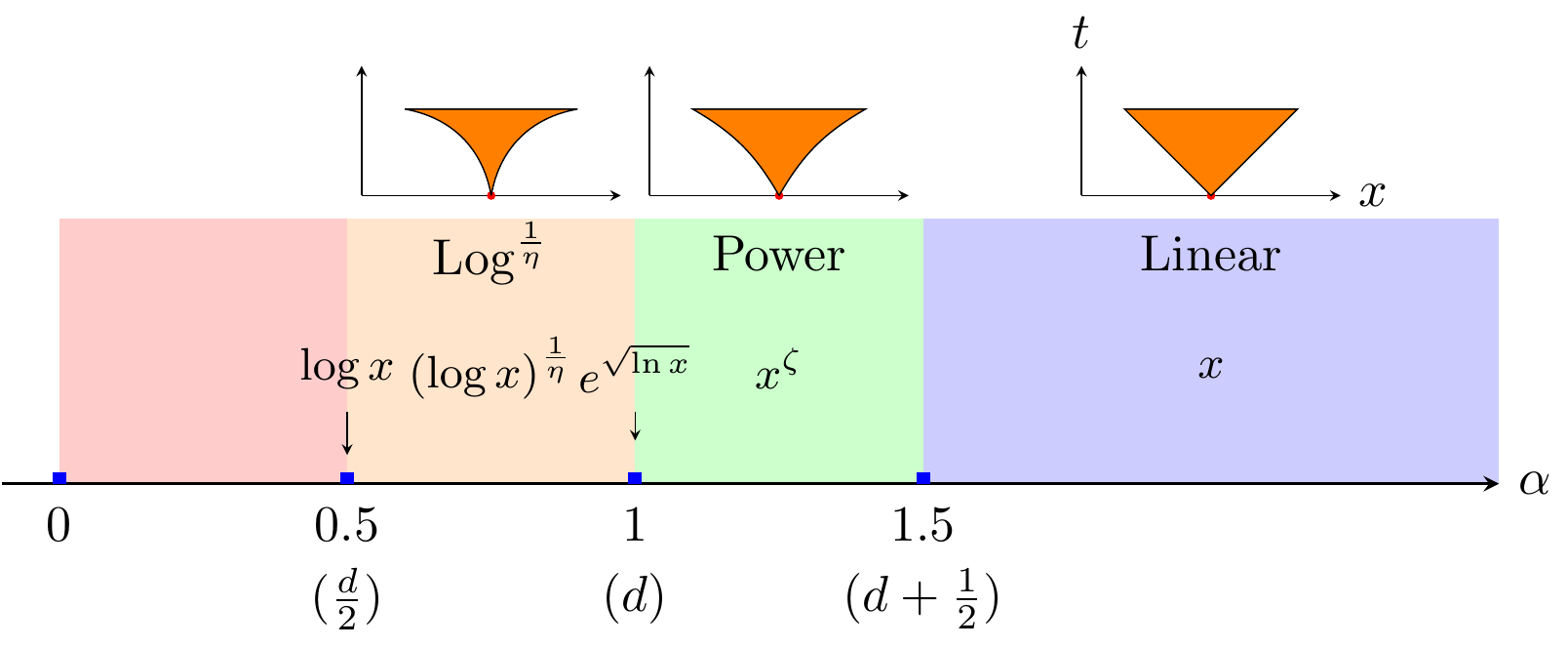}
\caption{The phase diagrams of the (butterfly) light cones for long-range coupled quantum dot at $N = 1$ (reproduction from Fig. 1 of Ref.~\onlinecite{zhou_operator_2020}, also see \cite{hallatschek_acceleration_2014} and \cite{chatterjee_multiple_2013}). Critical values of $\alpha$ in one dimension are labeled on the $\alpha$ axis, with $d$-dimensional results in the parenthesis.}
\label{fig:phase_diagram}
\end{figure}

In our previous works~\cite{chen_quantum_2018,zhou_operator_2020,xu_locality_2018}, the unitary time evolution of the OTOC is modeled as a stochastic height growth model. We substantiate this proposal by modeling the (power-law) interactions as independent Brownian motions. Then the dynamics of the OTOC is exactly a stochastic height growth process and we obtain the master equation. The phase diagram for $N = 1$ is solved and shown in Fig.~\ref{fig:phase_diagram}. Although the randomness is put by hand, we argue that the dephasing effect caused by quantum chaos can supply sufficient pseudo-randomness such that the classical stochastic model is valid in asymptotic long time. The light cone scalings of the $N = 1$ model has been numerically checked for a long-range Hamiltonian spin model in Ref.~\onlinecite{zhou_operator_2020}.

\begin{figure}[h]
\centering
\subfigure[]{
  \label{fig:model_1}
  \includegraphics[width=0.9\columnwidth]{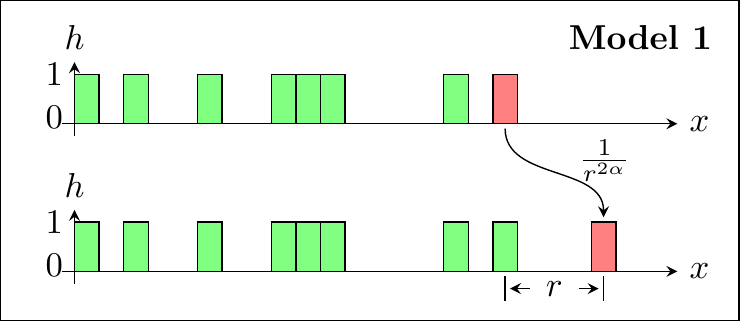}
}
\subfigure[]{
  \label{fig:model_N}
  \includegraphics[width=0.9\columnwidth]{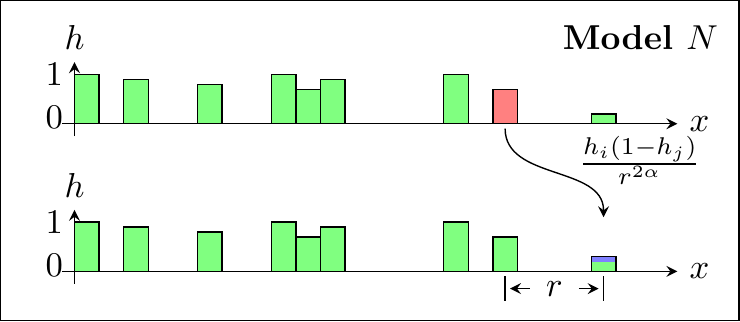}
}
\caption{The effective stochastic height growth model for the operator spreading in coupled quantum dot. Figures show the transition rate in terms of the height configurations. (a) \mone when $N = 1$. (b) \modelN for general $N$.}
\label{fig:height_model}
\end{figure}

The stochastic height model for general $N$ is defined as follows. There is a reduced height variable $h$, defined on each site that takes discrete values among $\{0, \frac{1}{N}, \frac{2}{N}, \cdots, 1\}$ (we could have defined the height to have integer values from $0$ to $N$, but the reduced height is more convenient in the continuum description below). A single site operator (non-identity) corresponds to an initial configuration with height $1$ at site $0$ and height $0$ elsewhere. The time evolution with long-range interaction performs the following. At each time slice, site $i$ contributes a rate of $\frac{3}{4} N h_i ( 1 - h_j ) D_{ij}  $ (no summation on $i, j$) to increase the height of site $j$ by $\frac{1}{N}$, and a rate of $\frac{1}{4}N h_i h_j D_{ij}$ (no summation) to decrease the height by $\frac{1}{N}$. The coefficients $D_{ij}$ decays as $\frac{1}{|i-j|^{2\alpha}}$. This is illustrated in Fig.~\ref{fig:height_model}. Microscopically, such a stochastic model can be derived exactly from the unitary Brownian circuit by random averaging~\cite{chen_quantum_2018,xu_locality_2018}.

We denote the model for $N=1$ as \mone and more generally as \modelN. In \modelN, the height takes $N+1$ discrete values between 0 and 1, while the height for \mone is binary. Their transition rates are also different.

The model can be motivated from the operator growth of the time evolved operators. The Heisenberg evolution equation tells us that only the commutator with the interaction terms can extend the support of the operator to a new spin (or retreat from that spin). The numerical coefficients $\frac{3}{4}$ and $\frac{1}{4}$ in the transition rates above is a feature of the spin-$\frac{1}{2}$ degrees of freedom---there are three Pauli matrices and one identity operators, so the rate to increase the height is $3$ times the rate to decrease the height. If we were to take $q$-dimensional spins, then the numerical value would be $(1 - \frac{1}{q^2})$ and $\frac{1}{q^2}$ respectively.
The light cone structure for different values of $q$ only differ by non-universal constants in the scaling functions.
Hence we can take $q \rightarrow \infty$, meaning only taking the rate for height increase, and write down the master equation for the height probability distribution $f(\vec{h}, t)$:
\begin{equation}
\label{eq:master_eq}
\begin{aligned}
  &\partial_t f( \vec{h}, t ) = \\
 \sum_i& N[h_i - \frac{1}{N} + \sum_{j\neq i} D_{ij} h_{j} ] ( 1 - h_i + \frac{1}{N} ) f( \vec{h} - \frac{1}{N}\vec{e}_i , t )  \\
 &-N [h_i  + \sum_{j} D_{ij} h_j ] ( 1 - h_i ) f( \vec{h}  , t )
\end{aligned}
\end{equation}
where $D_{ij}$ describes the (super)diffusion between different sites,
\begin{equation}
\label{eq:D_ij}
D_{ij}
 = \left\lbrace
   \begin{aligned}
     & \delta_{i\pm 1,j}  & \quad \text{nearest neighbor interaction} \\
     & \frac{1}{|i-j|^{2\alpha}} & \quad \text{long-range interaction} \\
   \end{aligned} \right.
\end{equation}
The equation is almost self-explanatory. The coefficient $(h_i - \frac{1}{N})( 1 - h_i + \frac{1}{N} ) $ and $(\sum_{j\neq i} D_{ij} h_{j} )( 1 - h_i + \frac{1}{N} )$ in front of $f( \vec{h} - \frac{1}{N}\vec{e}_i , t ) $ correspond to the rate of increasing the heights at the same and different sites. The other term with $f( \vec{h}  , t )$ serves to conserve the total probability.

The OTOC will typically behave as the average height on each site with respect to this height distribution. The height profile of $h = 1$ generally expands. Thus the locus $h( x_{\rm LC}(t), t)$ defines the light-cone structure.

For long-range interacting systems, several different light-cone scalings occur, depending on the exponent $\alpha$ in the interaction. In Ref.~\onlinecite{zhou_operator_2020}, we worked out the exact phase diagram for the light cone scaling of \mone, see Fig.~\ref{fig:phase_diagram} (also see \cite{hallatschek_acceleration_2014} and \cite{chatterjee_multiple_2013} in the language of long-range dispersal and percolation).

For general $N$, schematically we have
\begin{equation}
\text{\mone} \le \text{\modelN} \le N \text{\mone}.
\end{equation}
This series of inequalities denotes that \modelN spreads faster than \mone, while spreads comparatively slower if we increase the rate of \mone by a factor $N$. These bounds are intuitively clear if we compare their rates. The lower bound is simpler. We restrict \modelN so that the height can only take values of $0$ and $\frac{1}{N}$. This gives a much smaller transition rate which meanwhile is exactly the rate of \mone. Hence \modelN must spreads faster than \mone. To obtain the upper bound, we modify the rule of \modelN, so that whenever a height change occurs, the height is increased by $1$ rather than $\frac{1}{N}$. This modification apparently speeds up the spreading, and the rate is exactly $N$ times the rate of \mone. With both the upper and lower bounds, the light cone scalings of \modelN will be identical to \mone as long as $N$ is finite, although the non-universal coefficients in front of the scaling could depend on $N$.

This argument does not directly address the broadening of the front in the regime of the linear light cone, but we expect the $N \gg 1$ to behave similarly. Specifically, in one spatial dimension ($d = 1$), the broadening has the form
\begin{equation}
\text{broadening} =
\left\lbrace
\begin{aligned}
  & \frac{1}{t^{2\alpha - 2}} & \quad  \alpha \in ( 1.5, 2 ) \\
  & t^{\frac{1}{2}} & \quad \alpha \in [2, \infty) \\
\end{aligned} \right. .
\end{equation}

\begin{figure}[h]
\centering
\subfigure[]{
  \label{fig:LC_alpha_14}
  \includegraphics[width=0.8\columnwidth]{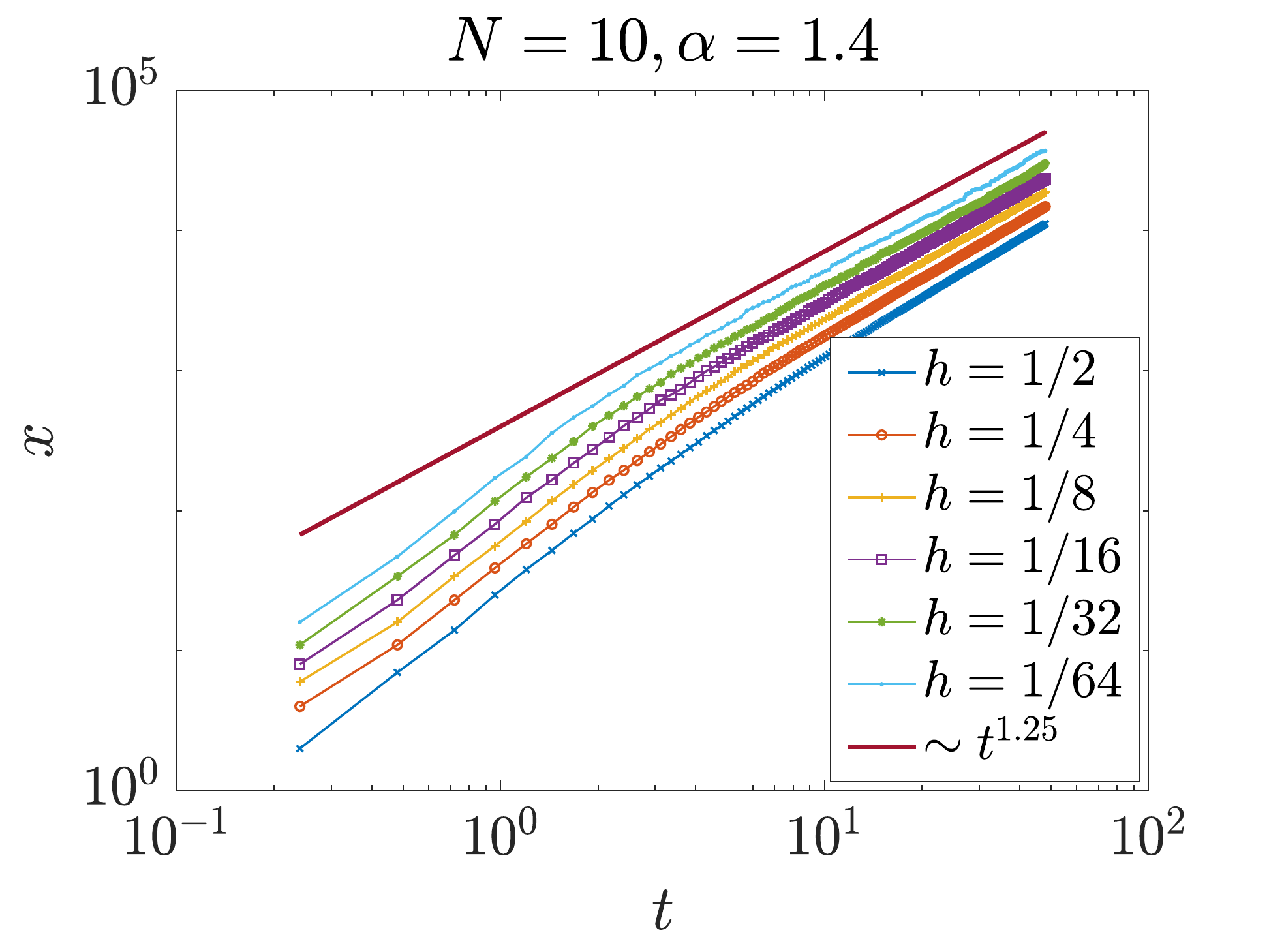}
}
\subfigure[]{
  \label{fig:Front_alpha_14}
  \includegraphics[width=0.8\columnwidth]{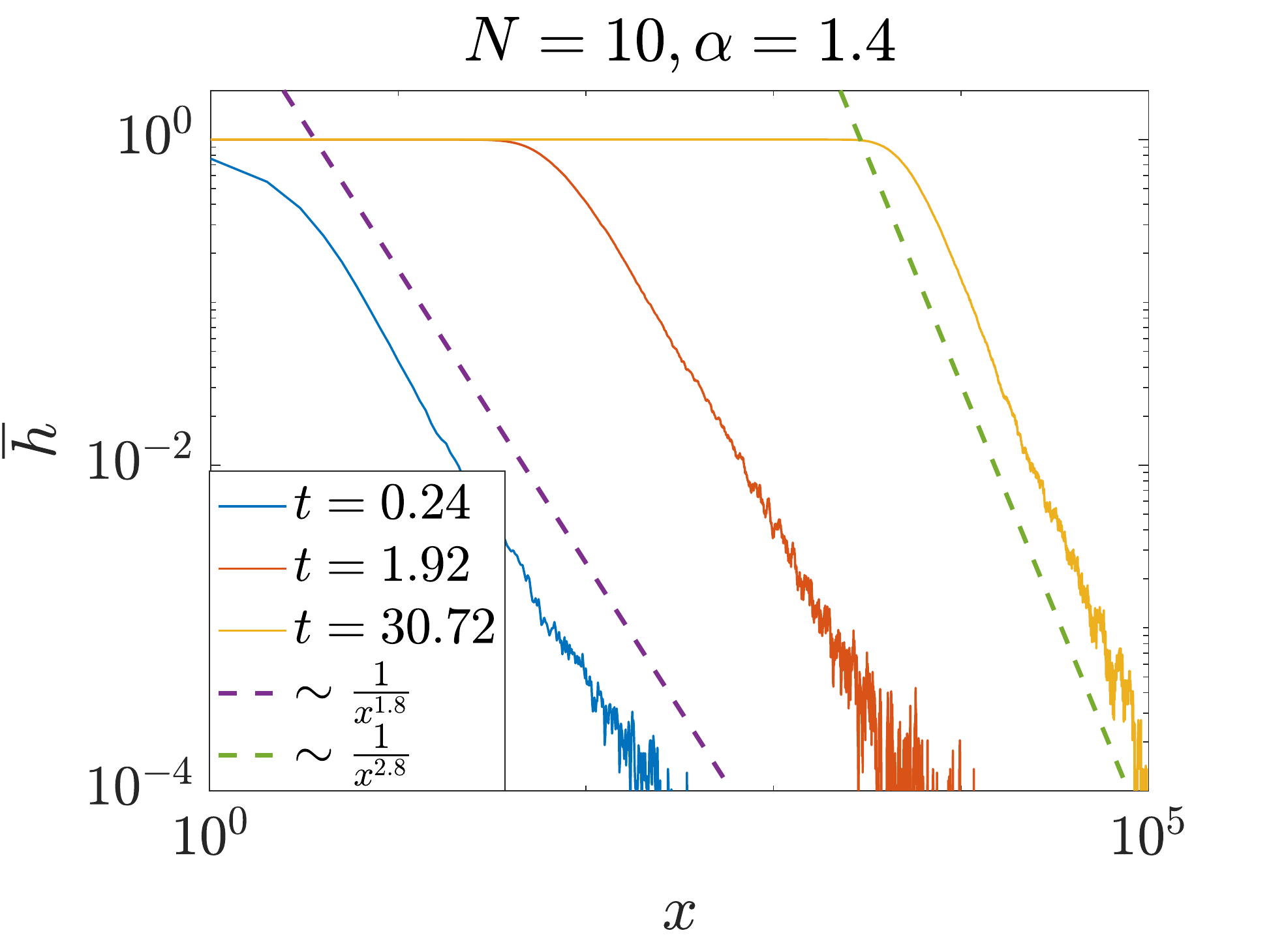}
}
\caption{The light cone scalings of \modelN (the Brownian circuit) at $\alpha = 1.4$. (a) The power-law fit of the light cone. (b) The tail of the height (i.e. the OTOC) has reached a form $\frac{1}{x^{2\alpha}}$, similar to the $N = 1$, indicating its convergence.}
\label{fig:BC_alpha_14}
\end{figure}

\begin{figure}[h]
\centering
\subfigure[]{
  \label{fig:N_40alpha_18}
  \includegraphics[width=0.8\columnwidth]{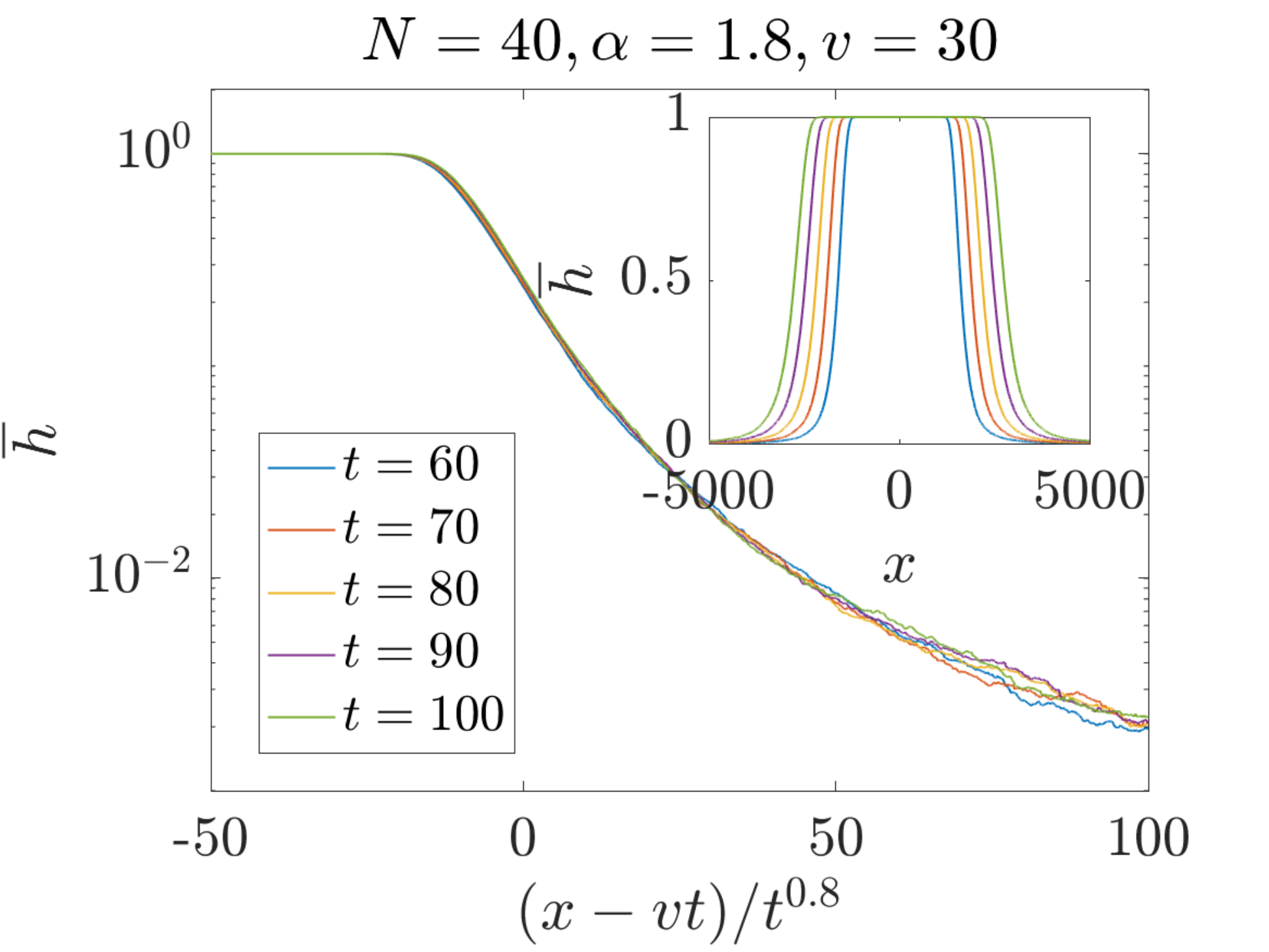}
}
\subfigure[]{
  \label{fig:N_40alpha_22}
  \includegraphics[width=0.8\columnwidth]{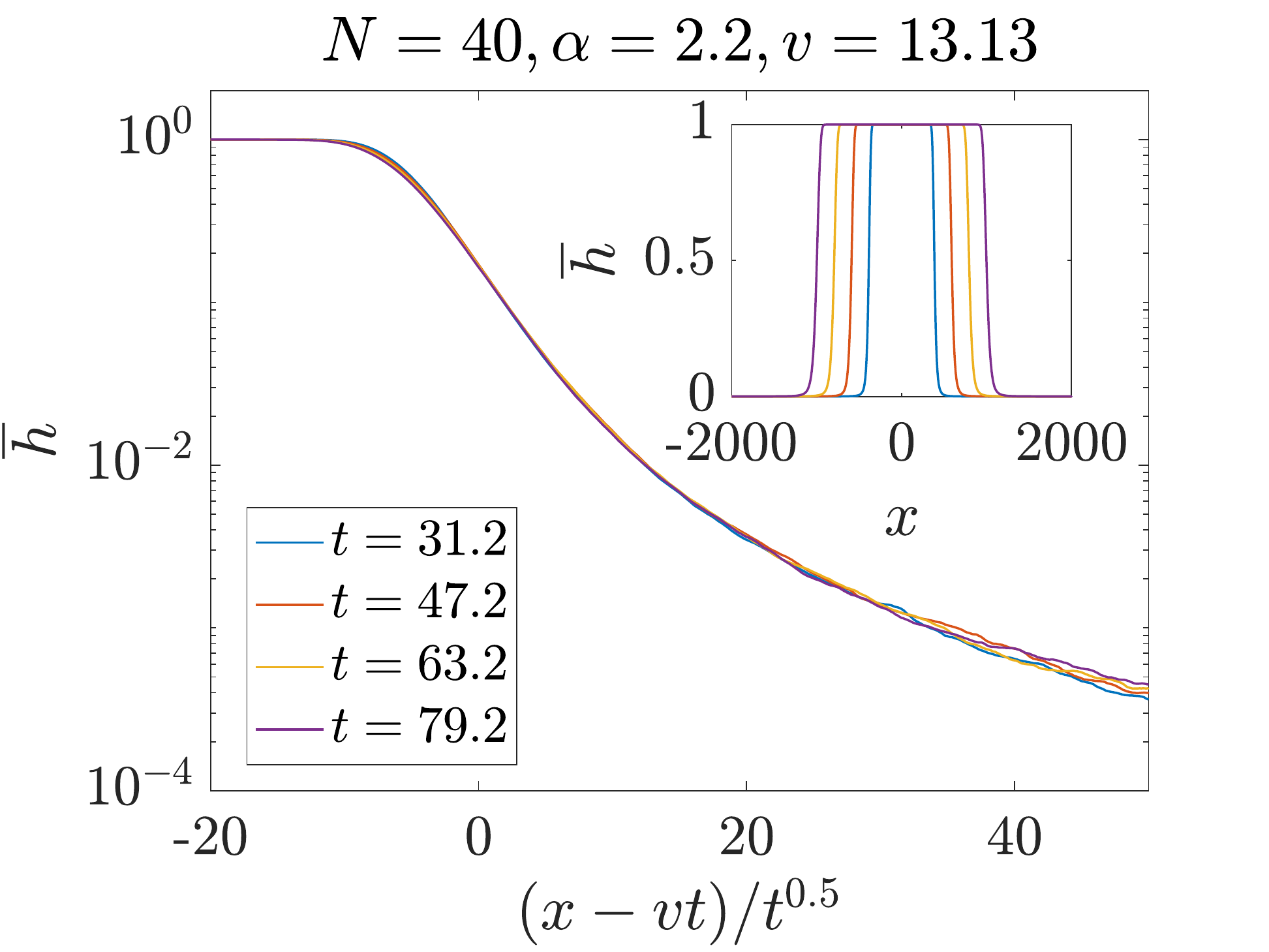}
}
\caption{Data collapses of \modelN (the Brownian circuit) for the linear-light-cone regimes ($\alpha > 1.5$) (a) $\alpha = 1.8$, fitted broadening $t^{0.8}$ (b) $\alpha = 2.2$, fitted broadening $t^{\frac{1}{2}}$. }
\label{fig:BC_collapse}
\end{figure}

We numerically verify the phase diagram by a Monte Carlo simulation of \modelN, see Fig.~\ref{fig:BC_alpha_14} and Fig.~\ref{fig:BC_collapse}. We sample a few $\alpha$s spanning the power-law and linear-light-cone regimes. In Fig.~\ref{fig:BC_alpha_14}, we show that the light cone has a power-law scaling $t^{\frac{1}{2\alpha - 2} } = t^{1.25}$ (or $x^{0.8}$) for $\alpha = 1.4$. In Fig.~\ref{fig:BC_collapse}, we show that there is a linear light cone for both $\alpha = 1.8$ and $\alpha = 2.2$. While $\alpha = 2.2$ has a diffusive broadening, the wave front for $\alpha = 1.8$ has a superdiffusive broadening, although the measured the exponent $ 0.8$ is slightly different from the theoretical prediction $\frac{1}{\zeta} = \frac{1}{2\alpha - 2} = \frac{1}{1.6} = 0.625$ for $N = 1$.

\section{The Mean-field Derivation}
\label{sec:mean_field}

A convenient way to understand a large $N$ stochastic problem is to first solve the mean-field limit $(N = \infty)$ and then incorporate the fluctuations resulting from the $\frac{1}{N}$ corrections.

For the operator spreading problem, the mean-field equation is the FKPP equation~\cite{kolmogorov_investigation_1937,fisher_wave_1937}, with the diffusion term for local interactions and superdiffusion terms for long-range interactions. We call the latter the fractional FKPP equation~\cite{del-castillo-negrete_front_2003,brockmann_front_2007,mancinelli_superfast_2002}. As alluded to in the introduction, the fluctuation plays a significant role to reduce the mean-field exponential size light cone of the fractional FKPP equation to the various light cones shapes for $N = 1$ (Fig.~\ref{fig:height_model}).

In the following, we follow the standard approach and derive a mean-field equation with $\frac{1}{N}$ corrections. In Sec.~\ref{subsec:fluc_fkkp}, we review how this is done for the (local) FKPP equations and then in Sec.~\ref{subsec:fokker_plank_eq} we generalize to the long-range case. The correction consists of the cutoff terms following the conventional wisdom in Sec.~\ref{subsec:fluc_fkkp} and a non-local noise term from the $\frac{1}{N}$ expansion of the master equation. We will study the effects of the cutoff term and noise term separately in Sec.~\ref{sec:cut-off_th} and Sec.~\ref{sec:eff_of_noise}.

\subsection{Fluctuations in the local FKPP equation}
\label{subsec:fluc_fkkp}

The FKPP equation~\cite{kolmogorov_investigation_1937,fisher_wave_1937} can model a wide range of dynamical processes, during which a stable phase of $h = 1$ can erode the other unstable phase $h = 0$. The equation has the form\footnote{When the operator spreading is local, the diffusion term has the form $(1-h) \Delta h$, numerical results show that the difference does not affect the scalings. }
\begin{equation}
\label{eq:fkpp-2}
\partial_t h =  \Delta h + \gamma ( 1- h ) h
\end{equation}
where $\Delta$ is the Laplacian. The height variable $h$ can be viewed as the density of a biological species or a certain kind of particle in other context, which diffuses and has chain reactions to proliferate. Consequently, the equation contains a linear diffusion term and a logistic type reaction term with reaction rate $\gamma$. Through linearization and other means (e.g. duality relation), one can show that there is a traveling wave solution, where the phase of $h = 1$ moves with a constant speed $v$ towards the phase of $h = 0$~\cite{ablowitz_explicit_1979}. These dynamics are usually called the pulled dynamics, since the velocity of the wave is selected by the particular form of the decay of the wavefront~\cite{ablowitz_explicit_1979}. In particular, the FKPP equation has an exponential tail stretching between $h = 1$ and $h  = 0$, and the velocity is determined by the exponent of the decay~\cite{brunet_shift_1997,brunet_aspects_2016,ablowitz_explicit_1979,fisher_wave_1937,kolmogorov_investigation_1937}.

There are multiple routes to model the fluctuations. In the stochastic process, height is a random variable and has fluctuations. Calculating the mean and variance of this random variable brings in a white-noise term with strength $\sqrt{\frac{1}{N}}$, and Eq.~\eqref{eq:fkpp-2} becomes the noisy FKPP equation \eqref{eq:fkpp} appeared in the introduction:
\begin{equation}
\label{eq:noisy_fkpp_std}
\begin{aligned}
\partial_t h =& \Delta h + \gamma ( 1- h ) h   + \sqrt{\frac{(1- h)h }{N}} \eta
\end{aligned}
\end{equation}

In the context of OTOC, the noise represents the quantum fluctuations in the dynamics.

There is another potentially more important source of fluctuation---the discreteness of the height. The height can only take values in an integer multiple of $\frac{1}{N}$, which means there should be no reaction term below this cutoff. In principle it could be hidden in the $\frac{1}{N^2}$ or even higher order corrections. But an intuitive way to implement this is to study the cutoff-theory
\begin{equation}
\label{eq:local-cutoff-theory}
\partial_t h =  \Delta h +  \gamma ( 1- h ) h\theta( h - \frac{1}{N})
\end{equation}
Here the $\theta$ function implements the constraints by completely suppressing the reaction below $\frac{1}{N}$. Using this trick, Ref.~\onlinecite{brunet_shift_1997} successfully obtained the finite-$N$ corrections that slows down the mean-field velocity.

A theory that takes both of these finite-$N$ effects into account should therefore contain both terms,
\begin{equation}
\label{eq:fkpp-cutoff-noise}
\begin{aligned}
  \partial_t h =& \Delta h  + \gamma ( 1- h ) h\theta( h - \frac{1}{N})
+\sqrt{\frac{\gamma( 1- h )h }{N} } \eta.
\end{aligned}
\end{equation}
Ref.~\onlinecite{brunet_phenomenological_2006} took this point of view and developed a phenomenological theory in which the noise gives a pulse-like disturbance to the (tail of the) front. The disturbance increases the velocity and gives rise to a diffusive broadening of the wavefront, due to the sample to sample velocity fluctuation around its mean value.

To summarize, the finite-$N$ effects introduce fluctuations of the wavefront, which is taken care by a combinations of the cutoff scheme and noise terms in Eq.~\eqref{eq:fkpp-cutoff-noise}. This theory correctly reproduces the OTOC phenomenology in systems with local interaction: we observe a traveling wave whose front is broadened as $\sqrt{t}$.

\subsection{Fluctuations in the long-range FKPP equation}
\label{subsec:fokker_plank_eq}
A natural generalization of Eq.~\eqref{eq:fkpp-2} to the long-range interactions is the {\it fractional} FKPP equation
\begin{align}
\label{eq:lr_fkpp_simple_noiseless}
\partial_t h  =& \Delta^{\mu/2}h + \gamma h(1-h)
\end{align}
where $\Delta^{\mu/2}$ is the fractional derivative given by
\begin{equation}
\label{eq:f_deri}
 \frac{\partial}{\partial |x|^{\mu}} f(x) \propto - \int_{\infty}^{\infty} \frac{f(x) - f(y) }{|x - y|^{1+ \mu} } dy
\end{equation}
in real space and $|k|^{\mu}$ in the momentum space. $0< \mu < 2$ models the long-range interaction, while $\mu=2$ reduces to the regular Laplacian. The fractional FKPP equation is a minimal model for many superdiffusive stochastic processes at the mean-field limit. It is known that this equation produces exponentially accelerating wavefront, while a direct numerical simulations of the underlying stochastic process can produce linearly growing light cones. The discrepancy is due to missing fluctuation effects when $h$ is small. To account for the fluctuation, Ref.~\cite{brockmann_front_2007} includes both cutoff approximation and noise in the mean-field equation, similar to the short range case
\bea
\label{eq:lr_fkpp_simple}
\partial_t h  =& \Delta^{\mu/2}h + \gamma h(1-h)\theta(h-1/N) \\
&+ \sqrt{\frac{1}{N}\gamma h(1-h) }\eta(x,t).
\eea
They found that the cutoff approximation alone stops the wave front from exponentially accelerating and leads to a finite velocity of the front dynamics for $1<\mu<2$. The velocity scales with the cutoff as $N^{1/\mu}$. To determine whether Eq.~\eqref{eq:lr_fkpp_simple} is able to reproduce the operator dynamics in the long-range Brownian circuit, one need to extend the study in Ref.~\cite{brockmann_front_2007} to the regime $0<\mu<1$ as well as investigate the role of the noise term.

\subsection{Microscopic derivation of FKPP-like equation }
\label{sec:derivation}
Before studying Eq.~\eqref{eq:lr_fkpp_simple} in detail, it is instructive to see how such an equation can arise from the microscopic master equation Eq.~\eqref{eq:master_eq}. For clarity, define
\begin{equation}
  g_i(\vec{h}) = (1 -  h_i )  [ h_i +  \sum_{j\neq i} D_{ij}h_j ]
\end{equation}
so that the discrete space master equation can be written as
\bea
  \partial_t &f( \vec{h}, t ) = \sum_i- N [ g_i( \vec{h} )f(\vec{h},t)  -  g_i( \vec{h} - \frac{\vec{e}_i}{N} ) f( \vec{h} - \frac{\vec{e}_i}{N} , t )  ]
\eea
which manifestly conserves the total probability $\sum_{\vec{h}} f$.
In the continuum height limit, the $\frac{1}{N}$ expansion leads to the Fokker-Planck equation
\begin{equation}
\begin{aligned}
  \partial_t f( \vec{h}, t )
  = \sum_i - \partial_{h_i} (g_i f)+  \frac{1}{2} \frac{1}{N} \partial^2 _{h_i}( g_i f).
\end{aligned}
\end{equation}
From the standard relation between the Langevin equation and Fokker-Plank equation, the probability distribution $f(\vec{h}, t)$ truncated to order $\frac{1}{N}$ can be generated by the solutions of the stochastic equation
\begin{equation}
\begin{aligned}
\label{eq:fkpp_discrete}
  \partial_t h_i = g_i +\sqrt{\frac{g_i}{N}}\eta_i(t)
\end{aligned}
\end{equation}
where $\eta_i(t)$ are independent Gaussian noise at each site obeying $\braket{\eta_i(t)\eta_j(t')}=\delta_{ij}\delta(t-t')$.

When the interaction is long-ranged, i.e. $D_{ij}=1/|i-j|^{2\alpha}$, the spatial continuum limit gives
\begin{equation}
\label{eq:naive_theory}
\begin{aligned}
  \partial_t h =&   (1 - h)\Delta^{\frac{\mu}{2}} h  + \gamma ( 1- h ) h  \\
 & + \sqrt{\frac{ (1 - h)\Delta^{\frac{\mu}{2}}  h + \gamma ( 1- h ) h  }{N} }\eta
\end{aligned}
\end{equation}
where $\mu = 2\alpha - 1$. To account for the discreteness of the on-site operator weight, we further implement the cutoff approximation by replacing $h$ on the right hand side of the equation with $\tilde h = h\theta(h-1/N)$, which is set to zero below a hard cutoff $1/N$. Eq.~\eqref{eq:naive_theory} shares many key features with the conventional fractional FKPP equation in Eq.~\eqref{eq:lr_fkpp_simple}, including the superdiffusion kernel and local growth term. The $(1-h)$ factor in front of the superdiffusion kernel is negligible at $h\ll 1$, a regime that determines the pulled front dynamics. The main difference between the two equation is that now the noise term also includes the superdiffusion kernel.~(Another way to justify the factor $(1-h)$ in the superdiffusion term is that it is necessary to keep the noise real). We call this form of the noise in Eq.~\eqref{eq:naive_theory} as the long-range noise and the noise in Eq.~\eqref{eq:lr_fkpp_simple} as the local noise. These different forms of noise raise the question on how the noise affects the front dynamics and whether the long-range noise is necessary to produce the observed phenomenology of the OTOC operator dynamics at small $N$, including the phase diagram of the light cone shape and the front broadening.

Before proceeding, we note that the continuum limit is valid for $\mu<2$ or $\alpha<1.5$. When $\mu>2$, there is also a normal diffusion term $\Delta h$ in the continuum limit in addition to the subleading $\Delta^{\mu/2}$. One can also directly study the discrete long-range FKPP equation in Eq.~\eqref{eq:fkpp_discrete}. It is found in~\cite{zhou_operator_2018} the light cone given by $\partial_t h_i = g_i$ without the cutoff approximation and noise is always exponential~($x\sim e^{ \# t}$) even for arbitrarily large $\alpha$, in contrast with the phase diagram in Fig.~\ref{fig:phase_diagram}. This indicates the essential role of fluctuation beyond the mean-field description, which we approximate by introducing the cutoff and the noise.

In the next two sections, we will first study the cutoff approximation of Eq.~\eqref{eq:lr_fkpp_simple} and Eq.~\eqref{eq:naive_theory} without the noise and determine the phase diagram of the light cone. Then we will discuss the role of the noise on top of the cutoff approximation.

\section{Linear and power-law light cones in the cut-off theory}
\label{sec:cut-off_th}
In this section, we inspect and analyze the light cone structure of the cut-off theory, namely setting $\eta = 0$ in Eq.~\eqref{eq:lr_fkpp_simple}, neglecting its noise. In deriving and displaying the results, we set aside the OTOC interpretation and solve the phase diagrams in terms of the FKPP parameter $\mu$ rather than the interaction parameter $\alpha$ from the OTOC problem. We analytically find a critical point $\mu  = 1$ that separates the power-law and linear light cones on its two sides.

\subsection{An effective model of the cutoff theory}
\label{subsec:eff_model}
The cutoff theory takes account of the discreteness of the height variable in the regime $h\sim 1/N$ via a cut-off function $\theta(h-1/N)$. Due to the cutoff, when below $1/N$, the height variable can only be grown by the superdiffusion term (the long-range hopping from other sites), not from the on-site reaction term. Such a mechanism can significantly delay the wavefront propagation because the height variable starting with 0 has to wait for sufficient hopping from other sites to exceed $h=1/N$ before proliferation.

On the level of cutoff approximation, Eq.~\eqref{eq:lr_fkpp_simple} and Eq.~\eqref{eq:naive_theory}  (with $\eta = 0$) are quite similar. They are equivalent when $h\sim 1/N$, a regime that is important to determine the front dynamics. The factor $(1-h)$ in front of the superdiffusion kernel in Eq.~\eqref{eq:naive_theory} only becomes significant at $h\sim 1$, which we expect cannot qualitatively affect the front dynamics.

\begin{figure}[h]
\centering
\includegraphics[width=0.8\columnwidth]{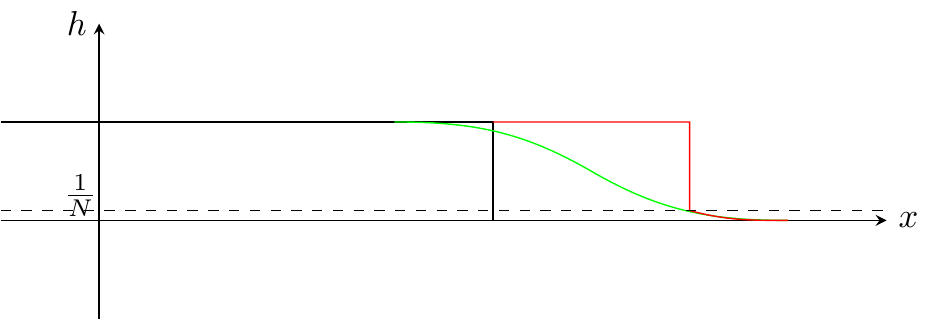}
\caption{The two-step iterative process as the $\gamma = \infty$ of the cut-off theory. We start with an semi-infinite domain (black), which will be formed automatically in the long time. Step 1 (green): evolve the height by the superdiffusion kernel for time $\Delta t$. Step 2 (red): set $h> \frac{1}{N} $ to be $1$.}
\label{fig:leap_eff}
\end{figure}

To estimate the light cone structure of this cutoff theory in the asymptotic limit, we take a further simplification in the wavefront dynamics. The basic idea is to replace the broad wavefront (the green curve in Fig.~\ref{fig:leap_eff}) by a fully filled sharp front (the red curve) above $1/N$. This can be realized if $\gamma$ is large so that the reaction time to form $h = \frac{1}{N}$ to reach a maximal height is negligible (more precisely independent of system size and time). Hence starting from a semi-infinite domain, the propagation of the wavefront in time $\Delta t$ is approximately an iteration of the following two steps (see Fig.~\ref{fig:leap_eff}):
\begin{enumerate}
\item The superdiffusion kernel evolves the profile for $\Delta t$. It produces the green profile (in the first step) in Fig.~\ref{fig:leap_eff}.\footnote{According to the equation, the height increment should also be multiplied by $(1-h)$, but due to the instant growth in the 2nd step, the factor $(1-h)$ only changes the regime where $h < \frac{1}{N}$. It is then in the interval of $(1 - \frac{1}{N}, 1])$. Hence approximate it to be $1$ everywhere. }
\item The reaction term sets all heights with $h> \frac{1}{N} $ to be $1$ (red curve in Fig.~\ref{fig:leap_eff}).
\end{enumerate}
To match the continuous process of Eq.~\eqref{eq:local-cutoff-theory}, the time interval $\Delta t$ should roughly be the local scrambling time. It is proportional to $\frac{1}{\gamma}\log N$, which is finite (and independent of system size and time) when $N$ is fixed. In principle the iteration and the process in Eq.~\eqref{eq:local-cutoff-theory} are strictly identical when $\gamma \rightarrow \infty$ and $\Delta t \rightarrow 0$. When we take finite $\Delta t$, we ignore the fact that the two steps in the iteration actually occurs simultaneously, and we ignore the superdiffusion that occurs in the region below the cutoff. But we believe these differences are immaterial: we expect to get the same light cone scalings even when they are ignored.

\begin{figure}[h]
\centering
\includegraphics[width=0.8\columnwidth]{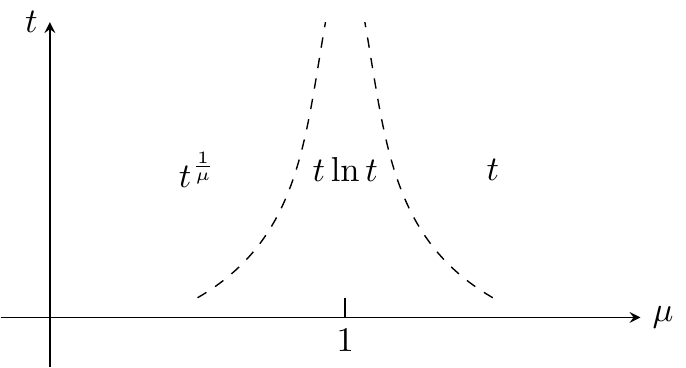}
\caption{The light cone phase diagrams for the two-step iterative process and the cutoff-theory. Dashed line represents a timescale of $e^{\frac{1}{| \mu - 1|}}$, below which the light cone is dominated by the marginal scaling $t\ln t$ at $\mu = 1$. Above this timescale, the light cone scalings on both sides of $\mu = 1$ will converge to power-law and linear shapes respectively. }
\label{fig:cutoff_phase_diagram}
\end{figure}

With these assumptions, we can quantitatively compute the light cone scalings of the simplified dynamics. It is clear that the increment of height for region below the cutoff is the same for each superdiffusion process, since we always evolve a semi-infinite domain. The increment brought by the superdiffusion is proportional to
\begin{equation}
\begin{aligned}
  &-  \int_{-\infty}^0  \frac{h(x) - h(y)}{ | x - y|^{1 + \mu } }  dy =  -  \int_{-\infty}^0  \frac{0 - 1}{ | x - y|^{1 + \mu } } dy \\
&\propto \frac{1}{x^{\mu}} \quad  \text{for } x > 0
\end{aligned}
\end{equation}
for a site with an initial distance $x$ to the $h = 1$ domain.

For the sake of the light cone scalings, we can conveniently take the proportionality constant to be $1$. We monitor the location with threshold height $h = \frac{1}{N}$ to trace the light cone. Let these locations at discrete time step $t$ to be $\ell(t)$. The height at position $x$ in $t$ steps is
\begin{equation}
\label{eq:h_x_t_iter}
h(x,t) = \sum_{\tau = 1}^t \frac{1}{(x - \ell(\tau)) ^{\mu} } \Delta  t.
\end{equation}
We use $h( \ell(t+1), t ) = \frac{1}{N} $ to solve $\ell(t + 1)$, which generates a recursive relation
\begin{equation}
\label{eq:l_t}
\sum_{\tau = 1}^t \frac{1}{(\ell(t+1) - \ell(\tau)) ^{\mu} }  = \frac{1}{\Delta t N} =  \text{constant}
\end{equation}
The light cone scaling functions are thus self-consistent solutions of Eq.~\eqref{eq:l_t}.

We first try the linear light cone ansatz: $\ell(t) = vt$:
\begin{equation}
\label{eq:mu_g_1}
\begin{aligned}
  &\sum_{\tau = 1}^t \frac{1}{ (\ell(t+1) - \ell(t))^\mu} = \frac{1}{v^\mu }  \sum_{\tau = 1}^t \frac{1}{( t+1  - \tau)^\mu  } \\
& \sim \frac{1}{v^\mu }  \int_0^t \frac{1}{( t + 1 - \tau)^\mu} d \tau
\end{aligned}
\end{equation}
The integral is convergent and of order $1$ so long as $ \mu > 1$. Therefore we have linear light cone when $\mu > 1$.

In regions with $0 < \mu < 1$, the light cone expands faster, and it is reasonable to try a power-function ansatz $\ell(t) \sim A t^{ \beta  } $:
\begin{equation}
\label{eq:mu_l_1}
\begin{aligned}
& \sum_{\tau = 1}^t \frac{1}{ (\ell(t+1) - \ell(\tau))^\mu} =  \frac{1}{A^\mu }\int_0^t \frac{1}{ ( (t +1)^\beta - \tau^\beta )^\mu} d\tau \\
&= \frac{1}{A^\mu } \frac{t}{ t^{\beta \mu }} \int_0^1 \frac{1}{ ( ( 1 +\frac{1}{t})^\beta - \tau^\beta )^\mu} d\tau
\end{aligned}
\end{equation}
The factor $t/t^{\beta \mu}$ enforces $\beta \mu = 1$ so that the result remains finite in the long-time limit.  One can further check that the integral converges when $0 < \mu < 1$. Therefore in this regime we have a power-law light cone $t^{\frac{1}{\mu}}$.

At $\mu  = 1$, a separate analysis is required. In the continuum limit, both the integrals in Eq.~\eqref{eq:mu_g_1} and Eq.~\eqref{eq:mu_l_1} converge. However, if we interchange the limit, taking $\mu \rightarrow 1^+$ first, then the integral in Eq.~\eqref{eq:mu_g_1} diverges. This means that the scaling function at $\mu  = 1$ should be parametrically faster (in terms of light cone expansion) than a linear function. On the other hand, the expression in Eq.~\eqref{eq:mu_l_1}  approaches zero if we take $\beta > 1$ for $\mu \rightarrow 1^+$. Hence the scaling function is also slower than a power function. The consistent solution is $\ell(t) = t \ln t$, see more detailed derivations in App.~\ref{app:consistent_mu_1}.

The regime close to $\mu = 1$ has different scaling behaviors for short and long times. Both integrals in Eq.~\eqref{eq:mu_g_1} and Eq.~\eqref{eq:mu_l_1} contains a factor of $\frac{1}{\mu - 1}$ after the integration. When $\mu \rightarrow 1$, the scaling functions on both sides approaches $\frac{t}{|1 - \mu|}$. The asymptotic linear and power-law scalings are only visible when $\frac{t}{|1 - \mu|} \gg t \ln t $. This sets a crossover timescale $e^{\frac{1}{|\mu - 1|}}$ below which we can only observe the marginal light cone scaling $t \ln t $ when $\mu $ is close but not exactly at $1$.

\begin{figure}[h]
\centering
\subfigure[]{
  \label{fig:cutoff_pow}
  \includegraphics[width=0.8\columnwidth]{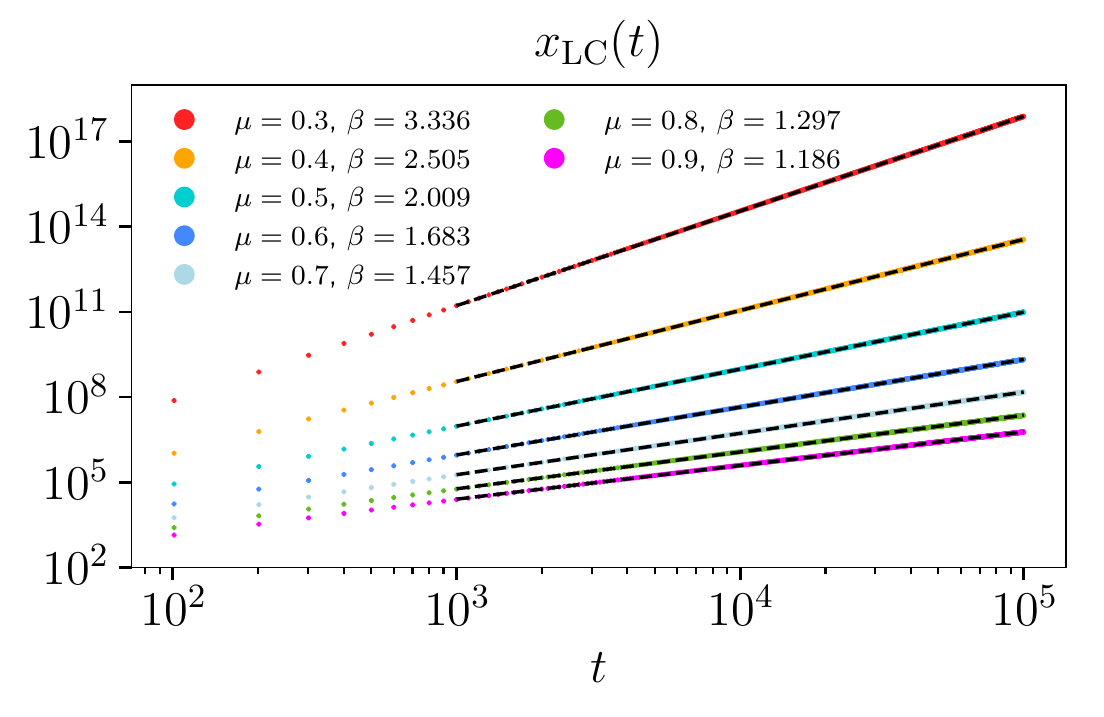}
}
\subfigure[]{
  \label{fig:cutoff_linear}
  \includegraphics[width=0.8\columnwidth]{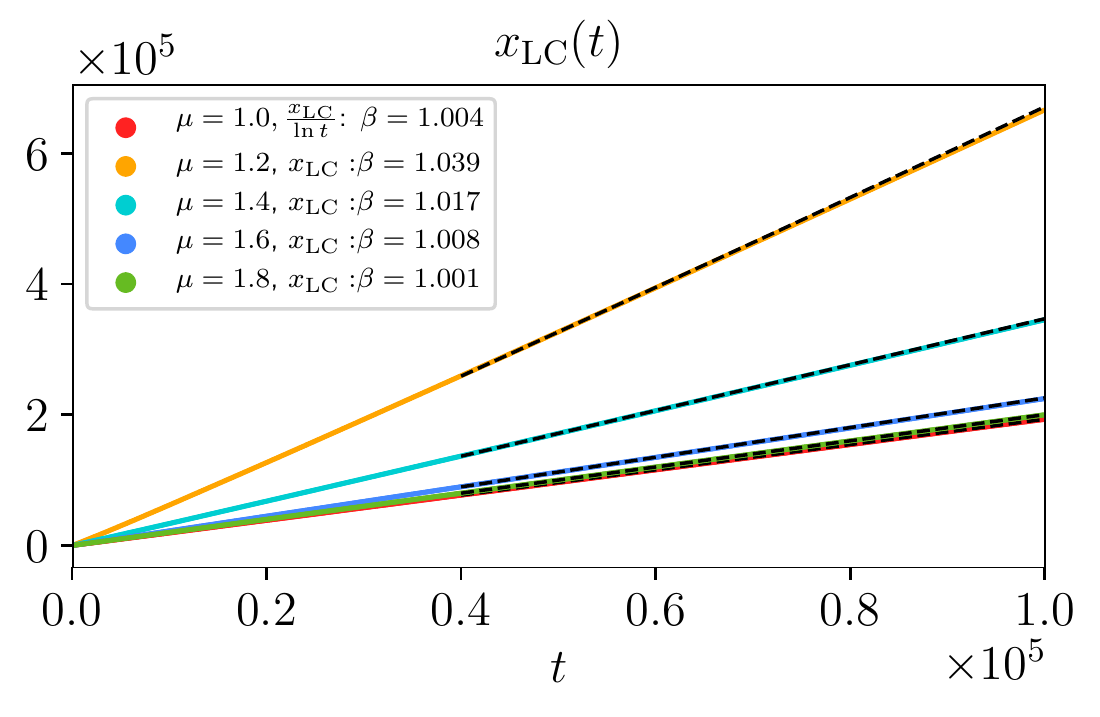}
}
\label{fig:cutoff_leap_num}
\caption{The numerical solutions $\ell(t)$ in the two-step iterative process. $\ell(t)$ labels the furtherest point where $h = 1$. The long time data is fitted to $A t^{\beta}$. (a) Power-law light cone regime when $0< \mu < 1$. (b) linear-light-cone regime when $\mu > 1$. $\mu = 1$ is fitted to $A t^{\beta} \ln t$.}
\end{figure}
We confirm this scalings by numerically solve $\ell(t)$ in the two-step iterative process. We choose the constant in Eq.~\eqref{eq:l_t} to be $\frac{1}{2}$. Given $\ell( \tau)$ for $\tau$ up to $t$, we search $x$ starting from $\ell(t) + 1$ in Eq.~\eqref{eq:h_x_t_iter}. We first double the increment until it overshoot, and then use binary search within an interval of $x$. In this way we find $\ell(t + 1)$.

The results are shown in Fig.~\ref{fig:cutoff_leap_num}. We fit $\ell(t)$ with $A t^{\beta}$ and find $\beta \approx 1$ for $\mu > 1$ (Fig.~\ref{fig:cutoff_linear}) and $\beta \approx \frac{1}{\mu}$ for $0< \mu < 1$ (Fig.~\ref{fig:cutoff_pow}). At $\mu  = 1$, the fit $At^{\beta} \ln t $ gives $\beta \approx 1$.

In the two step iterative process, we keep the tail generated by the superdiffusion in each step below height $\frac{1}{N}$. It is obvious that if we set the tail to zero in each step, we would end up with a linear light cone for all $\mu > 0$. By keeping the tail, we see that it speed up the propagation of the wavefront and give rise to a power-law light cone structure for $0 < \mu < 1$. This type of the $\frac{1}{N}$ correction thus plays a dominant rule in the power-law light cone regime.

The iterative process assumes an infinite $\gamma$. We also numerically solve Eq.~\eqref{eq:lr_fkpp_simple} and Eq.~\eqref{eq:naive_theory} for finite $\gamma$ and obtain the same scalings. As expected, the additional $(1-h)$ factor in Eq.~\eqref{eq:naive_theory} only modifies the velocity but does not change the scaling.

\subsection{Matching of the light cone phase diagram}
\label{subsec:match_light_cone}
In Sec.~\ref{sec:mean_field}, the Fokker-Plank equation (with leading order $\frac{1}{N}$ corrections) gives the relation $\mu = 2\alpha - 1$ that connects the OTOC physics and the FKPP equation. However, this is {\it not} compatible with our existing results. The critical point in 1d that separates the linear and power-law light cones is at $\alpha = 1.5$ (Tab.~\ref{tab:lc_scaling}). This translates to $\mu = 2$ on the FKPP side, which contradicts with the $\mu  = 1$ found in Sec.~\ref{subsec:eff_model}.
After inspecting the existing results, we propose that a more plausible relation should be a renormalized one $\mu  = 2\alpha - 2$. There are further evidences that suggest this relation.

One evidence is the power-law dependence of the linear velocity $v_B$ w.r.t. parameter $N$, the other is the marginal $\ln t$ scaling of velocity close to the critical point at $\mu = 1$. We compare the results of the cutoff theory (solutions for the effective model) and numerical results of \modelN. Both accept the renormalized relation $\mu  = 2\alpha - 2$.

These two quantities also provide us practical checks when noisy effect is considered in Sec.~\ref{sec:eff_of_noise}.

\subsubsection{Velocity $N$-dependence}
In the linear-light-cone regime ($\mu > 1$), previous literature~\cite{brockmann_front_2007} had derived a scaling relation between the linear velocity $v_B$ and $N$. We repeat the argument here. Assuming at time $t$, $h(x_0, t) = \frac{1}{N}$. Since there is no reaction term below the cutoff, the height increment at $x_0 + \Delta x$ is given completely by the superdiffusion,
\begin{equation}
\Delta t \int_{-\infty}^{x_0} \frac{1}{(x - \Delta x - x_0 )^{1 + \mu }}dx \sim \Delta t \frac{1}{\Delta x^{ \mu }}
\end{equation}
when this is equal to $\frac{1}{N}$ , we have $\Delta x = v_B \Delta t$. From this we have
\begin{equation}
  v_B \sim N^{\frac{1}{\mu}}.
\end{equation}

As we argue in Sec.~\ref{sec:otoc_finite_N}, \modelN $< N$ \mone. This means that if there is a linear velocity $v_B(N)$ for \modelN, it will be less than (or equal to, in the limiting process) to $N$ times the velocity of \mone. Thus the velocity $N$-dependence can not be larger than a linear dependence in $N$. Therefore transition between linear and power-law light cone should occur at $\mu = 1$. This result alone rejects $\mu = 2\alpha - 1$ but accepts $\mu = 2\alpha - 2$, since the transition point in \modelN is $\alpha = 1.5$.

\begin{figure}[h]
\centering
\subfigure[]{
  \label{fig:v_N_chaos}
  \includegraphics[width=0.8\columnwidth]{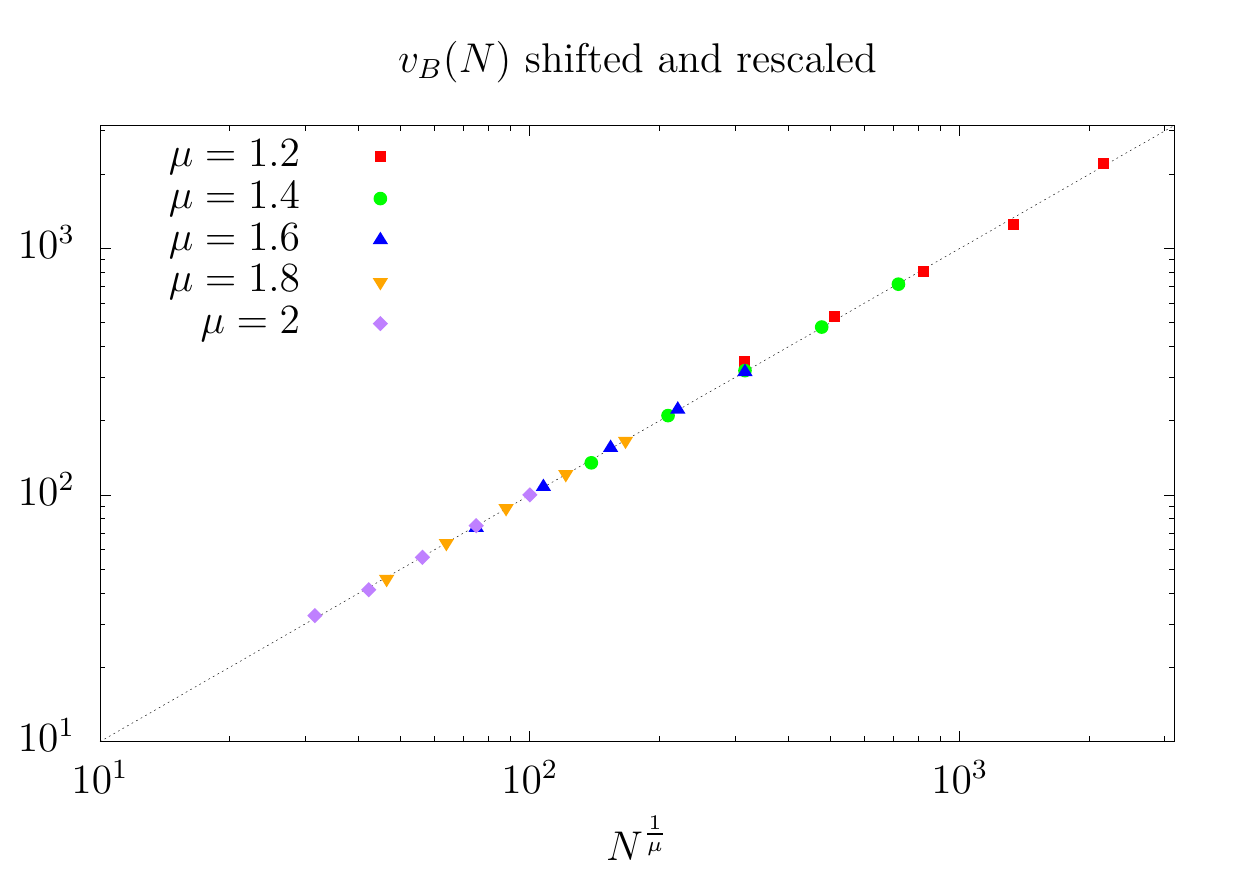}
}
\caption{Numerical verification of the velocity $N$-dependence for Eq.~\eqref{eq:local-cutoff-theory}. We fit the velocity as $v = a N^{\frac{1}{\mu} } + b$ and plot the rescaled and shifted velocity $(v-b)/a$. }
\label{fig:cutoff_v_N}
\end{figure}

We numerically verify this velocity $N$-dependence in the cutoff theory. Assuming a linear relation $v( \mu, N ) = a( \mu ) N^{\frac{1}{\mu}} + b( \mu ) $, we plot the rescaled and shifted velocity $(v-b)/a)$, see Fig.~\ref{fig:cutoff_v_N}. This reproduces the result in Ref.~\onlinecite{brockmann_front_2007}.

The numerical verification of $v_B \sim N^{\frac{1}{2\alpha - 2}}$ in \modelN (the Brownian circuit) is more challenging. The Monte Carlo approach to simulate \modelN is inherently easier for small $N$ rather than large $N$, as larger $N$ means larger velocity and larger system size to accommodate before it converges. As such, we are unable to produce as many data points as in Fig.~\ref{fig:cutoff_v_N}. Nevertheless we are able to confirm the results for $\alpha = 2$, for which the velocity scaling is close to $\sqrt{N}$.

\subsubsection{The marginal scaling}

At the end of Sec.~\ref{subsec:eff_model}, we discussed the parameter regimes $\mu \rightarrow 1$. Below a timescale $e^{\frac{1}{|\mu - 1|}}$, the light cone scaling is dominated by the one at $\mu  = 1$, i.e. $t \ln t$. In practice, the light cone will start off from $t \ln t$ and transit to the asymptotic light cones (linear for $\mu > 1$, power-law for $\mu < 1$ ) when $t \gg e^{\frac{1}{|\mu - 1|}}$. On the side of $\mu > 1$, the velocity will initially grow as $\ln t$ and eventually crosses over to a constant value.

This marginal scalings help us to mitigate the finite size effect and extract the transition point from power-law to linear light cones from relatively short timescales.

\begin{figure}[h]
\centering
\subfigure[]{
  \label{fig:BC_velocity}
  \includegraphics[width=0.8\columnwidth]{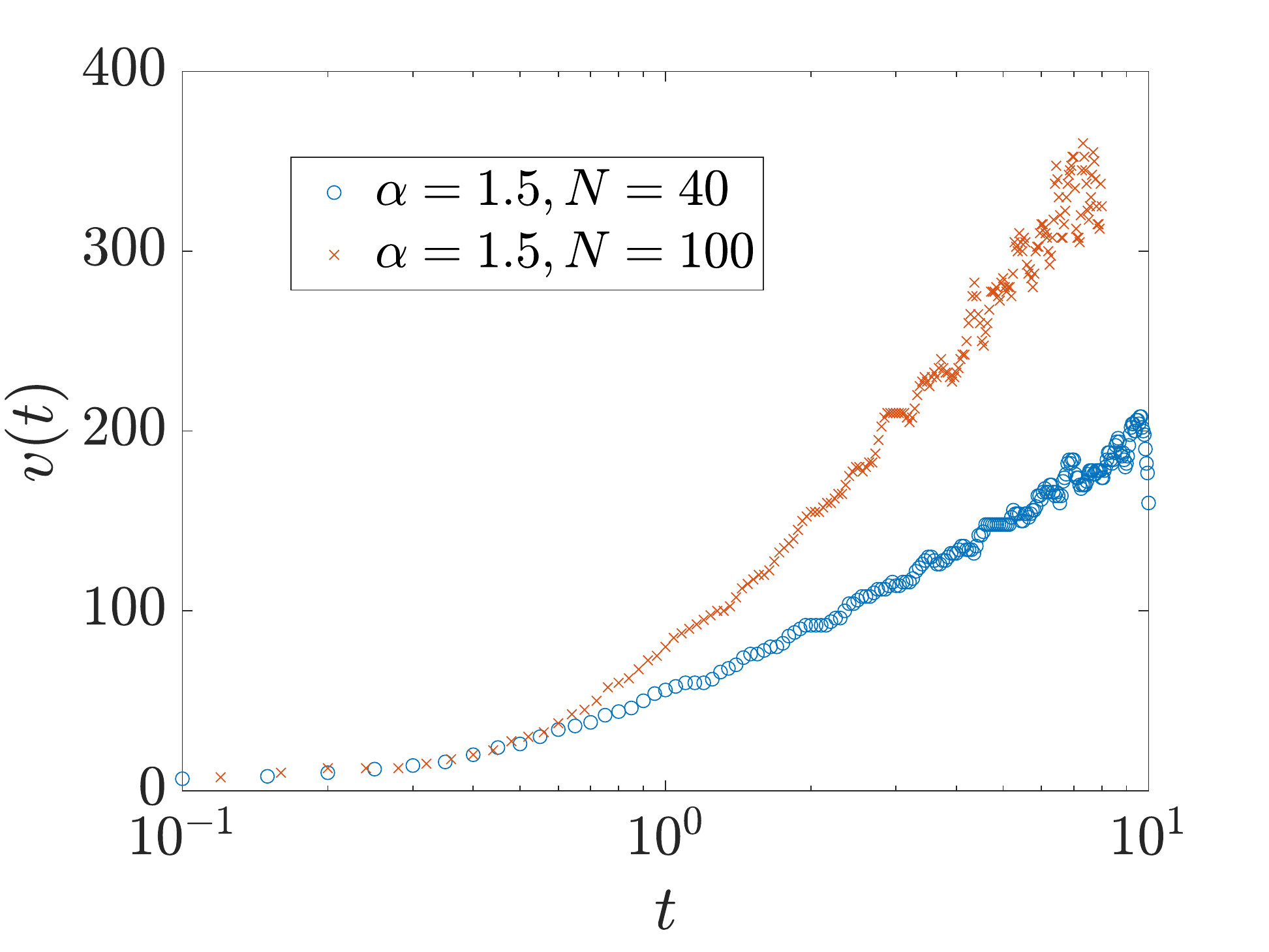}
}
\subfigure[]{
  \label{fig:BC_velocity_1}
  \includegraphics[width=0.8\columnwidth]{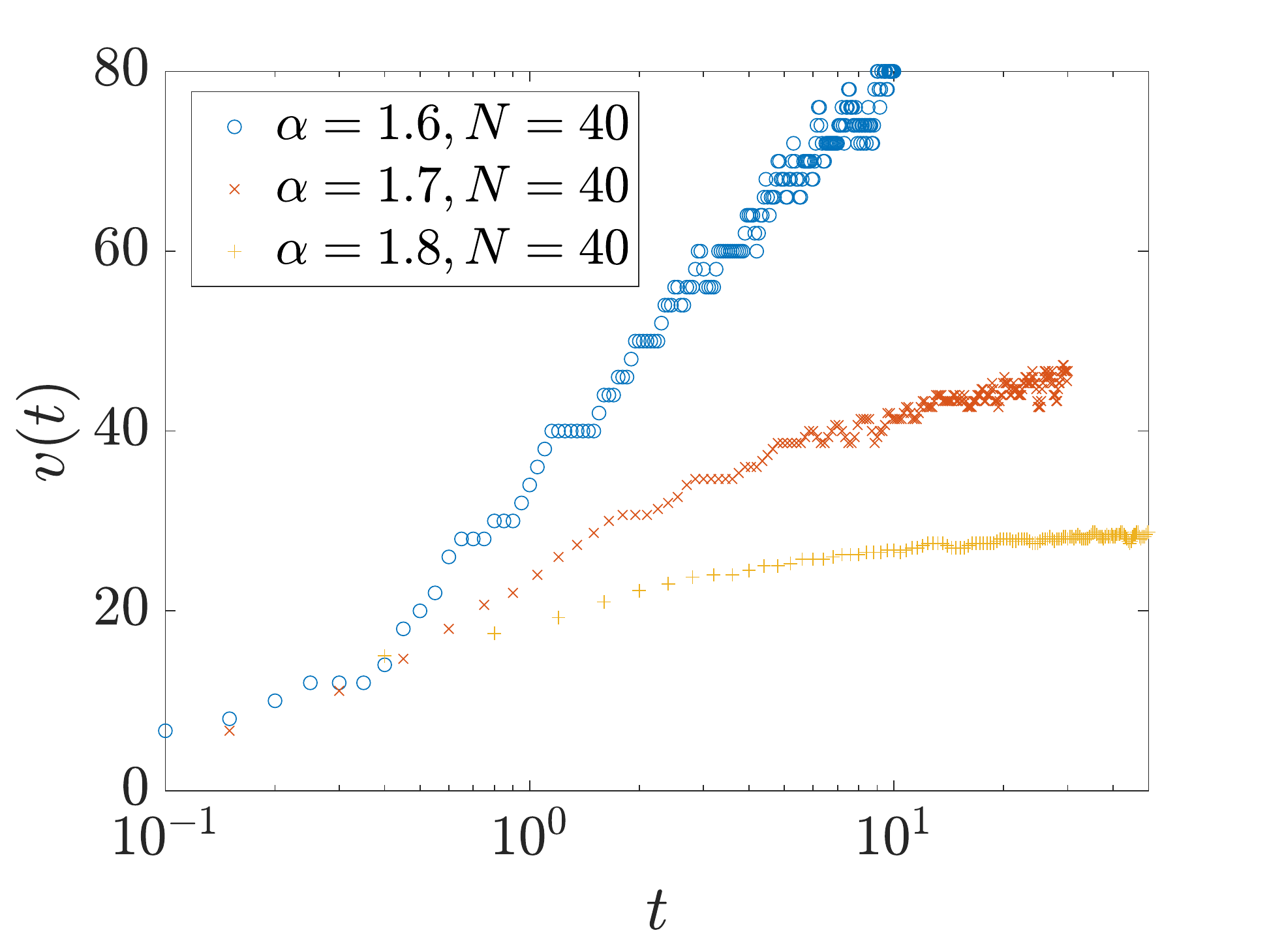}
}
\caption{Velocity $v(t)$ vs $t$ on the semi-log scale in \modelN (Brownian-circuit model). Here $v(t)\equiv d x_{\rm LC}/dt$. $x_{\rm LC}$ and $t$ is determined through $\overline{h(x,t)}=N/2$. When $\alpha=1.5$ [(a)], we observe $v(t)$ grows logarithmic in time. When $\alpha>1.5$ [(b)], especially in the case of $\alpha = 1.8$, we observe that $v(t)$ first grows logarithmically in time and then bends down as time evolves. We expect that it will saturate to a constant at late time.  In both plots, we take a domain wall initial condition in the calculation.}
\label{fig:v_log_t_brownian}
\end{figure}

We numerically calculate the velocity scaling for \modelN (Brownian circuit), see Fig.~\ref{fig:v_log_t_brownian}. The velocity does have a $\ln t$ short-time scaling when $\alpha$ is close to the critical point at $1.5$.

This confirms that the cutoff-theory (or the effective model) has captured not only the critical point, but also the short-time dynamics surrounding that regimes. This evidences further suggest to link the neighborhood of $\alpha = 1.5$ to $\mu = 1$, which is compatible with the renormalized relation $\mu = 2\alpha - 2$.

\section{Effects of Noise}
\label{sec:eff_of_noise}

In Sec.~\ref{sec:cut-off_th}, we showed that the cutoff theory of the FKPP equation has a transition between a power-law light cone and a linear light cone at $\mu=1$.
The transition matches the scaling of the OTOC in long-range systems under the identification $\mu=2\alpha-2$, as opposed to the na\"ive scaling $\mu=2\alpha-1$.
Nevertheless, the statistical features of the wavefront---specifically its broadening in time---are missing in the deterministic cutoff equation.

Simulation of the small-$N$ Brownian circuit in the linear-light-cone regime suggests that the wavefront of the OTOC can collapse into a scaling function of a single variable $(x-v_B t)/t^\xi$. When $1.5< \alpha <2$, the broadening exponent $\xi$ equals $1/(2 \alpha - 2)$, implying superdiffusive broadening, whereas the case where $\alpha > 2$ has diffusive broadening with $\xi = \frac{1}{2}$. The mapping $\mu = 2 \alpha - 2$ therefore suggests that the FKPP equation should have superdiffusive broadening with $\xi = \frac{1}{\mu}$ for $1 < \mu < 2$, and diffusive broadening $\xi = \frac{1}{2}$ for $\mu > 2$.

In order to connect to the small-$N$ Brownian-circuit picture and capture the fluctuation of the wavefront, it is necessary to include a noise term in the FKPP equation.
Indeed, in short-range systems, it has been shown that the FKPP equation with a local-noise term successfully reproduces the diffusive broadening of the wavefront \cite{Xu_Swingle_2018}.
For the long-range case, our expectation for the noise term is that it can produce a superdiffusively broadened wavefront for $1 < \mu < 2$.
In the following subsections, we experiment with different noise terms.

\subsection{Local noise and diffusive wavefront broadening}

The simplest choice is to use the noise term that leads to diffusive broadening in short-range interacting systems [Eq.~\eqref{eq:fkpp-cutoff-noise}]:
\begin{equation}
\label{eq:local-noise}
\sqrt{\frac{1}{N}\gamma h(1-h)}\eta(x,t),
\end{equation}
where $\gamma$ is the reaction strength and $\eta$ is the standard Gaussian noise.
We call \cref{eq:local-noise} the ``local noise'' term and note that while it is suppressed by $\frac{1}{N}$, when $h \sim \frac{1}{N}$, the noise is comparable with $h$ itself.
Hence, the noise is crucial at the wavefront, where $h \sim \frac{1}{N}$.

To investigate the effect of the local-noise term in \cref{eq:local-noise} on the front dynamics, we perform large-scale numerical of simulations Eq.~\eqref{eq:lr_fkpp_simple} on a lattice, given by
\begin{align}
\label{eq:lr_fkpp_simple_lattice}
\partial_t h_i =& \sum_{j\neq i} \Delta^{\mu/2}_{ij}h_j +\gamma h_i(1-h_i)\theta(h_i-\frac{1}{N})\\
& + \sqrt{\frac{1}{N}\gamma h_i(1-h_i)\theta(h_i-\frac{1}{N})}\eta_i(t) \\
\Delta^{\mu/2}_{ij} =& 1/|i-j|^{\mu+1}-\sum_k 1/|i-k|^{\mu+1}\delta_{ij},
\end{align}
where $\Delta_{ij}^{\mu/2}$ is the discrete superdiffusion kernel. In practice, we find that it is necessary to introduce the cutoff in the noise term as well. Otherwise, the noise term would lead to unphysical growth far ahead of the front and destroy the front dynamics. We integrate the differential equation for about 400 realizations and average the results to get the mean $h_i(t)$. The convergence of $h_i(t)$ to its asymptotic form becomes slow when $\mu \lesssim 1.2$. We use system sizes as large as $10^6$ to mitigate the effect of finite time, and $N$ is set to 100 for all the simulations.

\begin{figure}
\centering

\subfigure[]{
  \label{fig:v_marginal_local_noise}
  \includegraphics[width=0.9\columnwidth]{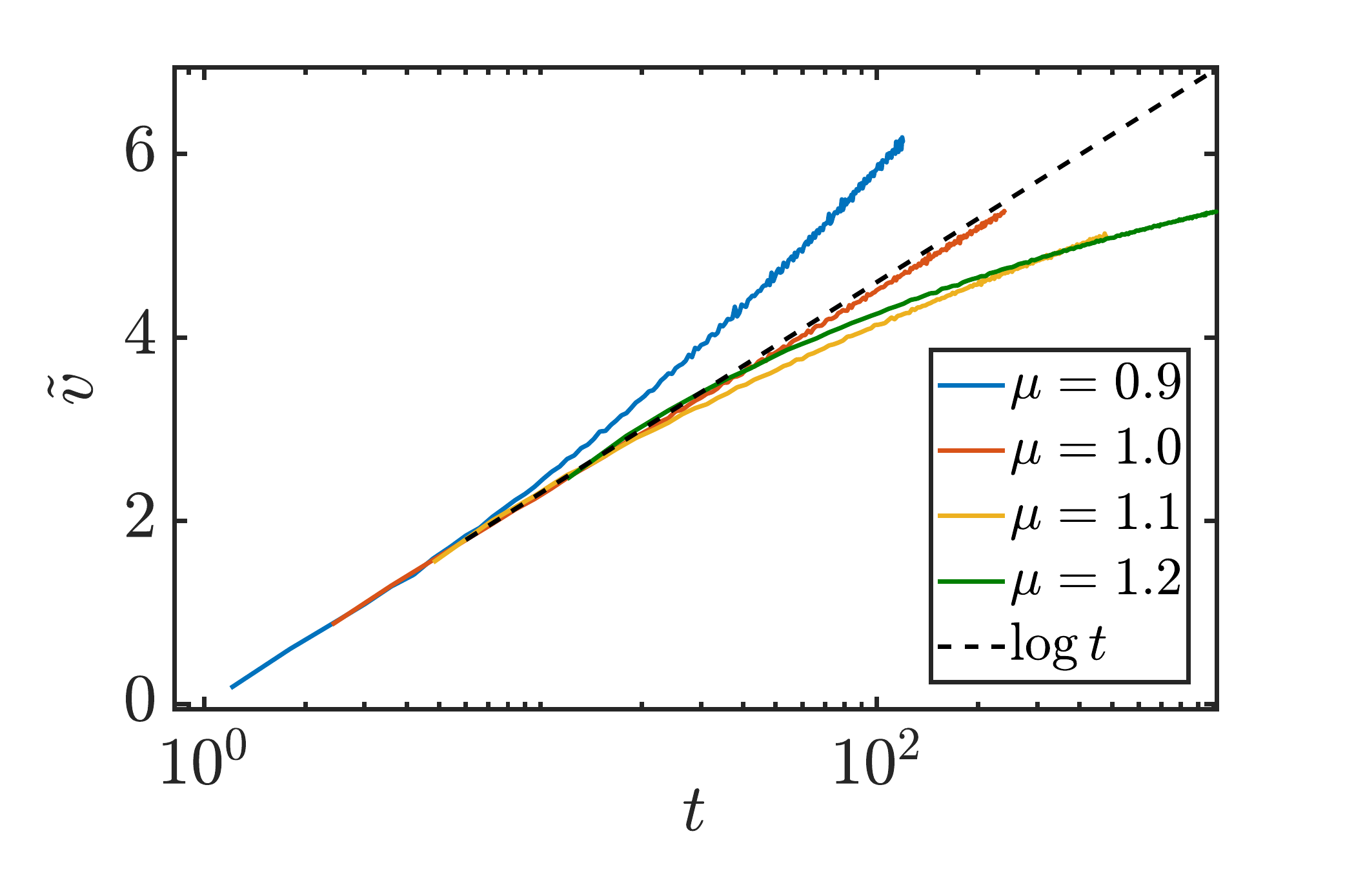}
}
\subfigure[]{
  \label{fig:bd_v_marginal_local_noise}
  \includegraphics[width=0.9\columnwidth]{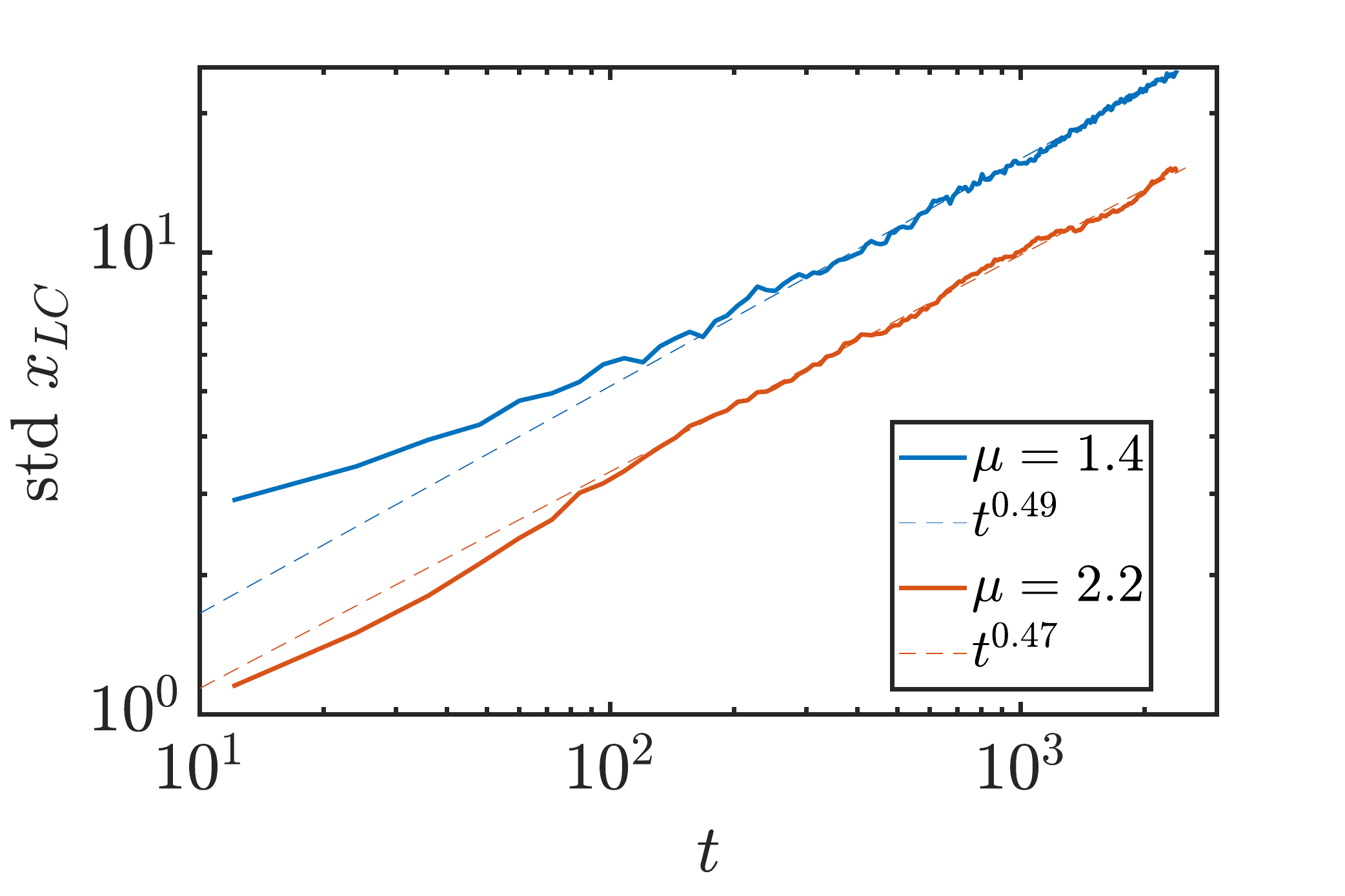}
}
\caption{(a) Velocity marginal scaling for the fractional FKPP equation with cutoff and local noise (Eq.~\eqref{eq:noisy_f_fkpp}). The data is averaged over 400 noise realizations. The velocities of different $\mu$ are shifted and rescaled as $\tilde v=a v+ b$ for comparison on the same scale. The shifting and rescaling do not change the behavior as a function of $t$. The linear dashed line is for comparison. (b) The wavefront broadening of $h_r(t)$ at $\mu=1.4$ and $\mu=2.2$ both show broadening close to diffusion, while superdiffusion at $\mu=1.4$ is expected from small $N$ analysis.}
\label{fig:fkpp_local_noise}
\end{figure}

We first check the marginal scaling of the velocity $v(t)$ in the vicinity of $\mu=1$. The results after averaging over 400 noise realizations are shown in Fig.~\ref{fig:fkpp_local_noise}(a). We find that the noise term in general increases the numerical value of the velocity, although the scaling of the velocity still matches the prediction from the cutoff theory.
In particular, we find that $v(t)\sim \log(t)$ at $\mu=1$.
This demonstrate that the local noise does not change the critical $\mu$ separating the linear and power-law light cones.

To extract the front broadening $\xi$, we calculate the standard deviation of the front position $x_{\rm LC}(t)$ for each noise realization, which is expected to scale as $\text{std}(x_{\rm LC})\sim t^\xi$. We pick two values of $\mu$: $\mu = 1.4$ and $\mu = 2.2$, and plot $\text{std}(x_{\rm LC})$ as a function of time in Fig.~\ref{fig:fkpp_local_noise}(b).
The straight lines on the log-log scale confirm the power-law scaling of $\text{std}(x_{\rm LC})$ and demonstrate that the noise term indeed induces wavefront broadening.
However, the slopes of the lines imply diffusive broadening ($\xi\sim 0.5$) for both $\mu = 1.4$ and $\mu = 2.2$, even though we expect $\mu = 1.4$ to broaden superdiffusively with a broadening exponent $\xi$ close to $\frac{1}{1.4}\approx 0.71$.
This result suggests that the local-noise term is not able to reproduce the superdiffusive broadening from the small-$N$ analysis.

\subsection{Long-range noise and superdiffusive wavefront broadening}
\begin{figure}
\centering

\subfigure[]{
  \label{fig:v_marginal_nonlocal_noise}
  \includegraphics[width=0.9\columnwidth]{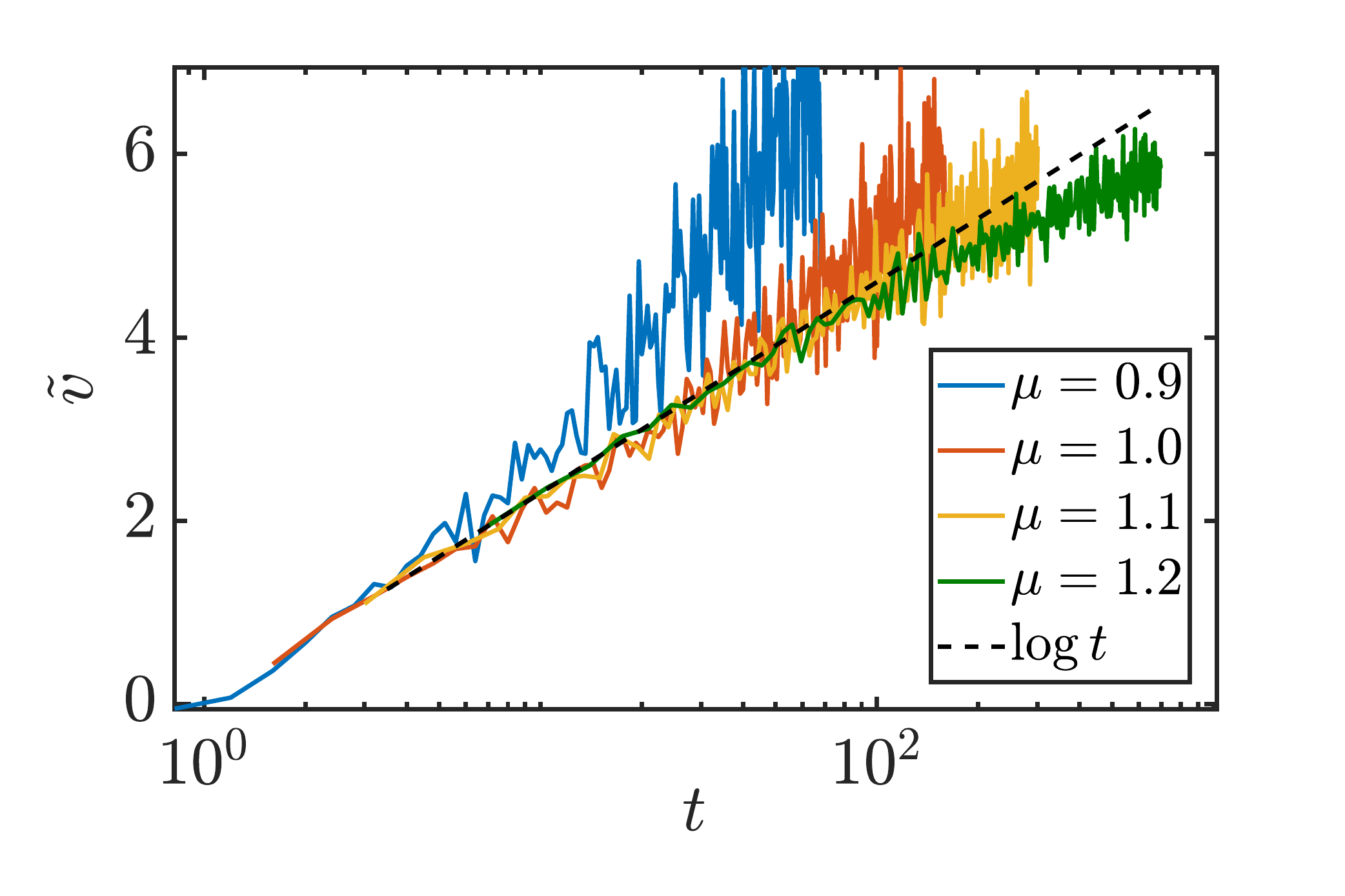}
}
\subfigure[]{
  \label{fig:bd_nonlocal_noise}
  \includegraphics[width=0.9\columnwidth]{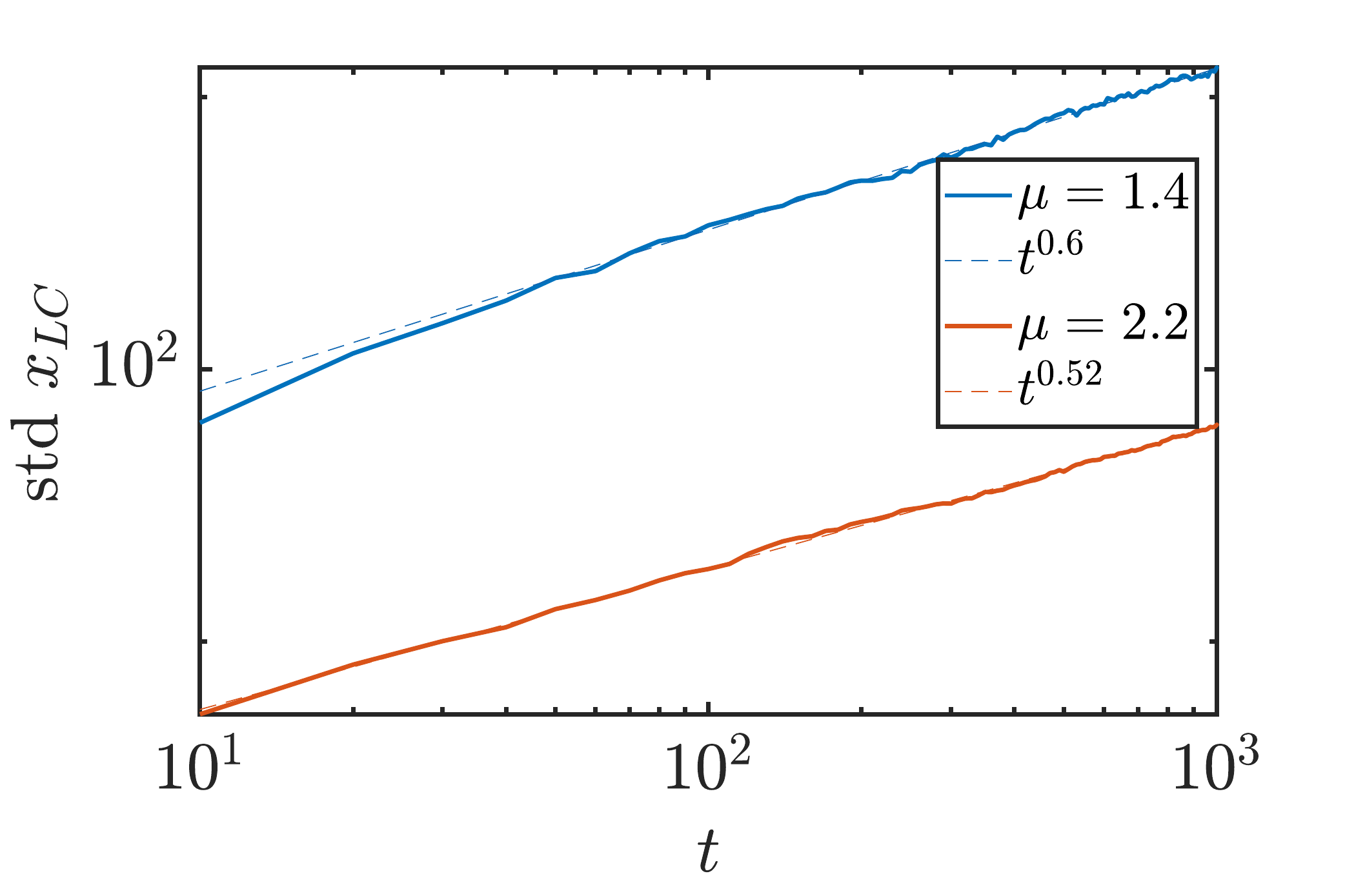}
}
\caption{Cutoff+Non-local noise. Velocity marginal scaling for na\"ive theory (fractional FKPP with cutoff and non-local noise, see Eq.~\eqref{eq:naive_theory}). The data is averaged over 400 noise realizations. The velocity is shifted and rescaled by the best linear fit against $\log t$ for the first ten data points. The critical point is close to $\mu = 1.1$.  }
\label{fig:fkpp_nonlocal_noise}
\end{figure}

Superdiffusive broadening indicates larger fluctuations than a diffusive wave front. We therefore turn to the long-range noise that arises from microscopic large $N$ limit of the master equation in Sec.~\ref{sec:derivation}, i.e. a noise term
\begin{equation}
    \sqrt{\frac{(1-h) \Delta^{\frac{\mu}{2}} + \gamma ( 1- h) h}{N}} \eta (x, t ).
\end{equation}
in Eq.~\eqref{eq:naive_theory}.

Numerically, we solve the discrete stochastic FKPP equation on a lattice in Eq.~\ref{eq:fkpp_discrete} with the cutoff approximation
\begin{align}
\label{eq:fkpp_lr_lattice}
&\partial_t h_i = \tilde g_i +\sqrt{\frac{\tilde g_i}{N}}\eta_i(t)\\
&\tilde g_i = (1 -  h_i )  [\gamma h_i \theta(h_i-\frac{1}{N})  + \sum_{j\neq i} \Delta_{ij}^{\mu/2}h_j\theta(h_j-\frac{1}{N}) ].
\end{align}
where the superdiffusive kernel $\Delta_{ij}^{\mu/2}$ is given in Eq.~\eqref{eq:lr_fkpp_simple_lattice}. The noise term requires $\tilde g$ to be positive and thus $\gamma > \sum\limits_{|i|>1} 1/|i|^{\mu+1}$.

In parallel with the study on the local noise, we first examine the marginal $\ln t$ scalings proposed in Sec.~\ref{subsec:match_light_cone} in the vicinity of $\mu=1$. The results are shown in Fig.~\ref{fig:fkpp_nonlocal_noise}.

Recall that the effective model (cutoff theory) without noise predicts a critical point of $\mu=1$ separating linear and power-law light, and that at the critical $\mu$ the velocity should grow as a logarithmic function of time indefinitely. By inspecting the curves in Fig.~\ref{fig:v_marginal_nonlocal_noise}, the critical point is in the range $1<\mu<1.1$. It is slightly different from the effective model prediction of $\mu = 1$, suggesting the non-local noise slightly shifts the critical point.

Now we explore the broadening effect induced by the long-range noise. Fig.~\ref{fig:fkpp_nonlocal_noise}(b) plots std$(x_{\rm LC}(t))$ on a log-log scale for both $\mu=1.4$ and $\mu=2.2$, similar to Fig.~\ref{fig:fkpp_local_noise}(b). In sharp contrast with the local-noise case, the data clearly demonstrates that the broadening exponent is superdiffusive for $\mu=1.4$ and is diffusive for $\mu=2.2$. However, the broadening exponent $\xi \sim 0.6$ is smaller than the expected value $1/\mu \sim 0.7$.

In summary, the local noise model in Eq.~\eqref{eq:lr_fkpp_simple_lattice} has the same critical $\mu=1$ as the noiseless model, and has diffusive broadening for $\mu>1$. On the other hand, the long-range noise model in Eq.~\eqref{eq:fkpp_lr_lattice} has slightly shifted critical $\mu$ and exhibit both superdiffusive broadening and diffusive broadening depending on the value of $\mu$, thus qualitatively capturing the phenomenology from the small $N$ analysis. However the precise value of the broadening exponent is different.

\begin{figure*}
\centering
\subfigure[]{
  \includegraphics[width=0.65\columnwidth]{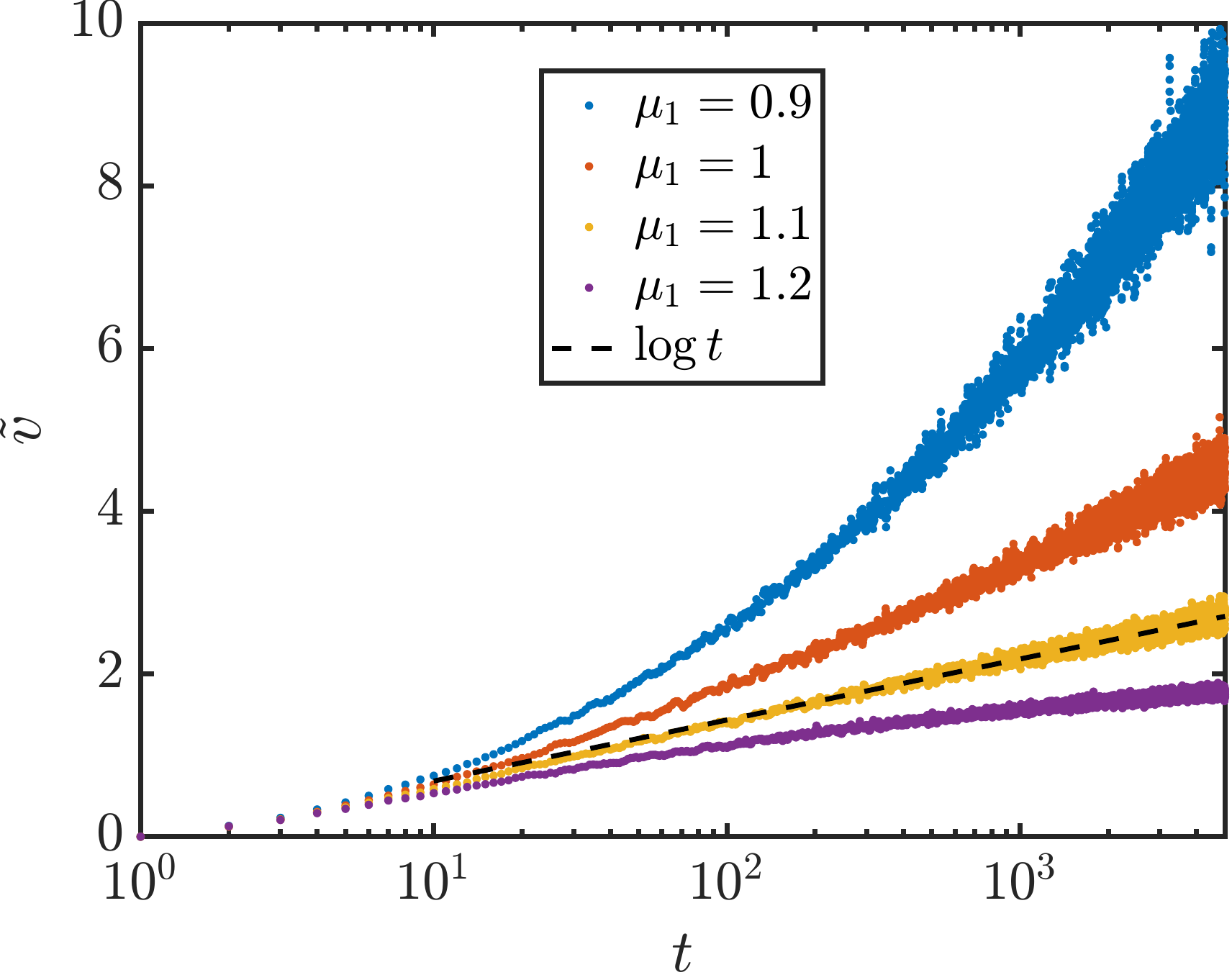}
  }
\subfigure[]{
  \includegraphics[width=0.65\columnwidth]{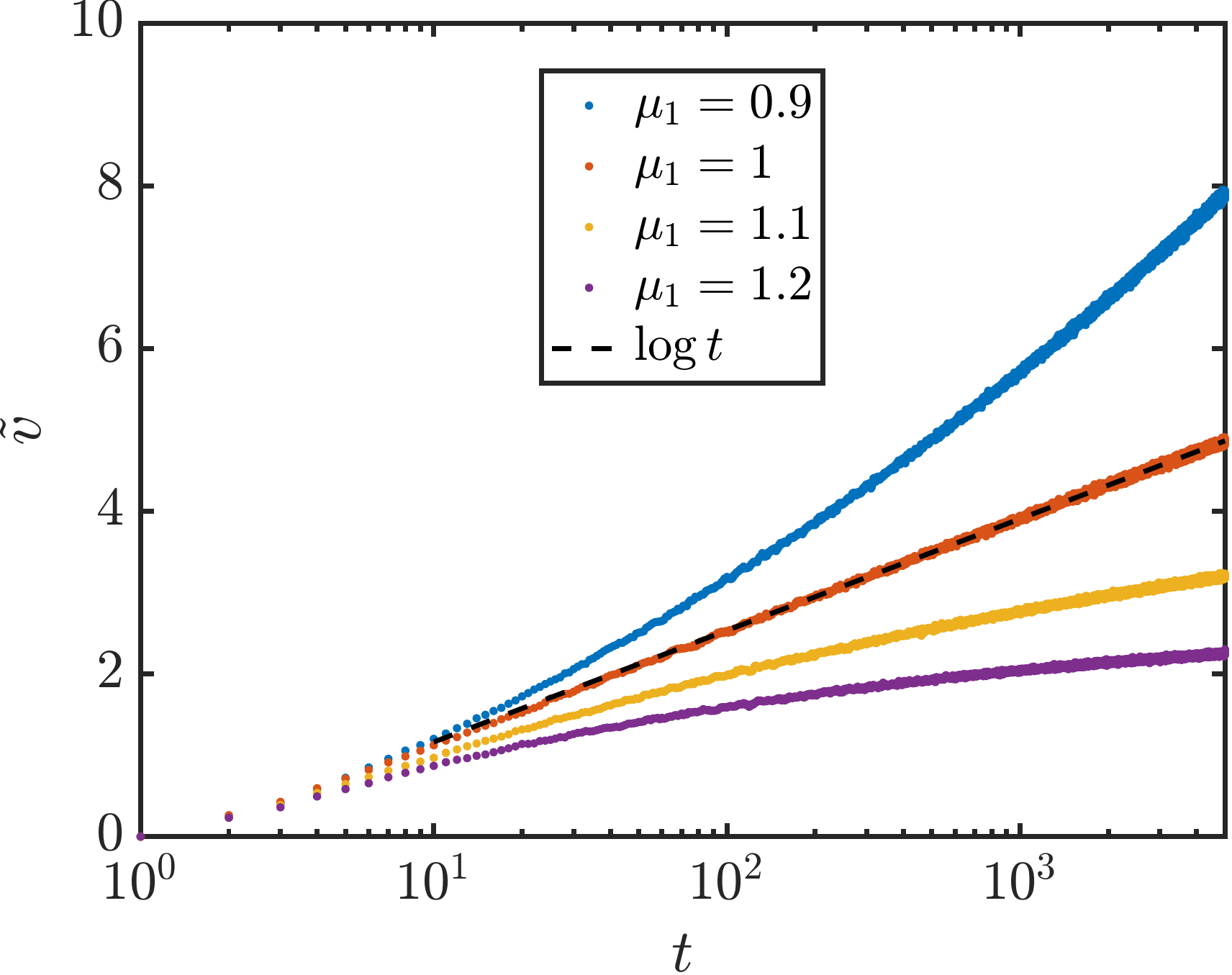}
  }
\subfigure[]{
  \includegraphics[width=0.65\columnwidth]{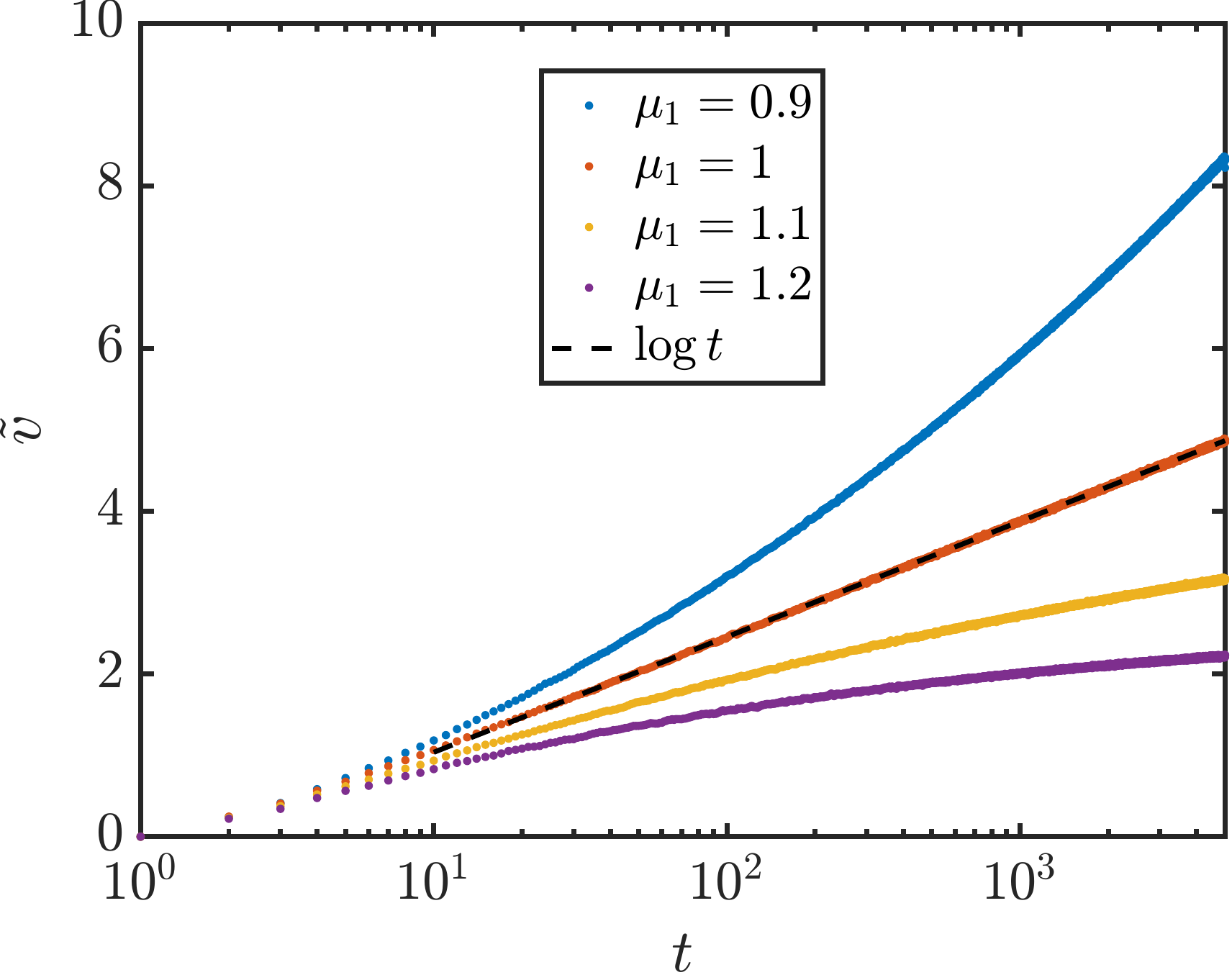}
  }
  \caption{Plots of the marginal scaling of the (rescaled) velocity $\tilde v$ as a function of time for the noisy effective model with (a) $\mu_2 = \mu_1$, (b) $\mu_2 = \mu_1 + \frac12$, (c) $\mu_2 = \mu_1+1$. For $\mu_1 = \mu_2$, the critical point when the marginal scaling of $\tilde v$ becomes linear ($\tilde v \propto \log t$) is when $\mu_1 = 1.1$ (yellow curve). For $\mu_2 = \mu_1 + \frac12$ and $\mu_2 = \mu_1+1$, the critical point returns to $\mu_1 = 1$ (red curve).}
  \label{fig:v_marginal_effective_model_noise}
\end{figure*}

\subsection{Noisy Effective Model}

Our studies on both the local noise and long-range noise demonstrate that the wavefront broadening is tied to the form of the noise. Furthermore, the noise may even affect the value of critical $\mu$ separating the linear light cone and power-law light cone. When $\mu$ is close to $1$, our numerical simulation is largely constrained by the system size and the number of time steps required to determine the light-cone contour. Therefore we return to the effective model with an additional noise term to explore the effects of different types of noise in the fractional FKPP equation.

In Sec.~\ref{subsec:eff_model}, we simplify the cutoff theory by taking $\gamma \rightarrow \infty$. Recall that the effective model considers the dynamics to be an iterative process with two steps: the operator spreads in space for a time $\Delta t$ and then all sites where the height function exceeds $\frac{1}{N}$ are set to 1.
To account for the effects of noise,
we add an additional action after the first step (evolving the profile by the superdiffusion kernel $\Delta^{\frac{\mu_1}{2}}$ for time $\Delta t$): suppose the height increase from $h$ to $h + \Delta h$, then we randomly change the height further to $h + \Delta h + \sqrt{\Delta h/N} r$, where $r$ is a standard Gaussian random variable with zero mean and variance $1$.
If the noise results in $h < 0$ for a given site, then we set $h$ to $0$. Here we deliberately choose two distinct indices--- setting $\mu = \mu_1$ for the deterministic term and (in general different) $\mu_2$ for the noise---to independently adjust the range of the noise.

For this noisy effective model, we numerically determined the critical point as well as the wavefront broadening for multiple values of $\mu_1$ and $\mu_2$.
In \cref{fig:v_marginal_effective_model_noise}, we simulate the noisy effective model for up to $5000$ steps and average over $\sim 1000$ samples.
When $\mu_1 = \mu_2$ [which is the case for Eq.~\eqref{eq:naive_theory}], we find that the critical point is close to $\mu_1 = 1.1$. When we adjust $\mu_2 = \mu_1 + \frac{1}{2}$ and $\mu_2 = \mu_1 + 1$, however, the critical point comes back to $\mu_1 = 1$.
This result suggests that using different exponents in the superdiffusion kernel for the deterministic and noise terms could potentially restore the critical point in the renormalized theory.

\begin{figure}[h]
  \includegraphics[width=0.9\columnwidth]{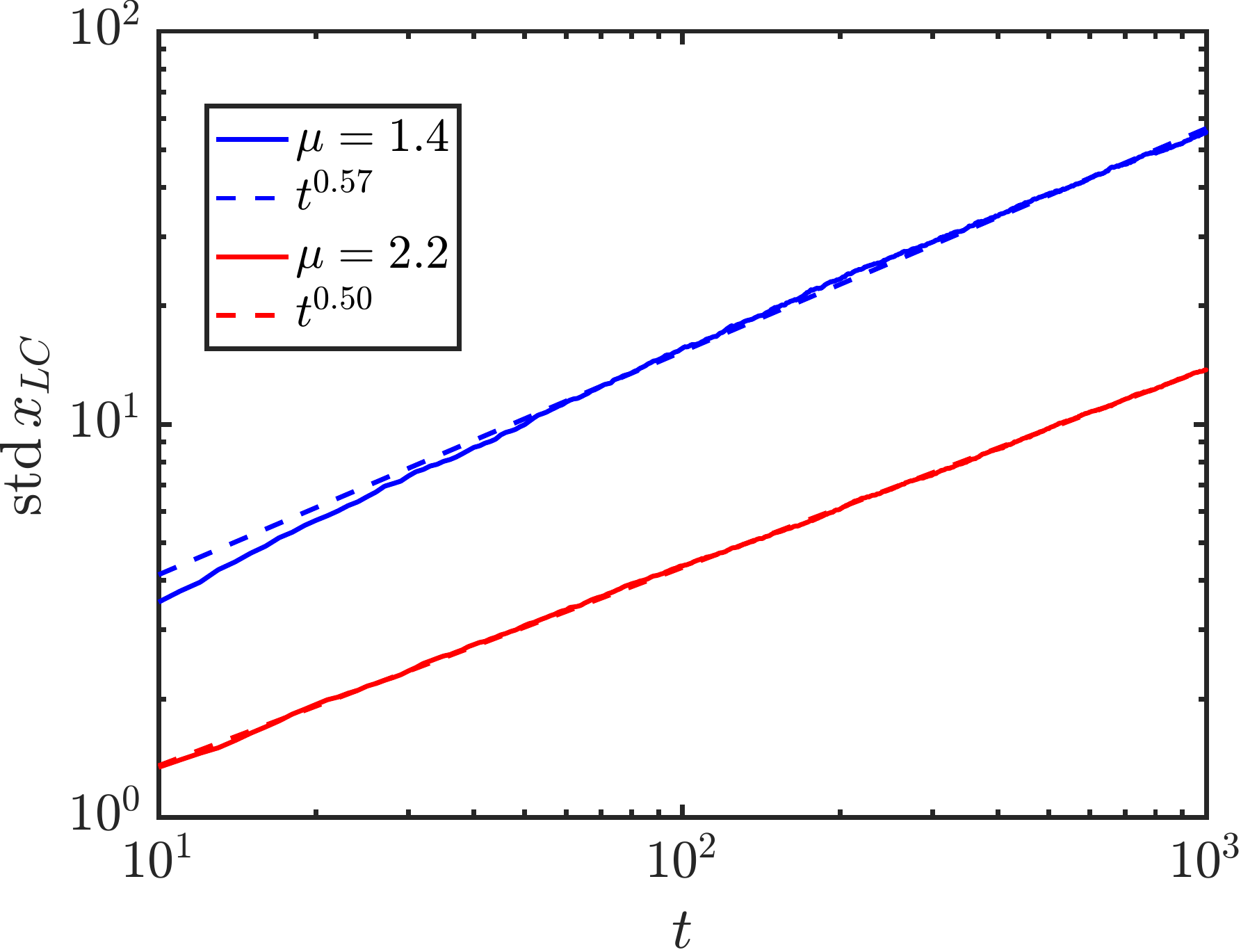}
  \caption{Plot of the wavefront broadening exponents for the noisy effective model with  $\mu_1 = \mu_2 = 1.4$ (blue) and $\mu_1 = \mu_2 = 2.2$ (red) in log-log scale. The dashed lines denote the best linear fit to the curves and show that the wavefront broadening is superdiffusive for $\mu_1 = \mu_2 = 1.4$ and diffusive for $\mu_1 = \mu_2 = 2.2$. While this qualitative behavior matches the small-$N$ phenomenology, the superdiffusive broadening exponent for $\mu = 1.4$ does not obey the exact $\xi \sim 1/\mu$ scaling expected from the small-$N$ prediction.}
  \label{fig:broadening_effective_model_noise}
\end{figure}

We then examine the broadening of the wavefront. As shown in \cref{fig:broadening_effective_model_noise}, the wavefront broadening is superdiffusive for $\mu_1 = \mu_2 = 1.4$ and diffusive for $\mu_1 = \mu_2 = 2.2$, which matches the small-$N$ prediction.
But, as with the fractional FKPP equation with long-range noise (cf. \cref{fig:fkpp_nonlocal_noise}), the noisy effective model gives $\xi \approx 0.57$ for $\mu_1 = 1.4$, which is slightly smaller than $\xi = 1/ \mu_1 \approx 0.71$ from the \modelN prediction. Similar phenomenology is observed for $\mu_2 = \mu_1 + \frac{1}{2}$ and $\mu_2 = \mu_1 + 1$, namely there are superdiffusive broadening, but $\xi$ is generally smaller than $1/\mu_1$.

Taken in totality, the results suggest that the use of long-range noise terms with a different exponent $\mu_2$ can restore the critical point back to $\mu =\mu_1 = 1$, and at the same time demonstrate superdiffusive broadening for $1< \mu_1 <2$. Nevertheless, the observed broadening exponent $\xi$ is smaller than $1/ \mu_1$.

\section{Discussion regarding the Renormalized Relation}
\label{sec:disc_renorm}
The renormalized relation $\mu = 2\alpha - 2$ reflects the collective effect of the finite-$N$ corrections, as contrast to the na\"ive relation $\mu = 2\alpha - 1$ in the leading order derivation of the $\frac{1}{N}$ correction. To understand why the na\"ive relation $\mu = 2\alpha - 1$ fails, let us consider \modelN, the stochastic process in $\Delta t$ produces a change of height at a site $r$ distance away (assuming semi-infinite initial condition) with probabilities
\begin{equation}
\label{eq:non-gaussian}
dh
=
\left\lbrace
\begin{aligned}
  & \frac{1}{N} & \quad \text{w.p.}\quad \frac{\Delta t}{| r|^{2\alpha-1}} \\
  & 0  & \quad \text{w.p.}\quad (1 - \frac{1}{| r|^{2\alpha-1}} \Delta t  ). \\
\end{aligned} \right.
\end{equation}
This is a very skewed distribution with a small probability to increase the height by $\frac{1}{N}$ and a large probability to remain the same. Among those instances in which the height actually reaches $\frac{1}{N}$, the strong local reaction can increase the height exponentially to $1$ in $\mathcal{O}(1)$ time, and the front can be pushed further forward. In the continuum equation, this physical noise is replaced by a Gaussian noise. In the approximation, the mean value of height growth is $\frac{1}{N} \frac{\Delta t}{| r|^{2\alpha-1}}$, the standard deviation is roughly $\frac{1}{N} \sqrt{\frac{\Delta t}{| r|^{2\alpha-1}}}$. As such, the original height growth of $\frac{1}{N}$ is roughly $\sqrt{\frac{| r|^{2\alpha-1}}{\Delta t}}$ away from the Gaussian noise average, so the Gaussian approximation significantly underestimates the probability for the height to increase by $\frac{1}{N}$. For fixed $r$, the central limit theorem can justify the Gaussian-noise approximation with enough repetitions of the noise process in Eq.~\ref{eq:non-gaussian}. However, with increasing $r$, the distribution in Eq.~\ref{eq:non-gaussian} becomes more and more skewed, and the number of repetitions for a good Gaussian approximation increases for larger $r$. Hence there are rare cases in which the noise in Eq.~\ref{eq:non-gaussian} increases a distant site to have height $\frac{1}{N}$, while the Gaussian noise approximation only produces a height much smaller than $1$. Because of the strong reaction combined with the superdiffusion, these instances can travel much faster than the average and quantitatively change the light cone scaling. In summary, the combination of the non-Gaussian noise and the long-range process makes \modelN (the Brownian circuit) much faster than the na\"ive theory.

\begin{figure}[h]
\centering
\includegraphics[width=\columnwidth]{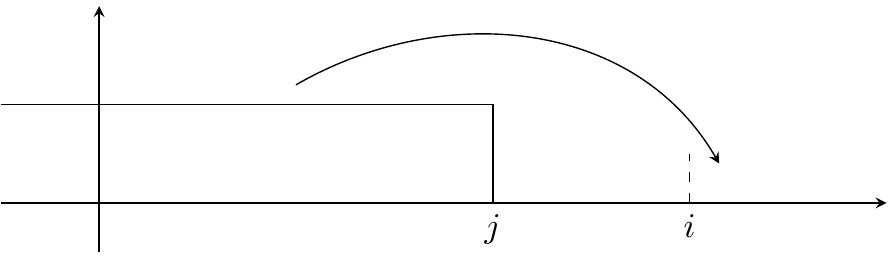}
\caption{Schematic picture of \monep}
\label{fig:monep}
\end{figure}

We give a heuristic argument about why the renormalized value of $\mu $ is $2\alpha - 2$. For the sake of presentation, we introduce \monep, which is a variant of \mone that played an important role in our treatment of $N = 1$ case in Ref.~\onlinecite{zhou_operator_2020}. In \mone, when a transition is made the height of another unoccupied site is increased from zero to one (Fig.~\ref{fig:model_1}). \monep follows the same rule for the transition rate, but whenever the height of the site is increased to $1$, we simultaneously increase the heights to one for {\it all the sites on its left}, see a schematic display in Fig.~\ref{fig:monep}. \mone and \monep have the same light cone structure for $\alpha > 1$. In fact, \monep can be a good approximation when the reaction rate $\gamma$ is large. In this case, once a site has height $1$, it can quickly spread and fill all the sites to its left.

Now in \monep, the height reaches $1$ whenever itself or any site on its right is occupied. The rate generated by a semi-infinite domain is therefore
\begin{equation}
\label{eq:jump_prob}
\sum_{x \ge 0, y\ge 0}  \frac{1}{|(i-x) -(j+y)|^{2\alpha} } \sim \frac{1}{|i - j|^{2\alpha - 2}}.
\end{equation}
Hence, effectively we have
\begin{equation}
 \mu = 2\alpha - 2
\end{equation}
in a speculative long wavelength theory of \monep. This is indeed the renormalized relation between $\mu$ and $\alpha$.

We can generalize the renormalized relation to higher dimensions. In $d$ spatial dimension, the fractional derivative can also be defined in Fourier space (cf. Eq.~\ref{eq:f_deri}, Sec. 2 of Ref.~\cite{kilbas_theory_2006}):
\begin{equation}
\label{eq:f_deri_d}
\begin{aligned}
 \frac{\partial}{\partial |x|^{\mu}} f(\vec{x}) &\propto - \int_{\mathbb{R}^d} |k|^{\mu} \tilde{f}(\vec{k} ) e^{i \vec{k} \cdot \vec{x} } d^d k \\
 &\propto
 \int_{\mathbb{R}^d} \frac{f(\vec{x}) - f(\vec{y})}{|\vec{x}- \vec{y} |^{\mu+d} } d^d y,
 \end{aligned}
\end{equation}
where $\tilde{f}(\vec{k} )$ is the Fourier transform of the square integrable function $f( \vec{x})$. By comparing with the mean field derivation, we would identify $2\alpha = \mu +d$, i.e. $\mu = 2\alpha - d$.

The tail generated at a distance $r$ from the center is $\frac{1}{r^\mu}$. Thus we can similarly establish an effective model in the radial direction. The two transition points---from power-law light cone to linear light cone and from power-law broadening to linear broadening---are at $\mu = 1$ and $\mu = 2$ respectively, which should correspond to $\alpha = d + \frac{1}{2}$ and $\alpha = d + 1$ in the $N = 1$ solution. Thus, only a renormalized relation $\mu = 2\alpha - 2d $ is consistent.
This can be alternatively understood by writing the jump probability in $d$ dimensions (c.f. Eq.~\ref{eq:jump_prob}):
\begin{equation}
\label{eq:jump_prob_d}
\sum_{|\vec{x}_0| \ge R, |\vec{y}_0| < R}  \frac{1}{|(\vec{x}-\vec{x_0}) -(\vec{y}-\vec{y_0})|^{2\alpha} } \sim \frac{1}{|\vec{x}-\vec{y}|^{2\alpha - 2d}}.
\end{equation}
This supports our choice of $\mu = 2\alpha - 2d$ as the renormalized exponent of the tail.
In summary, we predict that the renormalization of $\mu$ in higher dimensions is strong enough that it shifts $2\alpha - d$ to $2\alpha - 2d$.

\section{Conclusions}
\label{sec:disc}

In this paper, we study the large-$N$ scrambling physics in generic long-range interacting systems, where $N$ is the number of degrees of freedom on each spatial site. We first generalize the stochastic height model established for the $N = 1$ problem in our previous work~\cite{xu_locality_2018,chen_quantum_2018,zhou_operator_2020} to large (but finite) $N$, and conclude that the phase diagrams of the OTOC light-cone structures are the same (Fig.~\ref{fig:phase_diagram}).
Controlled by the power-law exponent $\alpha$ in the interaction, the system can have logarithmic, power-law and linear light cones.

At the other limit, the mean-field theory at $N = \infty$ is given by a fractional FKPP equation.
The equation only gives a logarithmic light cone (i.e. causal regions that extend exponentially in space) for any superdiffusive index $\mu$ in the equation.
Therefore, $\frac{1}{N}$ corrections are necessary to determine the correct light-cone structures.

Conventionally, for FKPP equation with a diffusive $\Delta h$ term, the $\frac{1}{N}$ corrections comes from a cut-off term, mimicking the discreteness of the underlying variable, as well as from a $\frac{1}{N}$ noise.
We therefore put in a cutoff function by hand and derive the noise term from the Fokker-Plank equation.
Here, the leading perturbative result suggests setting the superdiffusive index $\mu$ in the fractional FKPP equation to be $2\alpha - 1$.

We analytically study the cutoff theory without noise. Through a series of comparisons, we proposed that the relation between $\mu $ and $\alpha$ should be corrected to $\mu = 2\alpha - 2$, in order to match the physics on the two sides for $\alpha > 1$.
This results in $\mu = 1$ ($\alpha = 1.5$) being the critical point separating the linear and power-law light cones.
We verify this proposal by numerically simulating the cutoff theory in Eq.~\eqref{eq:local-cutoff-theory}.
In the linear-light-cone regime, $v$ scales as $N^{\frac{1}{\mu}}$ ($N^{\frac{1}{2 \alpha - 2}})$, and the short-time marginal scaling of velocity is $\ln t$ for $\mu$ close to $\mu = 1$ ($\alpha = 1.5$).

Finally, we experiment with different forms of the noise to reproduce the broadening of the wavefront. We simulate the following equation in an effective model:
\begin{equation}
\label{eq:noisy_f_fkpp}
\begin{aligned}
  \partial_t h =& (1-h) D_{\mu} \Delta^{\frac{\mu}{2}} h + \gamma h ( 1 - h) \theta (h - \frac{1}{N}) \\
&  + \sqrt{\frac{\gamma h ( 1 - h) + (1-h) D_{\mu_2} \Delta^{\frac{\mu_2}{2}} h }{N} } \eta.
\end{aligned}
\end{equation}
Here $\theta$ is the step function. We choose two parameters $\mu_1$ and $\mu_2$ to model potentially different decaying exponent of the super-diffusion kernel and the noise. The results are summarized in Tab.~\ref{tab:noise_term_res}.
\begin{table}[h]
    \centering
    \begin{tabular}{|c|c|c|}
        \hline
         Model & critical point & $\xi$ for $1 < \mu < 2$ \\ \hline
         FKPP + local noise & $\mu = 1$  & $\frac{1}{2}$\\ \hline
         FKPP + long-range noise & $1< \mu <1.1$ & $\frac{1}{2} < \xi < \frac{1}{\mu} $ \\ \hline
         effective, $\mu_1 = \mu_2$ & $\mu = 1.1$ & $\frac{1}{2} < \xi < \frac{1}{\mu} $ \\ \hline
         effective, $\mu_2 > \mu_1$ & $\mu = 1$ & $\frac{1}{2} < \xi < \frac{1}{\mu} $ \\ \hline
    \end{tabular}
    \caption{Critical points separating linear and power-law cone and wave front broadening exponents for different types of noise terms. }
    \label{tab:noise_term_res}
\end{table}

The best fit to the small-$N$ numerics seems to require long-range noise with index $\mu_2 > \mu_1$.
That way, both a critical point of $\mu_1 = 1$ and superdiffusive broadening when $1 < \mu_1 < 2$ are simultaneously obtained, although the broadening exponent $\xi$ is still smaller than the theoretical value of $\frac{1}{\mu_1}$.

We conclude that the fractional FKPP equation with cutoff and noise terms reproduces part of the phenomenology of the scrambling physics at finite-$N$.
Some aspects it reproduces with exact quantitative precision---for example, the critical points of the light cones of the cut-off theory once we identify $\mu = 2\alpha - 2$---but others it does so only approximately.
As examples of the latter, the equation misses the phase diagram in the range of $0 < \alpha < 1$, has a slight shift of the critical point from the theoretical value of $\mu = 1$ to $\mu = 1.1$ with some forms of noise, and slightly underestimates the broadening exponent.
For $d>1$, we propose $\mu = 2\alpha - 2d$ as the natural generalization for the renormalized exponent.

Barring finite-size and finite-time effects that are always present in numerics, one possible explanation for the discrepancy between our theory and the phenomenology is that the non-Gaussian nature of the noise in \modelN [Eq.~\eqref{eq:non-gaussian}] cannot be accounted by the renormalized relation $\mu = 2 \alpha - 2$ alone. These are collectively higher-order effects in $\frac{1}{N}$ that may not be fully captured by a noise term and cutoff function. We have defined the FKPP equation with two indices $\mu_1$ and $\mu_2$, one representing the index for the deterministic term and the other for the long-range noise. It would be interesting to study the phase diagram and broadening in the full parameter range of $\mu_1$ and $\mu_2$, rather than fixing $\mu_1 = \mu_2 =  2\alpha - 2$ upfront. We leave these as future works.



Finally, we list a few more open questions.
The first one would be to verify that the phase diagram generalizes to higher dimensions, where we predict that the renormalized exponent $\mu = 2\alpha-2$ generalizes to $\mu = 2\alpha - 2d$.
Second, in the literature of the fractional FKPP equation, the tails of the front play important roles in the pulled dynamics.
It would be interesting to understand from that point of view how that noise changes the tail scaling, thus leading to the change of the light cone structures.
Finally, we have not addressed the question of the short-time dynamics of the OTOCs.
In experiments, limited coherence times may not permit one to observe the asymptotic scalings predicted in this paper.
In future works, we hope to address the timescales that separate the asymptotic-time and short-time regimes.

\acknowledgements
TZ was supported by a postdoctoral fellowship from the Gordon and Betty Moore Foundation, under the EPiQS initiative, Grant GBMF4304, at the Kavli Institute for Theoretical Physics.
TZ is currently supported as a postdoctoral researcher from NTT Research Award AGMT DTD 9.24.20 and the Massachusetts Institute of Technology.

AYG was supported by the NSF Graduate Research Fellowship Program under Grant No.\ DGE-1840340. He also acknowledges funding by the DoE ASCR Quantum Testbed Pathfinder program (award No.~DE-SC0019040), ARO MURI, AFOSR, DoE QSA, NSF QLCI (award No.~OMA-2120757), DoE ASCR Accelerated Research in Quantum Computing program (award No.~DE-SC0020312), NSF PFCQC program, AFOSR MURI, and DARPA SAVaNT ADVENT.

BGS acknowledges support from the Simons Foundation via the It From Qubit Collaboration. This research is supported in part by the National Science Foundation under Grant No. NSF PHY-1748958. This work was supported by a grant to the KITP from the Simons Foundation (\#216179).

We acknowledge the University of Maryland supercomputing resources and advanced computing resources provided by Texas A\&M
High Performance Research Computing made available for conducting the numerical simulations in this work.

\bibliography{noisy_f_fkpp_paper}

\begin{thebibliography}{44}%
\makeatletter
\providecommand \@ifxundefined [1]{%
 \@ifx{#1\undefined}
}%
\providecommand \@ifnum [1]{%
 \ifnum #1\expandafter \@firstoftwo
 \else \expandafter \@secondoftwo
 \fi
}%
\providecommand \@ifx [1]{%
 \ifx #1\expandafter \@firstoftwo
 \else \expandafter \@secondoftwo
 \fi
}%
\providecommand \natexlab [1]{#1}%
\providecommand \enquote  [1]{``#1''}%
\providecommand \bibnamefont  [1]{#1}%
\providecommand \bibfnamefont [1]{#1}%
\providecommand \citenamefont [1]{#1}%
\providecommand \href@noop [0]{\@secondoftwo}%
\providecommand \href [0]{\begingroup \@sanitize@url \@href}%
\providecommand \@href[1]{\@@startlink{#1}\@@href}%
\providecommand \@@href[1]{\endgroup#1\@@endlink}%
\providecommand \@sanitize@url [0]{\catcode `\\12\catcode `\$12\catcode
  `\&12\catcode `\#12\catcode `\^12\catcode `\_12\catcode `\%12\relax}%
\providecommand \@@startlink[1]{}%
\providecommand \@@endlink[0]{}%
\providecommand \url  [0]{\begingroup\@sanitize@url \@url }%
\providecommand \@url [1]{\endgroup\@href {#1}{\urlprefix }}%
\providecommand \urlprefix  [0]{URL }%
\providecommand \Eprint [0]{\href }%
\providecommand \doibase [0]{https://doi.org/}%
\providecommand \selectlanguage [0]{\@gobble}%
\providecommand \bibinfo  [0]{\@secondoftwo}%
\providecommand \bibfield  [0]{\@secondoftwo}%
\providecommand \translation [1]{[#1]}%
\providecommand \BibitemOpen [0]{}%
\providecommand \bibitemStop [0]{}%
\providecommand \bibitemNoStop [0]{.\EOS\space}%
\providecommand \EOS [0]{\spacefactor3000\relax}%
\providecommand \BibitemShut  [1]{\csname bibitem#1\endcsname}%
\let\auto@bib@innerbib\@empty
\bibitem [{\citenamefont {Shenker}\ and\ \citenamefont
  {Stanford}(2014)}]{shenker_black_2014}%
  \BibitemOpen
  \bibfield  {author} {\bibinfo {author} {\bibfnamefont {S.~H.}\ \bibnamefont
  {Shenker}}\ and\ \bibinfo {author} {\bibfnamefont {D.}~\bibnamefont
  {Stanford}},\ }\bibfield  {title} {{\selectlanguage {en}\bibinfo {title}
  {Black holes and the butterfly effect}},\ }\href
  {https://doi.org/10.1007/JHEP03(2014)067} {\bibfield  {journal} {\bibinfo
  {journal} {Journal of High Energy Physics}\ }\textbf {\bibinfo {volume}
  {2014}},\ \bibinfo {pages} {67} (\bibinfo {year} {2014})}\BibitemShut
  {NoStop}%
\bibitem [{\citenamefont {Maldacena}\ \emph {et~al.}(2016)\citenamefont
  {Maldacena}, \citenamefont {Shenker},\ and\ \citenamefont
  {Stanford}}]{maldacena_bound_2015}%
  \BibitemOpen
  \bibfield  {author} {\bibinfo {author} {\bibfnamefont {J.}~\bibnamefont
  {Maldacena}}, \bibinfo {author} {\bibfnamefont {S.~H.}\ \bibnamefont
  {Shenker}},\ and\ \bibinfo {author} {\bibfnamefont {D.}~\bibnamefont
  {Stanford}},\ }\bibfield  {title} {\bibinfo {title} {A bound on chaos},\
  }\href {https://doi.org/10.1007/JHEP08(2016)106} {\bibfield  {journal}
  {\bibinfo  {journal} {Journal of High Energy Physics}\ }\textbf {\bibinfo
  {volume} {2016}},\ \bibinfo {pages} {106} (\bibinfo {year}
  {2016})}\BibitemShut {NoStop}%
\bibitem [{\citenamefont {Sachdev}\ and\ \citenamefont
  {Ye}(1993)}]{sachdev_gapless_1993}%
  \BibitemOpen
  \bibfield  {author} {\bibinfo {author} {\bibfnamefont {S.}~\bibnamefont
  {Sachdev}}\ and\ \bibinfo {author} {\bibfnamefont {J.}~\bibnamefont {Ye}},\
  }\bibfield  {title} {\bibinfo {title} {Gapless spin-fluid ground state in a
  random quantum {Heisenberg} magnet},\ }\href
  {https://link.aps.org/doi/10.1103/PhysRevLett.70.3339} {\bibfield  {journal}
  {\bibinfo  {journal} {Physical Review Letters}\ }\textbf {\bibinfo {volume}
  {70}},\ \bibinfo {pages} {3339} (\bibinfo {year} {1993})}\BibitemShut
  {NoStop}%
\bibitem [{\citenamefont {Kitaev}(2015)}]{kitaev2015}%
  \BibitemOpen
  \bibfield  {author} {\bibinfo {author} {\bibfnamefont {A.}~\bibnamefont
  {Kitaev}},\ }\href@noop {} {} (\bibinfo {year} {2015}),\ \bibinfo {note}
  {talks at KITP, April 7, 2015 and May 27, 2015}\BibitemShut {NoStop}%
\bibitem [{\citenamefont {Nahum}\ \emph {et~al.}(2018)\citenamefont {Nahum},
  \citenamefont {Vijay},\ and\ \citenamefont {Haah}}]{nahum_operator_2018}%
  \BibitemOpen
  \bibfield  {author} {\bibinfo {author} {\bibfnamefont {A.}~\bibnamefont
  {Nahum}}, \bibinfo {author} {\bibfnamefont {S.}~\bibnamefont {Vijay}},\ and\
  \bibinfo {author} {\bibfnamefont {J.}~\bibnamefont {Haah}},\ }\bibfield
  {title} {\bibinfo {title} {Operator {Spreading} in {Random} {Unitary}
  {Circuits}},\ }\href {https://link.aps.org/doi/10.1103/PhysRevX.8.021014}
  {\bibfield  {journal} {\bibinfo  {journal} {Physical Review X}\ }\textbf
  {\bibinfo {volume} {8}},\ \bibinfo {pages} {021014} (\bibinfo {year}
  {2018})}\BibitemShut {NoStop}%
\bibitem [{\citenamefont {von Keyserlingk}\ \emph {et~al.}(2018)\citenamefont
  {von Keyserlingk}, \citenamefont {Rakovszky}, \citenamefont {Pollmann},\ and\
  \citenamefont {Sondhi}}]{von_keyserlingk_operator_2018}%
  \BibitemOpen
  \bibfield  {author} {\bibinfo {author} {\bibfnamefont {C.}~\bibnamefont {von
  Keyserlingk}}, \bibinfo {author} {\bibfnamefont {T.}~\bibnamefont
  {Rakovszky}}, \bibinfo {author} {\bibfnamefont {F.}~\bibnamefont
  {Pollmann}},\ and\ \bibinfo {author} {\bibfnamefont {S.~L.}\ \bibnamefont
  {Sondhi}},\ }\bibfield  {title} {\bibinfo {title} {Operator {Hydrodynamics},
  {OTOCs}, and {Entanglement} {Growth} in {Systems} without {Conservation}
  {Laws}},\ }\href {https://link.aps.org/doi/10.1103/PhysRevX.8.021013}
  {\bibfield  {journal} {\bibinfo  {journal} {Physical Review X}\ }\textbf
  {\bibinfo {volume} {8}},\ \bibinfo {pages} {021013} (\bibinfo {year}
  {2018})}\BibitemShut {NoStop}%
\bibitem [{\citenamefont {Zhou}\ \emph {et~al.}(2020)\citenamefont {Zhou},
  \citenamefont {Xu}, \citenamefont {Chen}, \citenamefont {Guo},\ and\
  \citenamefont {Swingle}}]{zhou_operator_2020}%
  \BibitemOpen
  \bibfield  {author} {\bibinfo {author} {\bibfnamefont {T.}~\bibnamefont
  {Zhou}}, \bibinfo {author} {\bibfnamefont {S.}~\bibnamefont {Xu}}, \bibinfo
  {author} {\bibfnamefont {X.}~\bibnamefont {Chen}}, \bibinfo {author}
  {\bibfnamefont {A.}~\bibnamefont {Guo}},\ and\ \bibinfo {author}
  {\bibfnamefont {B.}~\bibnamefont {Swingle}},\ }\bibfield  {title} {\bibinfo
  {title} {Operator {{L}}\textbackslash{}'evy {{Flight}}: {{Light Cones}} in
  {{Chaotic Long}}-{{Range Interacting Systems}}},\ }\href
  {https://doi.org/10.1103/PhysRevLett.124.180601} {\bibfield  {journal}
  {\bibinfo  {journal} {Physical Review Letters}\ }\textbf {\bibinfo {volume}
  {124}},\ \bibinfo {pages} {180601} (\bibinfo {year} {2020})}\BibitemShut
  {NoStop}%
\bibitem [{\citenamefont {Xu}\ and\ \citenamefont
  {Swingle}(2018)}]{xu_locality_2018}%
  \BibitemOpen
  \bibfield  {author} {\bibinfo {author} {\bibfnamefont {S.}~\bibnamefont
  {Xu}}\ and\ \bibinfo {author} {\bibfnamefont {B.}~\bibnamefont {Swingle}},\
  }\bibfield  {title} {\bibinfo {title} {Locality, {Quantum} {Fluctuations},
  and {Scrambling}},\ }\href {http://arxiv.org/abs/1805.05376} {\bibfield
  {journal} {\bibinfo  {journal} {arXiv:1805.05376 [cond-mat, physics:hep-th,
  physics:quant-ph]}\ } (\bibinfo {year} {2018})},\ \bibinfo {note} {arXiv:
  1805.05376}\BibitemShut {NoStop}%
\bibitem [{\citenamefont {Chen}\ and\ \citenamefont
  {Zhou}(2019)}]{chen_quantum_2018}%
  \BibitemOpen
  \bibfield  {author} {\bibinfo {author} {\bibfnamefont {X.}~\bibnamefont
  {Chen}}\ and\ \bibinfo {author} {\bibfnamefont {T.}~\bibnamefont {Zhou}},\
  }\bibfield  {title} {\bibinfo {title} {Quantum chaos dynamics in long-range
  power law interaction systems},\ }\href
  {https://link.aps.org/doi/10.1103/PhysRevB.100.064305} {\bibfield  {journal}
  {\bibinfo  {journal} {Phys. Rev. B}\ }\textbf {\bibinfo {volume} {100}},\
  \bibinfo {pages} {064305} (\bibinfo {year} {2019})}\BibitemShut {NoStop}%
\bibitem [{\citenamefont {Zhou}\ and\ \citenamefont
  {Chen}(2018)}]{zhou_operator_2018}%
  \BibitemOpen
  \bibfield  {author} {\bibinfo {author} {\bibfnamefont {T.}~\bibnamefont
  {Zhou}}\ and\ \bibinfo {author} {\bibfnamefont {X.}~\bibnamefont {Chen}},\
  }\bibfield  {title} {\bibinfo {title} {Operator {Dynamics} in {Brownian}
  {Quantum} {Circuit}},\ }\href {http://arxiv.org/abs/1805.09307} {\bibfield
  {journal} {\bibinfo  {journal} {arXiv:1805.09307 [cond-mat, physics:hep-th]}\
  } (\bibinfo {year} {2018})},\ \bibinfo {note} {arXiv: 1805.09307}\BibitemShut
  {NoStop}%
\bibitem [{\citenamefont {Roberts}\ \emph {et~al.}(2018)\citenamefont
  {Roberts}, \citenamefont {Stanford},\ and\ \citenamefont
  {Streicher}}]{roberts_operator_2018}%
  \BibitemOpen
  \bibfield  {author} {\bibinfo {author} {\bibfnamefont {D.~A.}\ \bibnamefont
  {Roberts}}, \bibinfo {author} {\bibfnamefont {D.}~\bibnamefont {Stanford}},\
  and\ \bibinfo {author} {\bibfnamefont {A.}~\bibnamefont {Streicher}},\
  }\bibfield  {title} {{\selectlanguage {en}\bibinfo {title} {Operator growth
  in the {SYK} model}},\ }\href {https://doi.org/10.1007/JHEP06(2018)122}
  {\bibfield  {journal} {\bibinfo  {journal} {Journal of High Energy Physics}\
  }\textbf {\bibinfo {volume} {2018}},\ \bibinfo {pages} {122} (\bibinfo {year}
  {2018})}\BibitemShut {NoStop}%
\bibitem [{\citenamefont {Qi}\ and\ \citenamefont
  {Streicher}(2018)}]{qi_quantum_2018}%
  \BibitemOpen
  \bibfield  {author} {\bibinfo {author} {\bibfnamefont {X.-L.}\ \bibnamefont
  {Qi}}\ and\ \bibinfo {author} {\bibfnamefont {A.}~\bibnamefont {Streicher}},\
  }\bibfield  {title} {\bibinfo {title} {Quantum {{Epidemiology}}: {{Operator
  Growth}}, {{Thermal Effects}}, and {{SYK}}},\ }\href@noop {} {\bibfield
  {journal} {\bibinfo  {journal} {arXiv:1810.11958 [cond-mat, physics:hep-th,
  physics:quant-ph]}\ } (\bibinfo {year} {2018})},\ \Eprint
  {https://arxiv.org/abs/1810.11958} {arXiv:1810.11958 [cond-mat,
  physics:hep-th, physics:quant-ph]} \BibitemShut {NoStop}%
\bibitem [{\citenamefont {Fisher}(1937)}]{fisher_wave_1937}%
  \BibitemOpen
  \bibfield  {author} {\bibinfo {author} {\bibfnamefont {R.~A.}\ \bibnamefont
  {Fisher}},\ }\bibfield  {title} {{\selectlanguage {en}\bibinfo {title} {The
  {{Wave}} of {{Advance}} of {{Advantageous Genes}}}},\ }\href
  {https://doi.org/10.1111/j.1469-1809.1937.tb02153.x} {\bibfield  {journal}
  {\bibinfo  {journal} {Annals of Eugenics}\ }\textbf {\bibinfo {volume} {7}},\
  \bibinfo {pages} {355} (\bibinfo {year} {1937})},\ \bibinfo {note} {\_eprint:
  https://onlinelibrary.wiley.com/doi/pdf/10.1111/j.1469-1809.1937.tb02153.x}\BibitemShut
  {NoStop}%
\bibitem [{\citenamefont {Kolmogorov}\ \emph {et~al.}(1937)\citenamefont
  {Kolmogorov}, \citenamefont {Petrovsky},\ and\ \citenamefont
  {Piskunov}}]{kolmogorov_investigation_1937}%
  \BibitemOpen
  \bibfield  {author} {\bibinfo {author} {\bibfnamefont {A.}~\bibnamefont
  {Kolmogorov}}, \bibinfo {author} {\bibfnamefont {I.}~\bibnamefont
  {Petrovsky}},\ and\ \bibinfo {author} {\bibfnamefont {N.}~\bibnamefont
  {Piskunov}},\ }\bibfield  {title} {\bibinfo {title} {Investigation of the
  {{Equation}} of {{Diffusion Combined}} with {{Increasing}} of the
  {{Substance}} and {{Its Application}} to a {{Biology Problem}}},\ }\href@noop
  {} {\bibfield  {journal} {\bibinfo  {journal} {Bulletin of Moscow State
  University Series A: Mathematics and Mechanics}\ ,\ \bibinfo {pages} {1}}
  (\bibinfo {year} {1937})}\BibitemShut {NoStop}%
\bibitem [{\citenamefont {Brunet}(2016)}]{brunet_aspects_2016}%
  \BibitemOpen
  \bibfield  {author} {\bibinfo {author} {\bibfnamefont {{\'E}.}~\bibnamefont
  {Brunet}},\ }\bibfield  {title} {\bibinfo {title} {Some aspects of the
  {{Fisher}}-{{KPP}} equation and the branching {{Brownian}} motion}\
  }(\bibinfo {year} {2016})\BibitemShut {NoStop}%
\bibitem [{\citenamefont {Chen}\ and\ \citenamefont
  {Zhou}(2018)}]{chen_operator_2018}%
  \BibitemOpen
  \bibfield  {author} {\bibinfo {author} {\bibfnamefont {X.}~\bibnamefont
  {Chen}}\ and\ \bibinfo {author} {\bibfnamefont {T.}~\bibnamefont {Zhou}},\
  }\bibfield  {title} {\bibinfo {title} {Operator scrambling and quantum
  chaos},\ }\href@noop {} {\bibfield  {journal} {\bibinfo  {journal}
  {arXiv:1804.08655 [cond-mat, physics:hep-th, physics:quant-ph]}\ } (\bibinfo
  {year} {2018})},\ \Eprint {https://arxiv.org/abs/1804.08655}
  {arXiv:1804.08655 [cond-mat, physics:hep-th, physics:quant-ph]} \BibitemShut
  {NoStop}%
\bibitem [{\citenamefont {Rakovszky}\ \emph {et~al.}(2018)\citenamefont
  {Rakovszky}, \citenamefont {Pollmann},\ and\ \citenamefont {von
  Keyserlingk}}]{rakovszky_diffusive_2017}%
  \BibitemOpen
  \bibfield  {author} {\bibinfo {author} {\bibfnamefont {T.}~\bibnamefont
  {Rakovszky}}, \bibinfo {author} {\bibfnamefont {F.}~\bibnamefont
  {Pollmann}},\ and\ \bibinfo {author} {\bibfnamefont {C.~W.}\ \bibnamefont
  {von Keyserlingk}},\ }\bibfield  {title} {\bibinfo {title} {Diffusive
  hydrodynamics of out-of-time-ordered correlators with charge conservation},\
  }\href {https://link.aps.org/doi/10.1103/PhysRevX.8.031058} {\bibfield
  {journal} {\bibinfo  {journal} {Phys. Rev. X}\ }\textbf {\bibinfo {volume}
  {8}},\ \bibinfo {pages} {031058} (\bibinfo {year} {2018})}\BibitemShut
  {NoStop}%
\bibitem [{\citenamefont {Khemani}\ \emph {et~al.}(2018)\citenamefont
  {Khemani}, \citenamefont {Vishwanath},\ and\ \citenamefont
  {Huse}}]{khemani_operator_2017}%
  \BibitemOpen
  \bibfield  {author} {\bibinfo {author} {\bibfnamefont {V.}~\bibnamefont
  {Khemani}}, \bibinfo {author} {\bibfnamefont {A.}~\bibnamefont
  {Vishwanath}},\ and\ \bibinfo {author} {\bibfnamefont {D.~A.}\ \bibnamefont
  {Huse}},\ }\bibfield  {title} {\bibinfo {title} {Operator spreading and the
  emergence of dissipative hydrodynamics under unitary evolution with
  conservation laws},\ }\href
  {https://link.aps.org/doi/10.1103/PhysRevX.8.031057} {\bibfield  {journal}
  {\bibinfo  {journal} {Phys. Rev. X}\ }\textbf {\bibinfo {volume} {8}},\
  \bibinfo {pages} {031057} (\bibinfo {year} {2018})}\BibitemShut {NoStop}%
\bibitem [{\citenamefont {Aleiner}\ \emph {et~al.}(2016)\citenamefont
  {Aleiner}, \citenamefont {Faoro},\ and\ \citenamefont
  {Ioffe}}]{aleiner_microscopic_2016}%
  \BibitemOpen
  \bibfield  {author} {\bibinfo {author} {\bibfnamefont {I.~L.}\ \bibnamefont
  {Aleiner}}, \bibinfo {author} {\bibfnamefont {L.}~\bibnamefont {Faoro}},\
  and\ \bibinfo {author} {\bibfnamefont {L.~B.}\ \bibnamefont {Ioffe}},\
  }\bibfield  {title} {\bibinfo {title} {Microscopic model of quantum butterfly
  effect: Out-of-time-order correlators and traveling combustion waves},\
  }\href@noop {} {\bibfield  {journal} {\bibinfo  {journal} {arXiv:1609.01251
  [cond-mat, physics:hep-th, physics:quant-ph]}\ } (\bibinfo {year} {2016})},\
  \Eprint {https://arxiv.org/abs/1609.01251} {arXiv:1609.01251 [cond-mat,
  physics:hep-th, physics:quant-ph]} \BibitemShut {NoStop}%
\bibitem [{\citenamefont {Ablowitz}\ and\ \citenamefont
  {Zeppetella}(1979)}]{ablowitz_explicit_1979}%
  \BibitemOpen
  \bibfield  {author} {\bibinfo {author} {\bibfnamefont {M.~J.}\ \bibnamefont
  {Ablowitz}}\ and\ \bibinfo {author} {\bibfnamefont {A.}~\bibnamefont
  {Zeppetella}},\ }\bibfield  {title} {{\selectlanguage {en}\bibinfo {title}
  {Explicit solutions of {{Fisher}}'s equation for a special wave speed}},\
  }\href {https://doi.org/10.1016/S0092-8240(79)80020-8} {\bibfield  {journal}
  {\bibinfo  {journal} {Bulletin of Mathematical Biology}\ }\textbf {\bibinfo
  {volume} {41}},\ \bibinfo {pages} {835} (\bibinfo {year} {1979})}\BibitemShut
  {NoStop}%
\bibitem [{\citenamefont {Mi}\ \emph {et~al.}(2021)\citenamefont {Mi},
  \citenamefont {Roushan}, \citenamefont {Quintana}, \citenamefont
  {Mandr{\`a}}, \citenamefont {Marshall}, \citenamefont {Neill}, \citenamefont
  {Arute}, \citenamefont {Arya}, \citenamefont {Atalaya}, \citenamefont
  {Babbush}, \citenamefont {Bardin}, \citenamefont {Barends}, \citenamefont
  {Basso}, \citenamefont {Bengtsson}, \citenamefont {Boixo}, \citenamefont
  {Bourassa}, \citenamefont {Broughton}, \citenamefont {Buckley}, \citenamefont
  {Buell}, \citenamefont {Burkett}, \citenamefont {Bushnell}, \citenamefont
  {Chen}, \citenamefont {Chiaro}, \citenamefont {Collins}, \citenamefont
  {Courtney}, \citenamefont {Demura}, \citenamefont {Derk}, \citenamefont
  {Dunsworth}, \citenamefont {Eppens}, \citenamefont {Erickson}, \citenamefont
  {Farhi}, \citenamefont {Fowler}, \citenamefont {Foxen}, \citenamefont
  {Gidney}, \citenamefont {Giustina}, \citenamefont {Gross}, \citenamefont
  {Harrigan}, \citenamefont {Harrington}, \citenamefont {Hilton}, \citenamefont
  {Ho}, \citenamefont {Hong}, \citenamefont {Huang}, \citenamefont {Huggins},
  \citenamefont {Ioffe}, \citenamefont {Isakov}, \citenamefont {Jeffrey},
  \citenamefont {Jiang}, \citenamefont {Jones}, \citenamefont {Kafri},
  \citenamefont {Kelly}, \citenamefont {Kim}, \citenamefont {Kitaev},
  \citenamefont {Klimov}, \citenamefont {Korotkov}, \citenamefont {Kostritsa},
  \citenamefont {Landhuis}, \citenamefont {Laptev}, \citenamefont {Lucero},
  \citenamefont {Martin}, \citenamefont {McClean}, \citenamefont {McCourt},
  \citenamefont {McEwen}, \citenamefont {Megrant}, \citenamefont {Miao},
  \citenamefont {Mohseni}, \citenamefont {Montazeri}, \citenamefont
  {Mruczkiewicz}, \citenamefont {Mutus}, \citenamefont {Naaman}, \citenamefont
  {Neeley}, \citenamefont {Newman}, \citenamefont {Niu}, \citenamefont
  {O'Brien}, \citenamefont {Opremcak}, \citenamefont {Ostby}, \citenamefont
  {Pato}, \citenamefont {Petukhov}, \citenamefont {Redd}, \citenamefont
  {Rubin}, \citenamefont {Sank}, \citenamefont {Satzinger}, \citenamefont
  {Shvarts}, \citenamefont {Strain}, \citenamefont {Szalay}, \citenamefont
  {Trevithick}, \citenamefont {Villalonga}, \citenamefont {White},
  \citenamefont {Yao}, \citenamefont {Yeh}, \citenamefont {Zalcman},
  \citenamefont {Neven}, \citenamefont {Aleiner}, \citenamefont {Kechedzhi},
  \citenamefont {Smelyanskiy},\ and\ \citenamefont
  {Chen}}]{mi_information_2021-1}%
  \BibitemOpen
  \bibfield  {author} {\bibinfo {author} {\bibfnamefont {X.}~\bibnamefont
  {Mi}}, \bibinfo {author} {\bibfnamefont {P.}~\bibnamefont {Roushan}},
  \bibinfo {author} {\bibfnamefont {C.}~\bibnamefont {Quintana}}, \bibinfo
  {author} {\bibfnamefont {S.}~\bibnamefont {Mandr{\`a}}}, \bibinfo {author}
  {\bibfnamefont {J.}~\bibnamefont {Marshall}}, \bibinfo {author}
  {\bibfnamefont {C.}~\bibnamefont {Neill}}, \bibinfo {author} {\bibfnamefont
  {F.}~\bibnamefont {Arute}}, \bibinfo {author} {\bibfnamefont
  {K.}~\bibnamefont {Arya}}, \bibinfo {author} {\bibfnamefont {J.}~\bibnamefont
  {Atalaya}}, \bibinfo {author} {\bibfnamefont {R.}~\bibnamefont {Babbush}},
  \bibinfo {author} {\bibfnamefont {J.~C.}\ \bibnamefont {Bardin}}, \bibinfo
  {author} {\bibfnamefont {R.}~\bibnamefont {Barends}}, \bibinfo {author}
  {\bibfnamefont {J.}~\bibnamefont {Basso}}, \bibinfo {author} {\bibfnamefont
  {A.}~\bibnamefont {Bengtsson}}, \bibinfo {author} {\bibfnamefont
  {S.}~\bibnamefont {Boixo}}, \bibinfo {author} {\bibfnamefont
  {A.}~\bibnamefont {Bourassa}}, \bibinfo {author} {\bibfnamefont
  {M.}~\bibnamefont {Broughton}}, \bibinfo {author} {\bibfnamefont {B.~B.}\
  \bibnamefont {Buckley}}, \bibinfo {author} {\bibfnamefont {D.~A.}\
  \bibnamefont {Buell}}, \bibinfo {author} {\bibfnamefont {B.}~\bibnamefont
  {Burkett}}, \bibinfo {author} {\bibfnamefont {N.}~\bibnamefont {Bushnell}},
  \bibinfo {author} {\bibfnamefont {Z.}~\bibnamefont {Chen}}, \bibinfo {author}
  {\bibfnamefont {B.}~\bibnamefont {Chiaro}}, \bibinfo {author} {\bibfnamefont
  {R.}~\bibnamefont {Collins}}, \bibinfo {author} {\bibfnamefont
  {W.}~\bibnamefont {Courtney}}, \bibinfo {author} {\bibfnamefont
  {S.}~\bibnamefont {Demura}}, \bibinfo {author} {\bibfnamefont {A.~R.}\
  \bibnamefont {Derk}}, \bibinfo {author} {\bibfnamefont {A.}~\bibnamefont
  {Dunsworth}}, \bibinfo {author} {\bibfnamefont {D.}~\bibnamefont {Eppens}},
  \bibinfo {author} {\bibfnamefont {C.}~\bibnamefont {Erickson}}, \bibinfo
  {author} {\bibfnamefont {E.}~\bibnamefont {Farhi}}, \bibinfo {author}
  {\bibfnamefont {A.~G.}\ \bibnamefont {Fowler}}, \bibinfo {author}
  {\bibfnamefont {B.}~\bibnamefont {Foxen}}, \bibinfo {author} {\bibfnamefont
  {C.}~\bibnamefont {Gidney}}, \bibinfo {author} {\bibfnamefont
  {M.}~\bibnamefont {Giustina}}, \bibinfo {author} {\bibfnamefont {J.~A.}\
  \bibnamefont {Gross}}, \bibinfo {author} {\bibfnamefont {M.~P.}\ \bibnamefont
  {Harrigan}}, \bibinfo {author} {\bibfnamefont {S.~D.}\ \bibnamefont
  {Harrington}}, \bibinfo {author} {\bibfnamefont {J.}~\bibnamefont {Hilton}},
  \bibinfo {author} {\bibfnamefont {A.}~\bibnamefont {Ho}}, \bibinfo {author}
  {\bibfnamefont {S.}~\bibnamefont {Hong}}, \bibinfo {author} {\bibfnamefont
  {T.}~\bibnamefont {Huang}}, \bibinfo {author} {\bibfnamefont {W.~J.}\
  \bibnamefont {Huggins}}, \bibinfo {author} {\bibfnamefont {L.~B.}\
  \bibnamefont {Ioffe}}, \bibinfo {author} {\bibfnamefont {S.~V.}\ \bibnamefont
  {Isakov}}, \bibinfo {author} {\bibfnamefont {E.}~\bibnamefont {Jeffrey}},
  \bibinfo {author} {\bibfnamefont {Z.}~\bibnamefont {Jiang}}, \bibinfo
  {author} {\bibfnamefont {C.}~\bibnamefont {Jones}}, \bibinfo {author}
  {\bibfnamefont {D.}~\bibnamefont {Kafri}}, \bibinfo {author} {\bibfnamefont
  {J.}~\bibnamefont {Kelly}}, \bibinfo {author} {\bibfnamefont
  {S.}~\bibnamefont {Kim}}, \bibinfo {author} {\bibfnamefont {A.}~\bibnamefont
  {Kitaev}}, \bibinfo {author} {\bibfnamefont {P.~V.}\ \bibnamefont {Klimov}},
  \bibinfo {author} {\bibfnamefont {A.~N.}\ \bibnamefont {Korotkov}}, \bibinfo
  {author} {\bibfnamefont {F.}~\bibnamefont {Kostritsa}}, \bibinfo {author}
  {\bibfnamefont {D.}~\bibnamefont {Landhuis}}, \bibinfo {author}
  {\bibfnamefont {P.}~\bibnamefont {Laptev}}, \bibinfo {author} {\bibfnamefont
  {E.}~\bibnamefont {Lucero}}, \bibinfo {author} {\bibfnamefont
  {O.}~\bibnamefont {Martin}}, \bibinfo {author} {\bibfnamefont {J.~R.}\
  \bibnamefont {McClean}}, \bibinfo {author} {\bibfnamefont {T.}~\bibnamefont
  {McCourt}}, \bibinfo {author} {\bibfnamefont {M.}~\bibnamefont {McEwen}},
  \bibinfo {author} {\bibfnamefont {A.}~\bibnamefont {Megrant}}, \bibinfo
  {author} {\bibfnamefont {K.~C.}\ \bibnamefont {Miao}}, \bibinfo {author}
  {\bibfnamefont {M.}~\bibnamefont {Mohseni}}, \bibinfo {author} {\bibfnamefont
  {S.}~\bibnamefont {Montazeri}}, \bibinfo {author} {\bibfnamefont
  {W.}~\bibnamefont {Mruczkiewicz}}, \bibinfo {author} {\bibfnamefont
  {J.}~\bibnamefont {Mutus}}, \bibinfo {author} {\bibfnamefont
  {O.}~\bibnamefont {Naaman}}, \bibinfo {author} {\bibfnamefont
  {M.}~\bibnamefont {Neeley}}, \bibinfo {author} {\bibfnamefont
  {M.}~\bibnamefont {Newman}}, \bibinfo {author} {\bibfnamefont {M.~Y.}\
  \bibnamefont {Niu}}, \bibinfo {author} {\bibfnamefont {T.~E.}\ \bibnamefont
  {O'Brien}}, \bibinfo {author} {\bibfnamefont {A.}~\bibnamefont {Opremcak}},
  \bibinfo {author} {\bibfnamefont {E.}~\bibnamefont {Ostby}}, \bibinfo
  {author} {\bibfnamefont {B.}~\bibnamefont {Pato}}, \bibinfo {author}
  {\bibfnamefont {A.}~\bibnamefont {Petukhov}}, \bibinfo {author}
  {\bibfnamefont {N.}~\bibnamefont {Redd}}, \bibinfo {author} {\bibfnamefont
  {N.~C.}\ \bibnamefont {Rubin}}, \bibinfo {author} {\bibfnamefont
  {D.}~\bibnamefont {Sank}}, \bibinfo {author} {\bibfnamefont {K.~J.}\
  \bibnamefont {Satzinger}}, \bibinfo {author} {\bibfnamefont {V.}~\bibnamefont
  {Shvarts}}, \bibinfo {author} {\bibfnamefont {D.}~\bibnamefont {Strain}},
  \bibinfo {author} {\bibfnamefont {M.}~\bibnamefont {Szalay}}, \bibinfo
  {author} {\bibfnamefont {M.~D.}\ \bibnamefont {Trevithick}}, \bibinfo
  {author} {\bibfnamefont {B.}~\bibnamefont {Villalonga}}, \bibinfo {author}
  {\bibfnamefont {T.}~\bibnamefont {White}}, \bibinfo {author} {\bibfnamefont
  {Z.~J.}\ \bibnamefont {Yao}}, \bibinfo {author} {\bibfnamefont
  {P.}~\bibnamefont {Yeh}}, \bibinfo {author} {\bibfnamefont {A.}~\bibnamefont
  {Zalcman}}, \bibinfo {author} {\bibfnamefont {H.}~\bibnamefont {Neven}},
  \bibinfo {author} {\bibfnamefont {I.}~\bibnamefont {Aleiner}}, \bibinfo
  {author} {\bibfnamefont {K.}~\bibnamefont {Kechedzhi}}, \bibinfo {author}
  {\bibfnamefont {V.}~\bibnamefont {Smelyanskiy}},\ and\ \bibinfo {author}
  {\bibfnamefont {Y.}~\bibnamefont {Chen}},\ }\bibfield  {title} {\bibinfo
  {title} {Information scrambling in quantum circuits},\ }\href
  {https://doi.org/10.1126/science.abg5029} {\bibfield  {journal} {\bibinfo
  {journal} {Science}\ }\textbf {\bibinfo {volume} {374}},\ \bibinfo {pages}
  {1479} (\bibinfo {year} {2021})}\BibitemShut {NoStop}%
\bibitem [{\citenamefont {Li}\ \emph {et~al.}(2017)\citenamefont {Li},
  \citenamefont {Fan}, \citenamefont {Wang}, \citenamefont {Ye}, \citenamefont
  {Zeng}, \citenamefont {Zhai}, \citenamefont {Peng},\ and\ \citenamefont
  {Du}}]{li_measuring_2017}%
  \BibitemOpen
  \bibfield  {author} {\bibinfo {author} {\bibfnamefont {J.}~\bibnamefont
  {Li}}, \bibinfo {author} {\bibfnamefont {R.}~\bibnamefont {Fan}}, \bibinfo
  {author} {\bibfnamefont {H.}~\bibnamefont {Wang}}, \bibinfo {author}
  {\bibfnamefont {B.}~\bibnamefont {Ye}}, \bibinfo {author} {\bibfnamefont
  {B.}~\bibnamefont {Zeng}}, \bibinfo {author} {\bibfnamefont {H.}~\bibnamefont
  {Zhai}}, \bibinfo {author} {\bibfnamefont {X.}~\bibnamefont {Peng}},\ and\
  \bibinfo {author} {\bibfnamefont {J.}~\bibnamefont {Du}},\ }\bibfield
  {title} {\bibinfo {title} {Measuring {{Out-of-Time-Order Correlators}} on a
  {{Nuclear Magnetic Resonance Quantum Simulator}}},\ }\href
  {https://doi.org/10.1103/PhysRevX.7.031011} {\bibfield  {journal} {\bibinfo
  {journal} {Physical Review X}\ }\textbf {\bibinfo {volume} {7}},\ \bibinfo
  {pages} {031011} (\bibinfo {year} {2017})}\BibitemShut {NoStop}%
\bibitem [{\citenamefont {G{\"a}rttner}\ \emph {et~al.}(2017)\citenamefont
  {G{\"a}rttner}, \citenamefont {Bohnet}, \citenamefont {Safavi-Naini},
  \citenamefont {Wall}, \citenamefont {Bollinger},\ and\ \citenamefont
  {Rey}}]{Bollinger17}%
  \BibitemOpen
  \bibfield  {author} {\bibinfo {author} {\bibfnamefont {M.}~\bibnamefont
  {G{\"a}rttner}}, \bibinfo {author} {\bibfnamefont {J.~G.}\ \bibnamefont
  {Bohnet}}, \bibinfo {author} {\bibfnamefont {A.}~\bibnamefont
  {Safavi-Naini}}, \bibinfo {author} {\bibfnamefont {M.~L.}\ \bibnamefont
  {Wall}}, \bibinfo {author} {\bibfnamefont {J.~J.}\ \bibnamefont
  {Bollinger}},\ and\ \bibinfo {author} {\bibfnamefont {A.~M.}\ \bibnamefont
  {Rey}},\ }\bibfield  {title} {\bibinfo {title} {Measuring out-of-time-order
  correlations and multiple quantum spectra in a trapped-ion quantum magnet},\
  }\href {https://doi.org/10.1038/nphys4119} {\bibfield  {journal} {\bibinfo
  {journal} {Nat. Phys.}\ }\textbf {\bibinfo {volume} {13}},\ \bibinfo {pages}
  {781 EP } (\bibinfo {year} {2017})}\BibitemShut {NoStop}%
\bibitem [{\citenamefont {Landsman}\ \emph {et~al.}(2019)\citenamefont
  {Landsman}, \citenamefont {Figgatt}, \citenamefont {Schuster}, \citenamefont
  {Linke}, \citenamefont {Yoshida}, \citenamefont {Yao},\ and\ \citenamefont
  {Monroe}}]{landsman_verified_2019}%
  \BibitemOpen
  \bibfield  {author} {\bibinfo {author} {\bibfnamefont {K.~A.}\ \bibnamefont
  {Landsman}}, \bibinfo {author} {\bibfnamefont {C.}~\bibnamefont {Figgatt}},
  \bibinfo {author} {\bibfnamefont {T.}~\bibnamefont {Schuster}}, \bibinfo
  {author} {\bibfnamefont {N.~M.}\ \bibnamefont {Linke}}, \bibinfo {author}
  {\bibfnamefont {B.}~\bibnamefont {Yoshida}}, \bibinfo {author} {\bibfnamefont
  {N.~Y.}\ \bibnamefont {Yao}},\ and\ \bibinfo {author} {\bibfnamefont
  {C.}~\bibnamefont {Monroe}},\ }\bibfield  {title} {\bibinfo {title} {Verified
  quantum information scrambling},\ }\href
  {https://doi.org/10.1038/s41586-019-0952-6} {\bibfield  {journal} {\bibinfo
  {journal} {Nature}\ }\textbf {\bibinfo {volume} {567}},\ \bibinfo {pages}
  {61} (\bibinfo {year} {2019})}\BibitemShut {NoStop}%
\bibitem [{\citenamefont {Yao}\ \emph {et~al.}(2016)\citenamefont {Yao},
  \citenamefont {Grusdt}, \citenamefont {Swingle}, \citenamefont {Lukin},
  \citenamefont {{Stamper-Kurn}}, \citenamefont {Moore},\ and\ \citenamefont
  {Demler}}]{yao_interferometric_2016-1}%
  \BibitemOpen
  \bibfield  {author} {\bibinfo {author} {\bibfnamefont {N.~Y.}\ \bibnamefont
  {Yao}}, \bibinfo {author} {\bibfnamefont {F.}~\bibnamefont {Grusdt}},
  \bibinfo {author} {\bibfnamefont {B.}~\bibnamefont {Swingle}}, \bibinfo
  {author} {\bibfnamefont {M.~D.}\ \bibnamefont {Lukin}}, \bibinfo {author}
  {\bibfnamefont {D.~M.}\ \bibnamefont {{Stamper-Kurn}}}, \bibinfo {author}
  {\bibfnamefont {J.~E.}\ \bibnamefont {Moore}},\ and\ \bibinfo {author}
  {\bibfnamefont {E.~A.}\ \bibnamefont {Demler}},\ }\bibfield  {title}
  {\bibinfo {title} {Interferometric {{Approach}} to {{Probing Fast
  Scrambling}}},\ }\href@noop {} {\bibfield  {journal} {\bibinfo  {journal}
  {arXiv:1607.01801 [cond-mat, physics:hep-th, physics:quant-ph]}\ } (\bibinfo
  {year} {2016})},\ \Eprint {https://arxiv.org/abs/1607.01801}
  {arXiv:1607.01801 [cond-mat, physics:hep-th, physics:quant-ph]} \BibitemShut
  {NoStop}%
\bibitem [{\citenamefont {Meier}\ \emph {et~al.}(2019)\citenamefont {Meier},
  \citenamefont {Ang'ong'a}, \citenamefont {An},\ and\ \citenamefont
  {Gadway}}]{meier_exploring_2019}%
  \BibitemOpen
  \bibfield  {author} {\bibinfo {author} {\bibfnamefont {E.~J.}\ \bibnamefont
  {Meier}}, \bibinfo {author} {\bibfnamefont {J.}~\bibnamefont {Ang'ong'a}},
  \bibinfo {author} {\bibfnamefont {F.~A.}\ \bibnamefont {An}},\ and\ \bibinfo
  {author} {\bibfnamefont {B.}~\bibnamefont {Gadway}},\ }\bibfield  {title}
  {\bibinfo {title} {Exploring quantum signatures of chaos on a {{Floquet}}
  synthetic lattice},\ }\href {https://doi.org/10.1103/PhysRevA.100.013623}
  {\bibfield  {journal} {\bibinfo  {journal} {Physical Review A}\ }\textbf
  {\bibinfo {volume} {100}},\ \bibinfo {pages} {013623} (\bibinfo {year}
  {2019})},\ \Eprint {https://arxiv.org/abs/1705.06714} {arXiv:1705.06714}
  \BibitemShut {NoStop}%
\bibitem [{\citenamefont {Schnell}\ and\ \citenamefont
  {Spiess}(2001)}]{schnell_high-resolution_2001}%
  \BibitemOpen
  \bibfield  {author} {\bibinfo {author} {\bibfnamefont {I.}~\bibnamefont
  {Schnell}}\ and\ \bibinfo {author} {\bibfnamefont {H.~W.}\ \bibnamefont
  {Spiess}},\ }\bibfield  {title} {{\selectlanguage {en}\bibinfo {title}
  {High-{{Resolution 1H NMR Spectroscopy}} in the {{Solid State}}: {{Very Fast
  Sample Rotation}} and {{Multiple}}-{{Quantum Coherences}}}},\ }\href
  {https://doi.org/10.1006/jmre.2001.2336} {\bibfield  {journal} {\bibinfo
  {journal} {Journal of Magnetic Resonance}\ }\textbf {\bibinfo {volume}
  {151}},\ \bibinfo {pages} {153} (\bibinfo {year} {2001})}\BibitemShut
  {NoStop}%
\bibitem [{\citenamefont {S{\'a}nchez}\ \emph {et~al.}(2014)\citenamefont
  {S{\'a}nchez}, \citenamefont {Acosta}, \citenamefont {Levstein},
  \citenamefont {Pastawski},\ and\ \citenamefont
  {Chattah}}]{sanchez_clustering_2014}%
  \BibitemOpen
  \bibfield  {author} {\bibinfo {author} {\bibfnamefont {C.~M.}\ \bibnamefont
  {S{\'a}nchez}}, \bibinfo {author} {\bibfnamefont {R.~H.}\ \bibnamefont
  {Acosta}}, \bibinfo {author} {\bibfnamefont {P.~R.}\ \bibnamefont
  {Levstein}}, \bibinfo {author} {\bibfnamefont {H.~M.}\ \bibnamefont
  {Pastawski}},\ and\ \bibinfo {author} {\bibfnamefont {A.~K.}\ \bibnamefont
  {Chattah}},\ }\bibfield  {title} {\bibinfo {title} {Clustering and
  decoherence of correlated spins under double quantum dynamics},\ }\href
  {https://doi.org/10.1103/PhysRevA.90.042122} {\bibfield  {journal} {\bibinfo
  {journal} {Physical Review A}\ }\textbf {\bibinfo {volume} {90}},\ \bibinfo
  {pages} {042122} (\bibinfo {year} {2014})}\BibitemShut {NoStop}%
\bibitem [{\citenamefont {S{\'a}nchez}\ \emph {et~al.}(2020)\citenamefont
  {S{\'a}nchez}, \citenamefont {Chattah}, \citenamefont {Wei}, \citenamefont
  {Buljubasich}, \citenamefont {Cappellaro},\ and\ \citenamefont
  {Pastawski}}]{sanchez_perturbation_2020}%
  \BibitemOpen
  \bibfield  {author} {\bibinfo {author} {\bibfnamefont {C.~M.}\ \bibnamefont
  {S{\'a}nchez}}, \bibinfo {author} {\bibfnamefont {A.~K.}\ \bibnamefont
  {Chattah}}, \bibinfo {author} {\bibfnamefont {K.~X.}\ \bibnamefont {Wei}},
  \bibinfo {author} {\bibfnamefont {L.}~\bibnamefont {Buljubasich}}, \bibinfo
  {author} {\bibfnamefont {P.}~\bibnamefont {Cappellaro}},\ and\ \bibinfo
  {author} {\bibfnamefont {H.~M.}\ \bibnamefont {Pastawski}},\ }\bibfield
  {title} {\bibinfo {title} {Perturbation {{Independent Decay}} of the
  {{Loschmidt Echo}} in a {{Many}}-{{Body System}}},\ }\href
  {https://doi.org/10.1103/PhysRevLett.124.030601} {\bibfield  {journal}
  {\bibinfo  {journal} {Physical Review Letters}\ }\textbf {\bibinfo {volume}
  {124}},\ \bibinfo {pages} {030601} (\bibinfo {year} {2020})}\BibitemShut
  {NoStop}%
\bibitem [{\citenamefont {Joshi}\ \emph {et~al.}(2020)\citenamefont {Joshi},
  \citenamefont {Elben}, \citenamefont {Vermersch}, \citenamefont {Brydges},
  \citenamefont {Maier}, \citenamefont {Zoller}, \citenamefont {Blatt},\ and\
  \citenamefont {Roos}}]{joshi_quantum_2020}%
  \BibitemOpen
  \bibfield  {author} {\bibinfo {author} {\bibfnamefont {M.~K.}\ \bibnamefont
  {Joshi}}, \bibinfo {author} {\bibfnamefont {A.}~\bibnamefont {Elben}},
  \bibinfo {author} {\bibfnamefont {B.}~\bibnamefont {Vermersch}}, \bibinfo
  {author} {\bibfnamefont {T.}~\bibnamefont {Brydges}}, \bibinfo {author}
  {\bibfnamefont {C.}~\bibnamefont {Maier}}, \bibinfo {author} {\bibfnamefont
  {P.}~\bibnamefont {Zoller}}, \bibinfo {author} {\bibfnamefont
  {R.}~\bibnamefont {Blatt}},\ and\ \bibinfo {author} {\bibfnamefont {C.~F.}\
  \bibnamefont {Roos}},\ }\bibfield  {title} {\bibinfo {title} {Quantum
  {{Information Scrambling}} in a {{Trapped-Ion Quantum Simulator}} with
  {{Tunable Range Interactions}}},\ }\href
  {https://doi.org/10.1103/PhysRevLett.124.240505} {\bibfield  {journal}
  {\bibinfo  {journal} {Physical Review Letters}\ }\textbf {\bibinfo {volume}
  {124}},\ \bibinfo {pages} {240505} (\bibinfo {year} {2020})}\BibitemShut
  {NoStop}%
\bibitem [{\citenamefont {Choi}\ \emph {et~al.}(2017)\citenamefont {Choi},
  \citenamefont {Choi}, \citenamefont {Landig}, \citenamefont {Kucsko},
  \citenamefont {Zhou}, \citenamefont {Isoya}, \citenamefont {Jelezko},
  \citenamefont {Onoda}, \citenamefont {Sumiya}, \citenamefont {Khemani},
  \citenamefont {von Keyserlingk}, \citenamefont {Yao}, \citenamefont
  {Demler},\ and\ \citenamefont {Lukin}}]{Lukin17}%
  \BibitemOpen
  \bibfield  {author} {\bibinfo {author} {\bibfnamefont {S.}~\bibnamefont
  {Choi}}, \bibinfo {author} {\bibfnamefont {J.}~\bibnamefont {Choi}}, \bibinfo
  {author} {\bibfnamefont {R.}~\bibnamefont {Landig}}, \bibinfo {author}
  {\bibfnamefont {G.}~\bibnamefont {Kucsko}}, \bibinfo {author} {\bibfnamefont
  {H.}~\bibnamefont {Zhou}}, \bibinfo {author} {\bibfnamefont {J.}~\bibnamefont
  {Isoya}}, \bibinfo {author} {\bibfnamefont {F.}~\bibnamefont {Jelezko}},
  \bibinfo {author} {\bibfnamefont {S.}~\bibnamefont {Onoda}}, \bibinfo
  {author} {\bibfnamefont {H.}~\bibnamefont {Sumiya}}, \bibinfo {author}
  {\bibfnamefont {V.}~\bibnamefont {Khemani}}, \bibinfo {author} {\bibfnamefont
  {C.}~\bibnamefont {von Keyserlingk}}, \bibinfo {author} {\bibfnamefont
  {N.~Y.}\ \bibnamefont {Yao}}, \bibinfo {author} {\bibfnamefont
  {E.}~\bibnamefont {Demler}},\ and\ \bibinfo {author} {\bibfnamefont {M.~D.}\
  \bibnamefont {Lukin}},\ }\bibfield  {title} {\bibinfo {title} {Observation of
  discrete time-crystalline order in a disordered dipolar many-body system},\
  }\href {https://doi.org/10.1038/nature21426} {\bibfield  {journal} {\bibinfo
  {journal} {Nature}\ }\textbf {\bibinfo {volume} {543}},\ \bibinfo {pages}
  {221 EP } (\bibinfo {year} {2017})}\BibitemShut {NoStop}%
\bibitem [{\citenamefont {Blatt}\ and\ \citenamefont {Roos}(2012)}]{Blatt12}%
  \BibitemOpen
  \bibfield  {author} {\bibinfo {author} {\bibfnamefont {R.}~\bibnamefont
  {Blatt}}\ and\ \bibinfo {author} {\bibfnamefont {C.~F.}\ \bibnamefont
  {Roos}},\ }\bibfield  {title} {\bibinfo {title} {Quantum simulations with
  trapped ions},\ }\href {https://doi.org/10.1038/nphys2252} {\bibfield
  {journal} {\bibinfo  {journal} {Nature Physics}\ }\textbf {\bibinfo {volume}
  {8}},\ \bibinfo {pages} {277 EP } (\bibinfo {year} {2012})}\BibitemShut
  {NoStop}%
\bibitem [{\citenamefont {Britton}\ \emph {et~al.}(2012)\citenamefont
  {Britton}, \citenamefont {Sawyer}, \citenamefont {Keith}, \citenamefont
  {Wang}, \citenamefont {Freericks}, \citenamefont {Uys}, \citenamefont
  {Biercuk},\ and\ \citenamefont {Bollinger}}]{Britton12}%
  \BibitemOpen
  \bibfield  {author} {\bibinfo {author} {\bibfnamefont {J.~W.}\ \bibnamefont
  {Britton}}, \bibinfo {author} {\bibfnamefont {B.~C.}\ \bibnamefont {Sawyer}},
  \bibinfo {author} {\bibfnamefont {A.~C.}\ \bibnamefont {Keith}}, \bibinfo
  {author} {\bibfnamefont {C.~C.~J.}\ \bibnamefont {Wang}}, \bibinfo {author}
  {\bibfnamefont {J.~K.}\ \bibnamefont {Freericks}}, \bibinfo {author}
  {\bibfnamefont {H.}~\bibnamefont {Uys}}, \bibinfo {author} {\bibfnamefont
  {M.~J.}\ \bibnamefont {Biercuk}},\ and\ \bibinfo {author} {\bibfnamefont
  {J.~J.}\ \bibnamefont {Bollinger}},\ }\bibfield  {title} {\bibinfo {title}
  {{Engineered two-dimensional Ising interactions in a trapped-ion quantum
  simulator with hundreds of spins}},\ }\href
  {http://dx.doi.org/10.1038/nature10981} {\bibfield  {journal} {\bibinfo
  {journal} {Nature}\ }\textbf {\bibinfo {volume} {484}},\ \bibinfo {pages}
  {489} (\bibinfo {year} {2012})}\BibitemShut {NoStop}%
\bibitem [{\citenamefont {Brunet}\ \emph {et~al.}(2006)\citenamefont {Brunet},
  \citenamefont {Derrida}, \citenamefont {Mueller},\ and\ \citenamefont
  {Munier}}]{brunet_phenomenological_2006}%
  \BibitemOpen
  \bibfield  {author} {\bibinfo {author} {\bibfnamefont {E.}~\bibnamefont
  {Brunet}}, \bibinfo {author} {\bibfnamefont {B.}~\bibnamefont {Derrida}},
  \bibinfo {author} {\bibfnamefont {A.~H.}\ \bibnamefont {Mueller}},\ and\
  \bibinfo {author} {\bibfnamefont {S.}~\bibnamefont {Munier}},\ }\bibfield
  {title} {{\selectlanguage {en}\bibinfo {title} {Phenomenological theory
  giving the full statistics of the position of fluctuating pulled fronts}},\
  }\href {https://doi.org/10.1103/PhysRevE.73.056126} {\bibfield  {journal}
  {\bibinfo  {journal} {Physical Review E}\ }\textbf {\bibinfo {volume} {73}},\
  \bibinfo {pages} {056126} (\bibinfo {year} {2006})}\BibitemShut {NoStop}%
\bibitem [{\citenamefont {Brunet}\ and\ \citenamefont
  {Derrida}(1997)}]{brunet_shift_1997}%
  \BibitemOpen
  \bibfield  {author} {\bibinfo {author} {\bibfnamefont {E.}~\bibnamefont
  {Brunet}}\ and\ \bibinfo {author} {\bibfnamefont {B.}~\bibnamefont
  {Derrida}},\ }\bibfield  {title} {\bibinfo {title} {Shift in the velocity of
  a front due to a cutoff},\ }\href {https://doi.org/10.1103/PhysRevE.56.2597}
  {\bibfield  {journal} {\bibinfo  {journal} {Physical Review E}\ }\textbf
  {\bibinfo {volume} {56}},\ \bibinfo {pages} {2597} (\bibinfo {year}
  {1997})}\BibitemShut {NoStop}%
\bibitem [{\citenamefont {{del-Castillo-Negrete}}\ \emph
  {et~al.}(2003)\citenamefont {{del-Castillo-Negrete}}, \citenamefont
  {Carreras},\ and\ \citenamefont {Lynch}}]{del-castillo-negrete_front_2003}%
  \BibitemOpen
  \bibfield  {author} {\bibinfo {author} {\bibfnamefont {D.}~\bibnamefont
  {{del-Castillo-Negrete}}}, \bibinfo {author} {\bibfnamefont {B.~A.}\
  \bibnamefont {Carreras}},\ and\ \bibinfo {author} {\bibfnamefont {V.~E.}\
  \bibnamefont {Lynch}},\ }\bibfield  {title} {\bibinfo {title} {Front
  {{Dynamics}} in {{Reaction}}-{{Diffusion Systems}} with {{Levy Flights}}: {{A
  Fractional Diffusion Approach}}},\ }\href
  {https://doi.org/10.1103/PhysRevLett.91.018302} {\bibfield  {journal}
  {\bibinfo  {journal} {Physical Review Letters}\ }\textbf {\bibinfo {volume}
  {91}},\ \bibinfo {pages} {018302} (\bibinfo {year} {2003})}\BibitemShut
  {NoStop}%
\bibitem [{\citenamefont {Dumortier}\ \emph {et~al.}(2007)\citenamefont
  {Dumortier}, \citenamefont {Popovi{\'c}},\ and\ \citenamefont
  {Kaper}}]{dumortier_critical_2007}%
  \BibitemOpen
  \bibfield  {author} {\bibinfo {author} {\bibfnamefont {F.}~\bibnamefont
  {Dumortier}}, \bibinfo {author} {\bibfnamefont {N.}~\bibnamefont
  {Popovi{\'c}}},\ and\ \bibinfo {author} {\bibfnamefont {T.~J.}\ \bibnamefont
  {Kaper}},\ }\bibfield  {title} {{\selectlanguage {en}\bibinfo {title} {The
  critical wave speed for the
  {{Fisher}}\textendash{{Kolmogorov}}\textendash{{Petrowskii}}\textendash{{Piscounov}}
  equation with cut-off}},\ }\href {https://doi.org/10.1088/0951-7715/20/4/004}
  {\bibfield  {journal} {\bibinfo  {journal} {Nonlinearity}\ }\textbf {\bibinfo
  {volume} {20}},\ \bibinfo {pages} {855} (\bibinfo {year} {2007})}\BibitemShut
  {NoStop}%
\bibitem [{\citenamefont {Coulon}\ and\ \citenamefont
  {Roquejoffre}(2012)}]{coulon_transition_2012}%
  \BibitemOpen
  \bibfield  {author} {\bibinfo {author} {\bibfnamefont {A.-C.}\ \bibnamefont
  {Coulon}}\ and\ \bibinfo {author} {\bibfnamefont {J.-M.}\ \bibnamefont
  {Roquejoffre}},\ }\bibfield  {title} {\bibinfo {title} {Transition {{Between
  Linear}} and {{Exponential Propagation}} in {{Fisher}}-{{KPP Type
  Reaction}}-{{Diffusion Equations}}},\ }\href
  {https://doi.org/10.1080/03605302.2012.718024} {\bibfield  {journal}
  {\bibinfo  {journal} {Communications in Partial Differential Equations}\
  }\textbf {\bibinfo {volume} {37}},\ \bibinfo {pages} {2029} (\bibinfo {year}
  {2012})}\BibitemShut {NoStop}%
\bibitem [{\citenamefont {Brockmann}\ and\ \citenamefont
  {Hufnagel}(2007)}]{brockmann_front_2007}%
  \BibitemOpen
  \bibfield  {author} {\bibinfo {author} {\bibfnamefont {D.}~\bibnamefont
  {Brockmann}}\ and\ \bibinfo {author} {\bibfnamefont {L.}~\bibnamefont
  {Hufnagel}},\ }\bibfield  {title} {\bibinfo {title} {Front {{Propagation}} in
  {{Reaction}}-{{Superdiffusion Dynamics}}: {{Taming L}}\textbackslash{}'evy
  {{Flights}} with {{Fluctuations}}},\ }\href
  {https://doi.org/10.1103/PhysRevLett.98.178301} {\bibfield  {journal}
  {\bibinfo  {journal} {Physical Review Letters}\ }\textbf {\bibinfo {volume}
  {98}},\ \bibinfo {pages} {178301} (\bibinfo {year} {2007})}\BibitemShut
  {NoStop}%
\bibitem [{\citenamefont {Hallatschek}\ and\ \citenamefont
  {Fisher}(2014)}]{hallatschek_acceleration_2014}%
  \BibitemOpen
  \bibfield  {author} {\bibinfo {author} {\bibfnamefont {O.}~\bibnamefont
  {Hallatschek}}\ and\ \bibinfo {author} {\bibfnamefont {D.~S.}\ \bibnamefont
  {Fisher}},\ }\bibfield  {title} {{\selectlanguage {en}\bibinfo {title}
  {Acceleration of evolutionary spread by long-range dispersal}},\ }\href
  {https://www.pnas.org/content/111/46/E4911} {\bibfield  {journal} {\bibinfo
  {journal} {Proceedings of the National Academy of Sciences}\ }\textbf
  {\bibinfo {volume} {111}},\ \bibinfo {pages} {E4911} (\bibinfo {year}
  {2014})}\BibitemShut {NoStop}%
\bibitem [{\citenamefont {Chatterjee}\ and\ \citenamefont
  {Dey}(2013)}]{chatterjee_multiple_2013}%
  \BibitemOpen
  \bibfield  {author} {\bibinfo {author} {\bibfnamefont {S.}~\bibnamefont
  {Chatterjee}}\ and\ \bibinfo {author} {\bibfnamefont {P.~S.}\ \bibnamefont
  {Dey}},\ }\bibfield  {title} {\bibinfo {title} {Multiple phase transitions in
  long-range first-passage percolation on square lattices},\ }\href
  {https://arxiv.org/abs/1309.5757v2} {\bibfield  {journal} {\bibinfo
  {journal} {arXiv:1309.5757}\ } (\bibinfo {year} {2013})}\BibitemShut
  {NoStop}%
\bibitem [{\citenamefont {Mancinelli}\ \emph {et~al.}(2002)\citenamefont
  {Mancinelli}, \citenamefont {Vergni},\ and\ \citenamefont
  {Vulpiani}}]{mancinelli_superfast_2002}%
  \BibitemOpen
  \bibfield  {author} {\bibinfo {author} {\bibfnamefont {R.}~\bibnamefont
  {Mancinelli}}, \bibinfo {author} {\bibfnamefont {D.}~\bibnamefont {Vergni}},\
  and\ \bibinfo {author} {\bibfnamefont {A.}~\bibnamefont {Vulpiani}},\
  }\bibfield  {title} {{\selectlanguage {en}\bibinfo {title} {Superfast front
  propagation in reactive systems with non-{{Gaussian}} diffusion}},\ }\href
  {https://doi.org/10.1209/epl/i2002-00251-7} {\bibfield  {journal} {\bibinfo
  {journal} {Europhysics Letters}\ }\textbf {\bibinfo {volume} {60}},\ \bibinfo
  {pages} {532} (\bibinfo {year} {2002})}\BibitemShut {NoStop}%
\bibitem [{\citenamefont {{Xu}}\ and\ \citenamefont
  {{Swingle}}(2018)}]{Xu_Swingle_2018}%
  \BibitemOpen
  \bibfield  {author} {\bibinfo {author} {\bibfnamefont {S.}~\bibnamefont
  {{Xu}}}\ and\ \bibinfo {author} {\bibfnamefont {B.}~\bibnamefont
  {{Swingle}}},\ }\bibfield  {title} {\bibinfo {title} {{Accessing scrambling
  using matrix product operators}},\ }\href@noop {} {\bibfield  {journal}
  {\bibinfo  {journal} {arXiv e-prints}\ ,\ \bibinfo {eid} {arXiv:1802.00801}}
  (\bibinfo {year} {2018})},\ \Eprint {https://arxiv.org/abs/1802.00801}
  {arXiv:1802.00801 [quant-ph]} \BibitemShut {NoStop}%
\bibitem [{\citenamefont {Kilbas}\ \emph {et~al.}(2006)\citenamefont {Kilbas},
  \citenamefont {Srivastava},\ and\ \citenamefont
  {Trujillo}}]{kilbas_theory_2006}%
  \BibitemOpen
  \bibfield  {author} {\bibinfo {author} {\bibfnamefont {A.~A.}\ \bibnamefont
  {Kilbas}}, \bibinfo {author} {\bibfnamefont {H.~M.}\ \bibnamefont
  {Srivastava}},\ and\ \bibinfo {author} {\bibfnamefont {J.~J.}\ \bibnamefont
  {Trujillo}},\ }\href@noop {} {\emph {\bibinfo {title} {Theory and
  {{Applications}} of {{Fractional Differential Equations}}, {{Volume}} 204
  ({{North-Holland Mathematics Studies}})}}}\ (\bibinfo  {publisher} {{Elsevier
  Science Inc.}},\ \bibinfo {address} {{USA}},\ \bibinfo {year}
  {2006})\BibitemShut {NoStop}%
\end{thebibliography}%


%

\appendix

\section{Consistent solutions of the effective model at $\mu = 1$}
\label{app:consistent_mu_1}

In this appendix, we prove that the consistent solution to Eq.~\eqref{eq:l_t} at $\mu = 1$ is $\ell(t)$. In other words, we will show that
\begin{equation}
\lim_{ t \rightarrow \infty} \sum_{\tau = 1}^{t} \frac{1}{( t + 1) \ln ( t + 1 ) - \tau \ln \tau}
\end{equation}
exists and is a positive real number. We pull out a factor of $t$ and convert it into an integral
\begin{equation}
\begin{aligned}
  &\sum_{\tau = 1}^{T} \frac{1}{( t + 1) \ln ( t + 1 ) - \tau \ln \tau} \\
  =&  \sum_{\tau = 1}^{t} \frac{1}{t} \frac{1}{( 1 + \frac{1}{t}) \ln ( t + 1 ) - \frac{\tau}{t}  \ln \frac{\tau}{t} t} \\
  &\approx \int_{0}^{1} \frac{1}{( 1 + \frac{1}{t}) \ln ( t + 1 ) - \tau \ln \tau t} d\tau.
\end{aligned}
\end{equation}
We replace the lower limit of $\frac{1}{t}$ by $0$ because it only introduces errors of order $\mathcal{O}(\frac{1}{t})$. The possible singular point is at $\tau = 1$. To have a clearer view, we make a change $\tau \rightarrow 1 - \tau$; the integral becomes
\begin{equation}
\int_0^1 \frac{1}{( 1 + \frac{1}{t} ) \ln ( t + 1 ) - (1 - \tau) \ln ( 1 - \tau ) t  } d \tau.
\end{equation}
The denominator at small $\tau$ with the large $t$ limit is
\begin{equation}
\begin{aligned}
  &( 1 + \frac{1}{t} ) \ln t + ( 1 + \frac{1}{t} ) \ln ( 1 + \frac{1}{t} )  - \ln t + \tau ( \ln t + 1 ) \\
 &= \frac{\ln t}{ t}  + \tau (\ln t + 1 )  + \mathcal{O}(\frac{1}{t} ).
\end{aligned}
\end{equation}
The integration around $t = 0$ will give
\begin{equation}
  - \frac{1}{ \ln t + 1}  \ln ( \frac{ \ln t}{t} + \mathcal{O}(\frac{1}{t}) ) \rightarrow 1 \quad {\rm when} \quad  t \rightarrow \infty.
\end{equation}
Hence the sum converges when $\ell(t) \sim t \ln t $.

\end{document}